\documentclass[11pt]{article}
\usepackage{makecell}

\usepackage[preprint]{acl}

\usepackage{times}
\usepackage{latexsym}

\usepackage[T1]{fontenc}

\usepackage[utf8]{inputenc}
\usepackage{multirow}
\usepackage{microtype}
\usepackage{amsmath}
\usepackage{amssymb}
\usepackage{inconsolata}
\usepackage{booktabs}
\usepackage{tabularx}
\usepackage[table]{xcolor}

\usepackage{graphicx}
\usepackage{enumitem}
\usepackage{algorithm}
\usepackage{algorithmic}
\usepackage{soul}
\sethlcolor{yellow!30}

\title{MECoBench: A Systematic Study of Multimodal Agent Collaboration in Embodied Environments}

\author{
 \textbf{Qingyun Liu\textsuperscript{1\ensuremath{*}}},
 \textbf{Jiwen Zhang\textsuperscript{1\ensuremath{*}}},
 \textbf{Jingyi Hu\textsuperscript{1}},
 \textbf{Siyuan Wang\textsuperscript{3\ensuremath{\dagger}}},
 \textbf{Zhongyu Wei\textsuperscript{1,2\ensuremath{\dagger}}}
\\
 \textsuperscript{1}Fudan University,
 \textsuperscript{2}Shanghai Innovation Institute,
 \textsuperscript{3}The Chinese University of Hong Kong
\\
\normalsize{\texttt{\{qyliu25,jiwenzhang21,jyhu25\}@m.fudan.edu.cn}}\\
    \normalsize{\texttt{siyuanwang@cuhk.edu.hk, zywei@fudan.edu.cn}}
}

\begin{document}
\maketitle
\begingroup
\renewcommand{\thefootnote}{\fnsymbol{footnote}}
\footnotetext[1]{Equal contribution.}
\footnotetext[2]{Corresponding author.}
\endgroup
\begin{abstract}
Recent multimodal large language models (MLLMs) have strong potential as embodied agents, but their ability to collaborate in visually grounded environments remains underexplored. To address this gap, we introduce MECoBench, a multimodal embodied cooperation benchmark with an evaluation platform spanning diverse real-world tasks, two cooperation structures, and three collaboration modes. Through extensive experiments across various MLLMs, we summarize three key findings: (i) Collaboration generally improves embodied task completion, but its benefits depend on balancing collaborative gains against coordination complexity. (ii) Communication is essential to collaboration gains, while the best collaboration mode depends on team size and model capability. (iii) Moreover, collaboration improves robustness under noisy priors and exploration conditions. Generally, MECoBench provides a systematic testbed for understanding the mechanisms and limits of multimodal embodied collaboration. Code and dataset are available at \url{https:/github.com/q-i-n-g/MECoBench}.
\end{abstract}

\section{Introduction}

\begin{figure}[t]
\setlength{\abovecaptionskip}{0pt}
    \centering
    \includegraphics[width=0.95\linewidth]{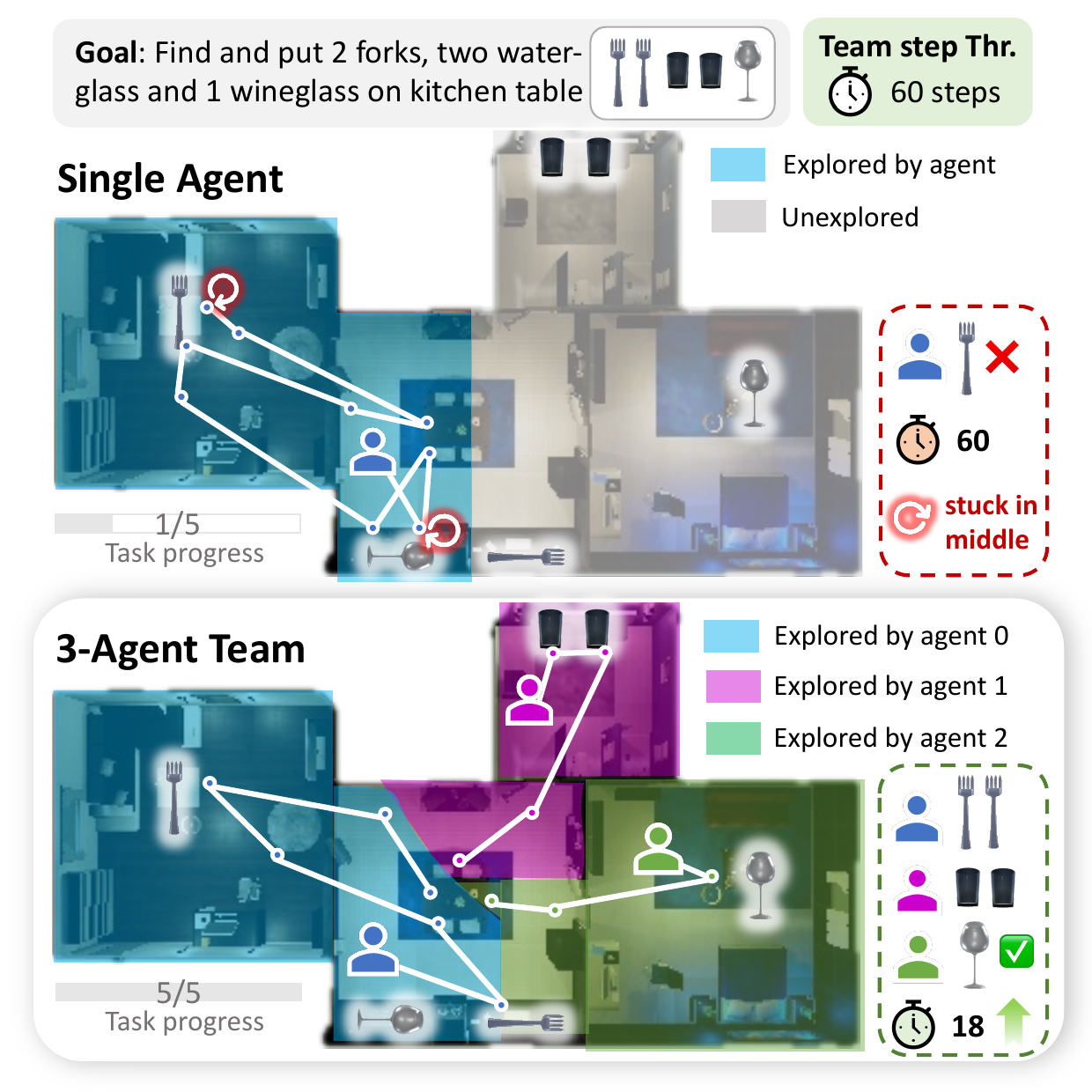}
     \caption{\textbf{An illustration of realistic scenarios}, where cross-modal multi-agent collaboration significantly improves the efficiency compared with single-agent.
     }
    \label{fig:intro}
\end{figure}

\begin{table*}[t]
\centering
\small
\setlength{\abovecaptionskip}{2pt}
\setlength{\tabcolsep}{3pt}
\renewcommand{\arraystretch}{1.12}
\caption{\textbf{Comparison with related benchmarks.}
Multimodal: Whether agents receive multimodal observations rather than
text-only observations.
Collaboration structure (\textit{Structure}): I and D denote independent
and interdependent collaboration.
Collaboration mode (\textit{Collab.}): I, C, and D denote isolated,
centralized, and decentralized collaboration.
Communication medium (\textit{Comm.}): T and V denote textual and visual communication.
``--'' indicates no explicit inter-agent communication.}
\label{tab:benchmark_comparison}

\begin{tabular}{lccccccc}
\toprule
\textbf{Benchmark}
& \textbf{Env.}
& \textbf{Multimodal}
& \textbf{Team size}
& \textbf{Structure}
& \textbf{Collab.}
& \textbf{Comm.}
& \textbf{\#Test cases} \\
\midrule

FurnMove~\cite{10.1007/978-3-030-58558-7_28}
& 3D & \checkmark & Fixed (2) & D & D & -- & 1,000 \\

CoELA~\cite{ICLR2024_54b8b4e0}
& 3D & \texttimes & Fixed (2) & I & D & T & 44 \\

RoCoBench~\cite{10610855}
& 3D & \texttimes & Fixed (2--3) & I/D & D & T & 6 \\

VillagerBench~\cite{dong-etal-2024-villageragent} & 3D & \texttimes & Variable (2--3) & I/D & C & -- & 225 \\

PARTNR~\cite{ICLR2025_a3cf318f}
& 3D & \texttimes & Fixed (2) & I & C/D & -- & 1,000 \\

TeamCraft~\cite{long2024teamcraftbenchmarkmultimodalmultiagent}
& 3D & \checkmark & Variable (2--4) & I/D & C/D & -- & 950 \\

Collab-Overcooked~\cite{sun-etal-2025-collab}
& 2D & \texttimes & Fixed (2) & D & D & T & 30 \\

MineCollab~\cite{white2025collaboratingactionactionmultiagent}
& 3D & \texttimes & Variable (2--5) & I & D & T & 184 \\

COOP$^2$~\cite{yang2026coop2definingobservingrepairing}
& 2D & \texttimes & Variable (3/6) & I/D & I/C/D & T & 36 \\

\midrule
\textbf{MECoBench}
& \textbf{3D}
& \checkmark
& \textbf{Variable (1--5)}
& \textbf{I/D}
& \textbf{I/C/D}
& \textbf{T/V}
& \textbf{192} \\

\bottomrule
\end{tabular}
\vspace{-4pt}
\end{table*}

Recent multimodal large language models (MLLMs)~\cite{anthropic2026claudeopus47systemcard,openai2026gpt54thinking,googledeepmind2026gemini31pro} have demonstrated strong vision-language understanding and reasoning capabilities, making them promising foundations for embodied agents in interactive environments~\cite{NEURIPS2023_4ec43957, Szot_2025_CVPR}.
While most existing MLLM-based studies focus on single-agent embodied intelligence~\cite{yang2025embodiedbench,liu2025visualagentbench,Zhang_2025_ICCV}, many real-world tasks such as household assistance require multiple agents to cooperate~\cite{sycara1998multiagent}, as shown in Figure~\ref{fig:intro}. 
Previous works on pure language-based multi-agent framework~\cite{schmidgall-etal-2025-agent,ICLR2024_6507b115} have proved that such cooperation could improve efficiency and overcome individual capability limits. However, the potential of multi-agent collaboration under multimodal embodied settings remains an underexplored area. 

Answering this is non-trivial, because embodied multi-agent collaboration differs fundamentally from text-based collaboration. Existing multi-agent benchmarks rely on textual task descriptions, shared symbolic states, or pre-processed observations, reducing collaboration to plan coordination or dialogue~\cite{zhu-etal-2025-multiagentbench, sun-etal-2025-collab,ICLR2024_54b8b4e0,yang2026coop2definingobservingrepairing,agashe-etal-2025-llm}. By contrast, embodied agents must coordinate from partial, dynamically changing visual observations while exploring environments, resolving perceptual inconsistencies, avoiding spatial conflicts, and acting under physical constraints~\cite{Shridhar_2020_CVPR}. This tightly couples perception, exploration, communication, and coordination, leading to failure modes and emergent behaviors that text-based benchmarks cannot capture~\cite{Feng2026}. Therefore, it is valuable to systematically study MLLM-based multi-agent collaboration in embodied settings, calling for a new benchmark that supports visually grounded interaction and diverse cooperation patterns.

To address this gap, we introduce MECoBench, a \textbf{m}ultimodal \textbf{e}mbodied \textbf{co}operation benchmark together with an evaluation platform for systematically analyzing visually grounded embodied multi-agent cooperation. 
MECoBench contains diverse embodied tasks covering eight types of common real-world activities, and introduces two cooperation structures to support realistic evaluation. In parallel cooperation, agents can work on different subtasks simultaneously, whereas in space-constrained sequential cooperation, agents must coordinate in a chained manner to complete the task.
Beyond task construction, our platform supports controlled variation along multiple dimensions, including collaboration mode, team size and exploration difficulty. This design enables fine-grained analysis of not only whether collaboration helps, but also when, why, and through which mechanisms it improves embodied task solving.

Based on MECoBench, our work aims to answer three research questions progressively: 
(i) To what extent does multi-agent collaboration benefit MLLMs on embodied tasks, and how does this benefit change as the team size scales? (ii) How do different collaboration modes and communication mechanisms shape collaboration outcomes? (iii) Does collaboration remain effective and robust under challenging information conditions? To answer these questions, we evaluate a wide range of MLLMs across model families and parameter scales, and obtain the following findings:
\begin{figure*}[t]
\setlength{\abovecaptionskip}{-4pt}
    \centering
    \includegraphics[width=0.9\textwidth]{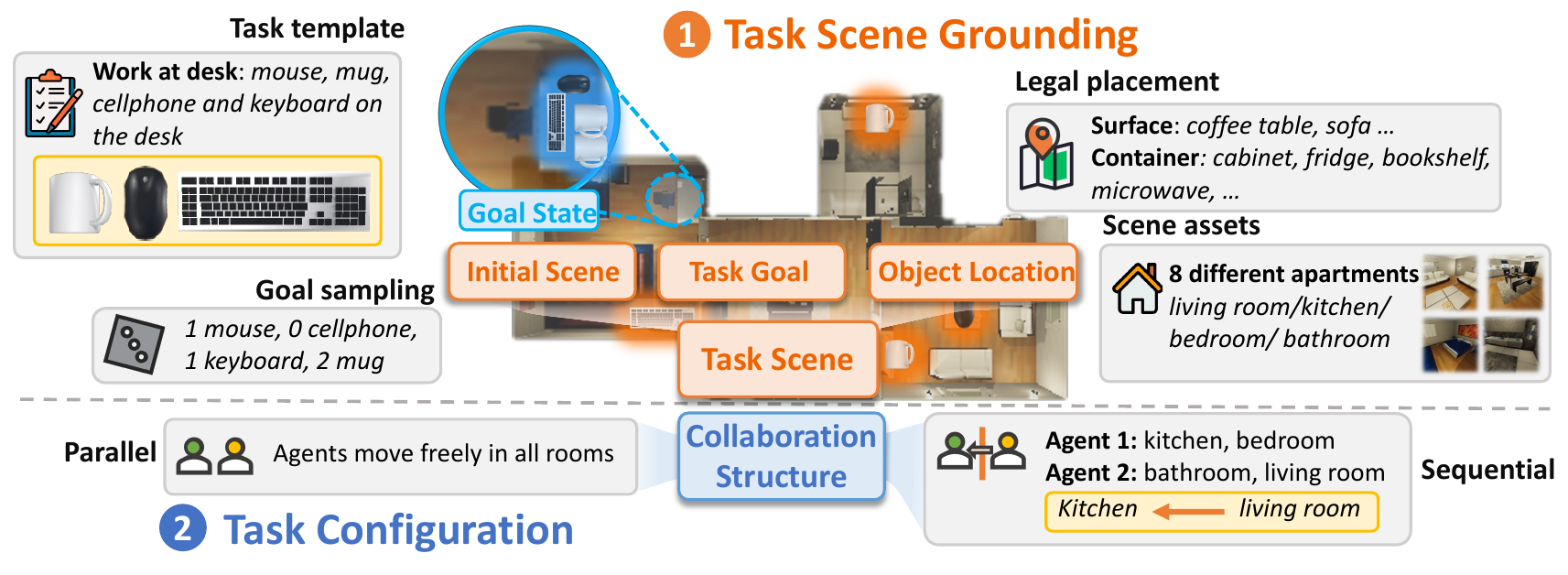}
    \caption{\textbf{Data construction pipeline of MECoBench}. 
    Each task is first grounded from a high-level task into a concrete scene, then set collaboration configuration for parallel or sequential execution.}
    \label{fig:task_generation}
\vspace{-8pt}
\end{figure*}

\begin{itemize}[leftmargin=*,partopsep=0pt,topsep=0pt]
\setlength{\itemsep}{0pt}
\setlength{\parsep}{0pt}
\setlength{\parskip}{0pt}
    \item \textbf{Multi-agent collaboration generally improves embodied task completion, but its benefits depend on balancing gains against coordination complexity.} Most MLLMs exhibit basic collaborative capability and benefit from a second agent participation. As the team grows further, however, performance saturates and then degrades, forming an inverted-U trend and indicating that moderate team sizes provide the best trade-off between parallelism and coordination overhead. Encouragingly, larger teams remain more robust as task complexity increases.
    
    \item \textbf{Communication is the key driver of collaboration gains, and the best mode is contingent on both team size and model capability.} Removing communication consistently degrades performance, with 
    substantially larger drops on sequential tasks and larger teams. Centralized coordination is most effective in small teams but saturates earlier than decentralized coordination as the team scales. Beyond textual messaging, shared memory yields more efficient collaboration and benefits loosely coupled parallel tasks, while vision-augmented leadership improves coordination efficiency---provided the leader model is capable enough to leverage visual input.
    
    \item \textbf{Collaboration further enhances robustness under challenging information conditions.} Two-agent collaboration still improves task completion when location priors are removed and agents must explore from scratch, and yields its largest relative gains under noisy priors, showing that multi-agent teams can compensate for misleading information through communication and distributed exploration. Available task information further amplifies the benefits of collaboration.
\end{itemize}

\section{MECoBench}
To investigate MLLM-based multi-agent collaboration under embodied settings, we propose \textbf{MECoBench}, a comprehensive benchmark with an interactive evaluation platform. MECoBench is built upon VirtualHome~\cite{Puig_2018_CVPR}, a realistic household simulator covering a wide range of everyday scenarios, and spans different collaboration structures, variable team sizes, diverse collaboration modes and communication mechanisms. Table~\ref{tab:benchmark_comparison} compares MECoBench with existing embodied multi-agent collaboration benchmarks.

\subsection{Benchmark Construction}
To cover common real-world activities, we define eight semantic task templates (detailed definition in Table~\ref{tab:task_templates}), including five adapted from WAH~\cite{puig2021watchandhelp} and three newly designed. Beyond semantic diversity, we introduce two collaboration structure to capture different forms of spatial accessibility in the real world. In the \textit{\textbf{parallel}} setting, all agents can freely access all rooms and complete subgoals independently and concurrently, modeling cases where individuals operate in a shared space. In the \textit{\textbf{sequential}} setting, agents are assigned to disjoint room zones, making inter-agent object transfers necessary and reflecting situations where individuals have access to different regions and must coordinate across spatial boundaries.

The final task is constructed through a two-stage pipeline shown in Figure~\ref{fig:task_generation}. In the scene grounding stage, we firstly sample subgoals from a task template, and then randomly place the selected goal objects at legal initial locations. We define a broad range of surfaces and containers to enable diverse and realistic object distributions (listed in Table~\ref{tab:legal-placement}). After a task scene is instantiated, we further assign its collaboration structure. Under sequential settings, the apartment is partitioned into disjoint room zones based on the initial locations of task objects and their target goal locations, and different zones are assigned to different agents. 

Overall, MECoBench consists of 96 tasks, each evaluated under both parallel and sequential setups, yielding 192 test cases in total. Each task contains 2–7 subgoals. The scenes are uniformly distributed over both the number of subgoals and the eight task templates to ensure balanced task complexity and scenario diversity.
Details of construction and statistic are in Appendix~\ref{app:benchmark}.

\subsection{Evaluation Platform}
\begin{figure}[t]
\setlength{\abovecaptionskip}{0pt}
    \centering
    \includegraphics[width=0.8\linewidth]{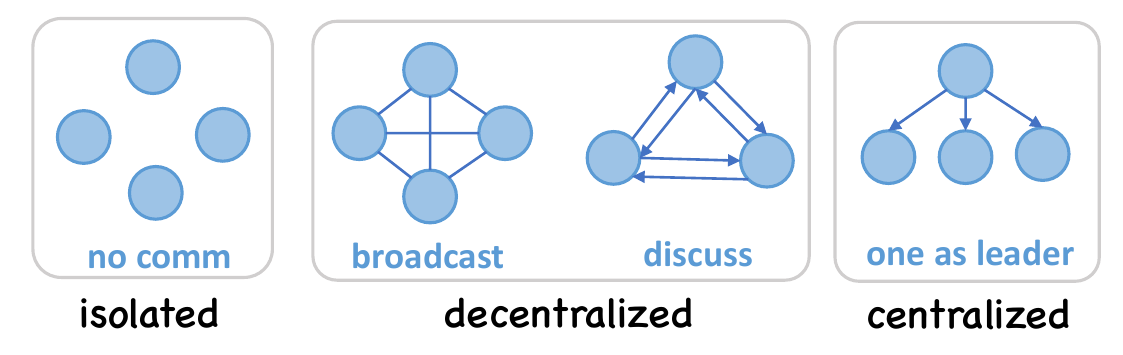}
    \caption{\textbf{Illustration of four protocols under three collaboration modes}.}
    \label{fig:2-1-protocol}
\vspace{-8pt}
\end{figure}
\begin{figure*}[!t]
\setlength{\abovecaptionskip}{0pt}
    \centering
    \includegraphics[width=0.9\textwidth]{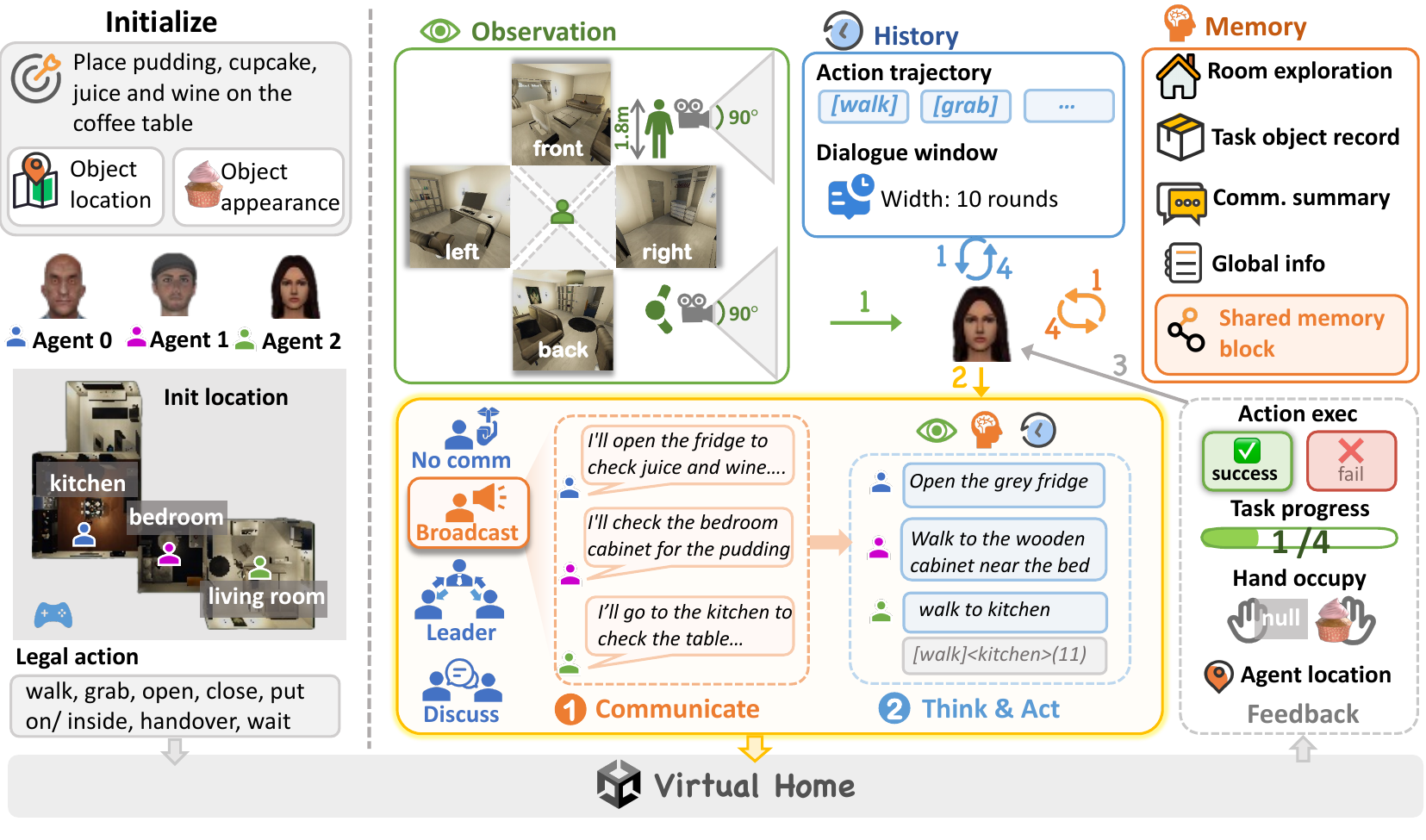}
    \caption{\textbf{Overall workflow of evaluation.} Agents receive task goals and prior information, perceive the environment, communicate under different protocols, reason with history and memory, and execute actions with feedback.}
    \label{fig:workflow}
\vspace{-8pt}
\end{figure*}

The platform supports a flexible number of agents and provides three collaboration modes with four basic protocols to enable systematic evaluation.

\paragraph{Collaboration Mode}
To analyze how different coordination structures shape multi-agent performance, we design three collaboration modes implemented with four protocols as summarized in Figure~\ref{fig:2-1-protocol}.
The \textbf{isolated} mode \textit{disables communication}, requiring agents to act independently. 
The \textbf{decentralized} mode enables peer-level coordination through \textit{broadcast} or \textit{discussion}. In broadcast, each agent shares one message with all others before action; in discussion, agents communicate sequentially and can continue for another round when consensus is not reached. 
The \textbf{centralized} mode assigns one agent as \textit{leader} to aggregate worker reports and assign actions to the team.

\paragraph{Evaluation Pipeline}

As illustrated in Figure~\ref{fig:workflow}, we formulate the evaluation as an iterative \emph{observe--communicate--act} workflow. At the beginning of each task,
the environment is initialized with the task scene graph and $N$ agents. Each agent is placed in its assigned initial room with the provided task goal $G$, the available prior information $I$, and the shared action space $A$ (see Appendix~\ref{app:action}).
At each timestep $t$, each agent $i$ obtains a panoramic environment observation $O_t^i$ from its first-person view. To support long-horizon execution over dozens of steps, the agent reasons not only over its current observation $O_t^i$, but also over its action history $h_{a,t}^i$, recent dialogue history $h_{d,t}$, self-maintained memory $M_t^i$, and initial information $I$. Depending on the chosen collaboration protocol $P$, each agent $i$ first generates the communication message $c_t^i$ as
\begin{equation}
c_t^i = P_i(O_t^i, h_{a,t}^i, h_{d,t}, M_t^i, G, I).
\end{equation}
The messages from all agents form the communication context $C_t = \{c_t^1, c_t^2, \ldots, c_t^N\}$.
After communication, each agent generates its next action conditioned on the
current communication context:
\begin{equation}
a_t^i = \pi_i(O_t^i, C_t, h_{a,t}^i, M_t, G, I, A),
\end{equation}
where $\pi_i$ denotes the action policy of agent $i$. The selected actions are executed in simulator, which returns execution feedback $F_t$ and resulting environment changes. Finally, the feedback is used to update the history and memory. This workflow continues until the task is completed or the maximum number of steps is reached.

\section{Experiment Setup}

\subsection{Implementation Details} For parallel tasks, we compare single-agent and two-agent execution under the same effective 60-step budget, defined as the total number of steps across all agents. Thus, for an n-agent team, each agent is allocated 60/n steps. 
For sequential tasks, we use a fixed two-agent setup with an 80-step budget. 
Unless otherwise specified, we use the broadcast collaboration protocol in all multi-agent settings. 

By default, we provide the task object appearance and possible location list to agents as the prior information $I$ at initialization. Since MECoBench focuses on collaboration rather than object search, this prior reduces localization difficulty and better isolates cooperative ability. More detials are in Appendix~\ref{app:eval-setup}.

\subsection{Evaluation Metrics}
We evaluate multi-agent collaboration from three aspects: \textbf{(1) Effectiveness}: \textit{Success Rate (SR)}, defined as the proportion of successful cases, and \textit{Completion Rate (CR)}, defined as the average fraction of completed subgoals across cases. \textbf{(2) Efficiency}: \textit{Step-CR AUC (AUC)}, measuring the area under the completion-rate curve over equivalent steps, and \textit{Average Token Cost per Step}. \textbf{(3) Collaboration quality}: \textit{Division of Labor (DOL)}, \textit{Conflict Action Rate (CAR)}, and \textit{Handover Failure Rate (HFR)}. DOL measures the distribution of subgoal completion across agents, calculated as
\begin{equation}
s_i=\frac{\mathrm{count}_i}{\sum_j \mathrm{count}_j}, \quad
\mathrm{DOL}=\frac{1-\sum_i s_i^2}{1-\frac{1}{n}}.
\end{equation}
where \(n>1\) is the number of agents, and \(s_i\) denotes the fraction of subgoals completed by agent \(i\) among all completed subgoals.
CAR measures the average fraction of conflicting actions between agents, such as grabbing or opening the same object. HFR measures the average fraction of failed handovers in sequential tasks.

\subsection{Evaluated Models} We evaluate a diverse set of both closed-source and open-source models.
The closed-source models include state-of-art model GPT-5.4~\cite{openai2026gpt54thinking}, Gemini-3.1-Pro~\cite{googledeepmind2026gemini31pro}, and GPT-5-mini~\cite{openai2025gpt5mini}
The open-source models include Qwen3-VL~\cite{bai2025qwen3vltechnicalreport}, Qwen3.5~\cite{qwen3.5}, Gemma4~\cite{googledeepmind2026gemma4modelcard}, InternVL 3.5~\cite{wang2025internvl35advancingopensourcemultimodal}, Llama4~\cite{meta2025llama4modelcard} and GLM 4.6V~\cite{vteam2025glm45vglm41vthinkingversatilemultimodal}, covering model sizes from 8B to 235B parameters. Appendix~\ref{app:model} lists all models evaluated.

\section{Will Multi-agent Collaboration Benefit Embodied Task Completion?}
\label{sec:exp-1}

\subsection{Do Models Know How To Collaborate? }
\begin{figure}[h]
\setlength{\abovecaptionskip}{0pt}
    \centering
    \includegraphics[width=1\linewidth]{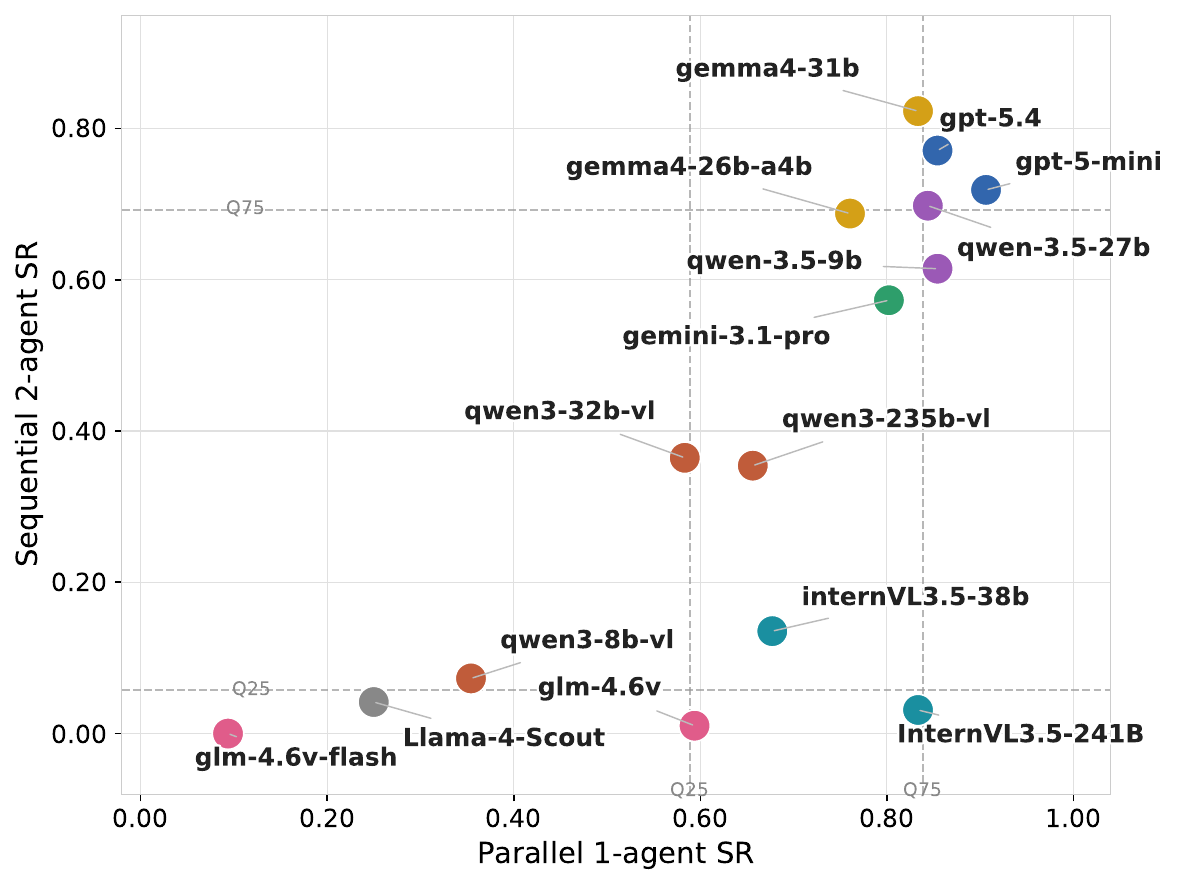}
    \caption{\textbf{Comparison of model performance between parallel single-agent and sequential two-agent settings.} Dashed lines indicate quartile thresholds.}
    \label{fig:1-1-overall}
\vspace{-8pt}
\end{figure}
\paragraph{Can models collaborate when collaboration is necessary?} We compare the model performance under single- and two-agent settings, using parallel single-agent and sequential two-agent results as proxies for individual and collaborative ability, respectively.
As demonstrated in Figure~\ref{fig:1-1-overall}, single-agent performance on parallel tasks correlates positively with two-agent collaboration performance on sequential tasks, while the sequential success rates are generally lower. This suggests that spatial constraints make collaboration harder than individual execution, while \textbf{most models still exhibit basic collaborative capability}. Among these models, GPT-5.4 and GPT-5-mini achieve the best overall performance, and Gemma4 series performs particularly well on collaboration tasks. In contrast, InternVL3.5 performs well individually but struggles to cooperate, revealing a gap between individual execution and group collaboration.

\begin{figure}[t]
\setlength{\abovecaptionskip}{0pt}
    \centering
    \includegraphics[width=1\linewidth]{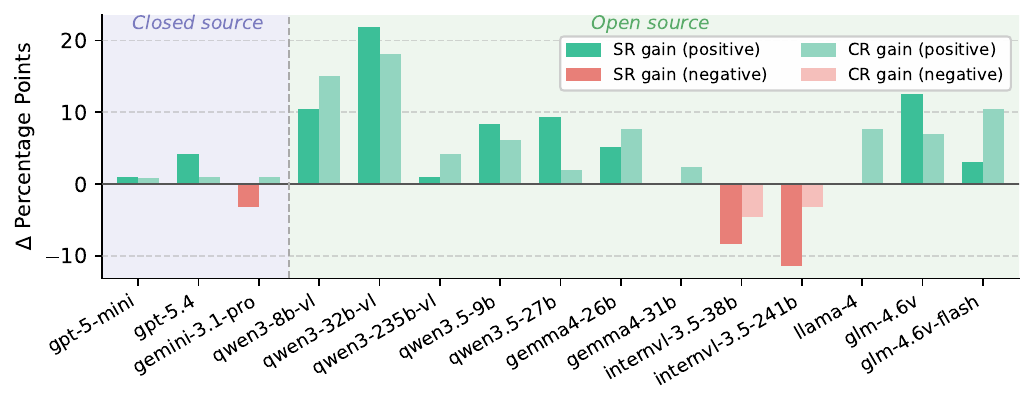}
    \caption{\textbf{Performance change from 1-agent to 2-agent under parallel settings.} Bars show the absolute change in SR and CR (percentage points)}
    \label{fig:1-1-gain}
\vspace{-8pt}
\end{figure}

\paragraph{Can models benefit from collaboration when it is optional?} Compared with single agent, Figure~\ref{fig:1-1-gain} shows that \textbf{most models benefit from the participation of a second agent}. 
The improvement is particularly large for weak and mid-level models, likely because their single-agent baselines leave more room for gains. Notably, InternVL3.5 series exhibits a clear performance drop, further confirming its weakness in multi-agent coordination. Detailed results and analyses are provided in Appendix~\ref{app:main_results}.

\subsection{Do More Agents Always Help?}
\begin{figure}[h]
\setlength{\abovecaptionskip}{0pt}
    \centering
     \includegraphics[width=1\linewidth]{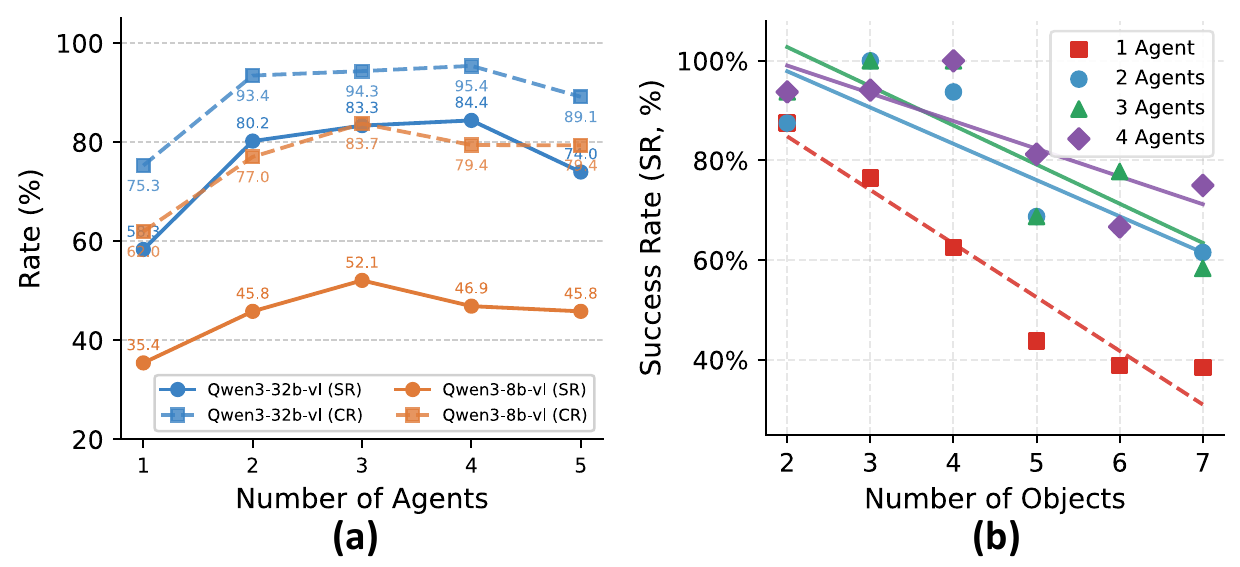}
    \caption{\textbf{Team size scaling effect.} (a) Performance curve of different team size. (b) SR of Qwen3-32B-VL versus the \#objects, with fitted trend lines for different team sizes. }
    \label{fig:agent-number-complexity}
\vspace{-12pt}
\end{figure}

To investigate this, we increase the number of agents from one to five under parallel settings with fixed effective step budget.
Results are shown in Figure~\ref{fig:agent-number-complexity}(a), where \textbf{collaboration performance generally follows an inverted-U-shaped trend}: adding agents initially improves performance, but further scaling leads to saturation and then degradation. 
Overall, moderate team sizes strike the best balance between collaboration benefits and costs.

We further analyze the effect of task complexity using object count as a proxy. Figure~\ref{fig:agent-number-complexity}(b) shows that success rates decline with increasing object count, due to longer trajectories and greater planning difficulty (see Appendix~\ref{app:obj-num} for full results across all models and task conditions). Nevertheless, \textbf{multi-agent settings maintain more robust performance under high-complexity tasks}, highlighting the value of collaboration in complex embodied scenarios.

\subsection{How Do Agents Collaborate?}
\label{sec:how-agent-collab}

\begin{figure*}[t]
\setlength{\abovecaptionskip}{0pt}
    \centering
    \includegraphics[width=0.9\linewidth]{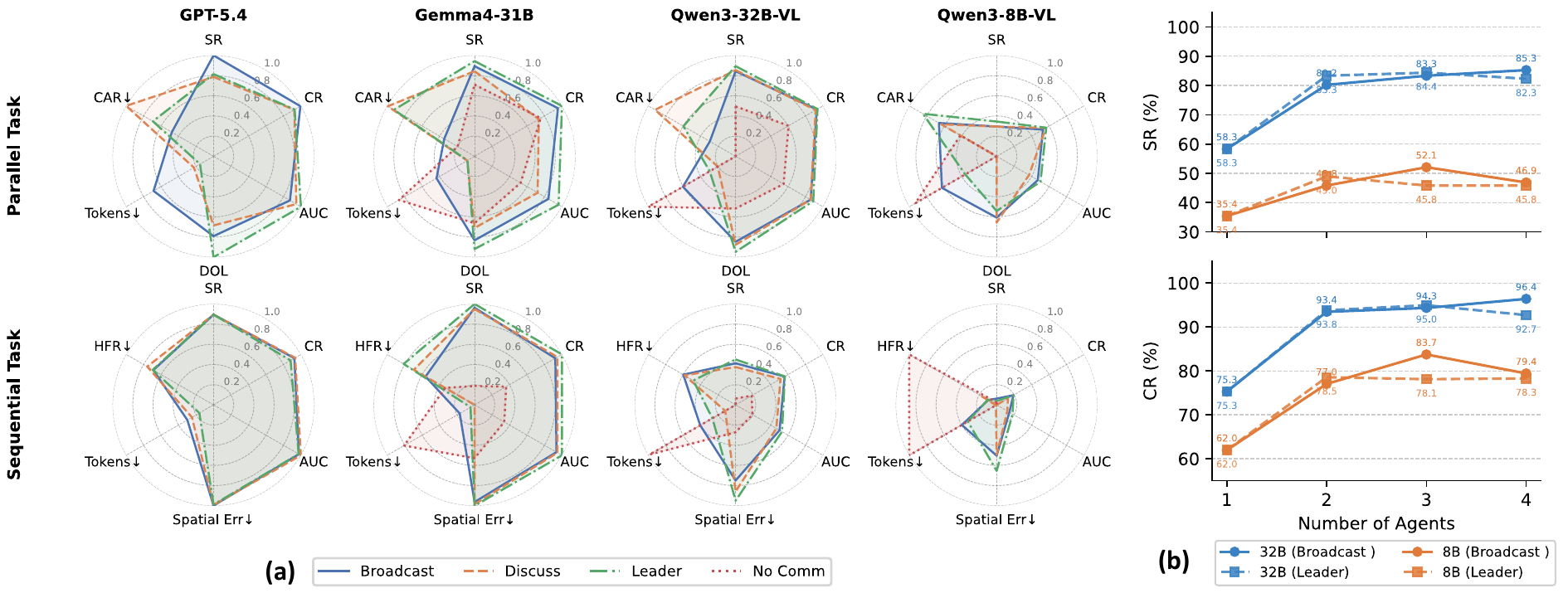}
    \caption{\textbf{Comparison of collaboration modes.} (a) Performance profiles in 2-agent setup on parallel and sequential tasks. (b) SR and CR trends with increasing team size under decentralized broadcast and centralized leader-based.}
    \label{fig:comm-mode-analysis}
\vspace{-8pt}
\end{figure*}
We delve deeper into fine-grained collaborative behaviors to understand how agents collaborate in practice. 
These behaviors are identified from the agents' communication and action trajectories using keyword-based matching rules, and are organized into three categories, as reported in Table~\ref{tab:micro_behavior}. Examples are provided in Appendix~\ref{app:4.3}.

\paragraph{Information Sharing} Agents share their intended plans (\textit{Info-Plan}), current locations (\textit{Info-Loc}), and detected task-relevant objects (\textit{Info-Obj}). \textbf{Plan and location sharing are frequent and consistently beneficial}, while object discovery reporting shows mixed effects.

\paragraph{Task Coordination} Action-level delegation (\textit{Act-Del}) issues concrete executable requests, such as asking another agent to grab, check, or place, and brings the strongest gains in both parallel and sequential tasks. In contrast, task assignment (\textit{Task-Assign}) provides coarser responsibility divisions, such as assigning rooms, areas, or subtasks, and offers only marginal benefit in parallel tasks while hurting sequential ones. \textbf{More specific and actionable communication is easier to translate into successful execution.}

\paragraph{Alignment and Correction} Coordination query (\textit{Coord-Q.}) refers to questions that synchronize teammates' states or locations. It occurs more often in sequential tasks where object transfer requires repeated position checks and serves as an indicator of task and coordination difficulty. \textit{Correct.} denotes behavioral correction through self-reflection or peer feedback. Despite its negative SR correlation, it consistently improves CR in both settings, suggesting that \textbf{correction serves as an error-recovery mechanism that improves task progress even when full success is unrecoverable.}
More analysis provided in Appendix~\ref{app:4.3}.

\begin{table}[t]
\centering
\setlength{\abovecaptionskip}{2pt}
\setlength{\tabcolsep}{3pt}
\scriptsize
\caption{\textbf{Fine-grained collaborative behaviors across all models.} Freq. denotes the percentage of cases containing
each behavior. $\Delta$SR and $\Delta$CR are percentage-point differences
between cases with and without the behavior.}
\resizebox{\columnwidth}{!}{%
\begin{tabular}{@{}lcccccc@{}}
\toprule
\multirow{2}{*}{\textbf{Behav.}}
& \multicolumn{3}{c}{\textbf{Parallel}}
& \multicolumn{3}{c}{\textbf{Sequential}} \\
\cmidrule(lr){2-4}\cmidrule(lr){5-7}
& \textbf{Freq.} & $\boldsymbol{\Delta}$\textbf{SR}
& $\boldsymbol{\Delta}$\textbf{CR}
& \textbf{Freq.} & $\boldsymbol{\Delta}$\textbf{SR}
& $\boldsymbol{\Delta}$\textbf{CR} \\
\midrule
Info-Plan
& 75.6\% & +4.6$\uparrow$  & +2.8$\uparrow$
& 86.3\% & +16.9$\uparrow$ & +18.0$\uparrow$ \\

Act-Del.
& 57.1\% & +12.2$\uparrow$ & +7.7$\uparrow$
& 54.3\% & +13.1$\uparrow$ & +19.2$\uparrow$ \\

Info-Loc.
& 47.3\% & +9.1$\uparrow$  & +6.9$\uparrow$
& 68.0\% & +7.6$\uparrow$  & +13.3$\uparrow$ \\

Info-Obj.
& 25.9\% & +5.0$\uparrow$  & +3.9$\uparrow$
& 16.3\% & -6.7$\downarrow$ & -1.6$\downarrow$ \\

Coord-Q.
& 17.6\% & -10.1$\downarrow$ & -1.8$\downarrow$
& 48.5\% & -15.5$\downarrow$ & -6.7$\downarrow$ \\

Correct.
& 16.9\% & -6.7$\downarrow$ & +1.6$\uparrow$
& 6.9\%  & -11.3$\downarrow$ & +7.7$\uparrow$ \\

Task-Assign
& 15.7\% & +1.9$\approx$ & +2.9$\uparrow$
& 8.5\%  & -11.5$\downarrow$ & -4.9$\downarrow$ \\
\bottomrule
\end{tabular}
}
\label{tab:micro_behavior}
\vspace{-8pt}
\end{table}

\section{What Makes Multi-agent Collaboration Effective?}
\label{sec:exp-2}
We answer this question along two dimensions: \textbf{collaboration mode} and \textbf{communication mechanism}. We first examine which collaboration mode is most effective. We then study the role of communication by testing whether it is necessary and whether alternative mechanisms further improve collaboration.

\subsection{Which Collaboration Mode Works Best?}
\paragraph{Two-Agent Collaboration.} We compare different collaboration modes beyond basic broadcast mechanism. As shown in Figure~\ref{fig:comm-mode-analysis}(a), leader-based collaboration achieves the best overall performance for most models, suggesting that \textbf{centralized collaboration helps small teams allocate tasks and avoid spatial conflicts}. However, GPT-5.4 remains competitive with simple broadcast communication, indicating that stronger models may rely less on structured collaboration mode.

\paragraph{Scaling to Larger Teams.} We further examine how collaboration modes scale with team size on parallel tasks. Figure~\ref{fig:comm-mode-analysis}(b) illustrates that centralized coordination is effective in small teams but saturates as team size increases, while decentralized coordination becomes more competitive by avoiding the bottleneck of relying on a single leader to aggregate information and assign actions. Moreover, under centralized coordination, model capability matters, as Qwen3-32B-VL benefits from three agents, whereas Qwen3-8B-VL peaks at two agents and then degrades. Therefore, \textbf{the best collaboration mode depends on both team size and model capability}. Full results across all metrics and team sizes are provided in Appendix~\ref{app:collab-mode}.

\subsection{Is Communication Necessary?}
\begin{figure}[t]
\setlength{\abovecaptionskip}{0pt}
    \centering
    \includegraphics[width=\linewidth]{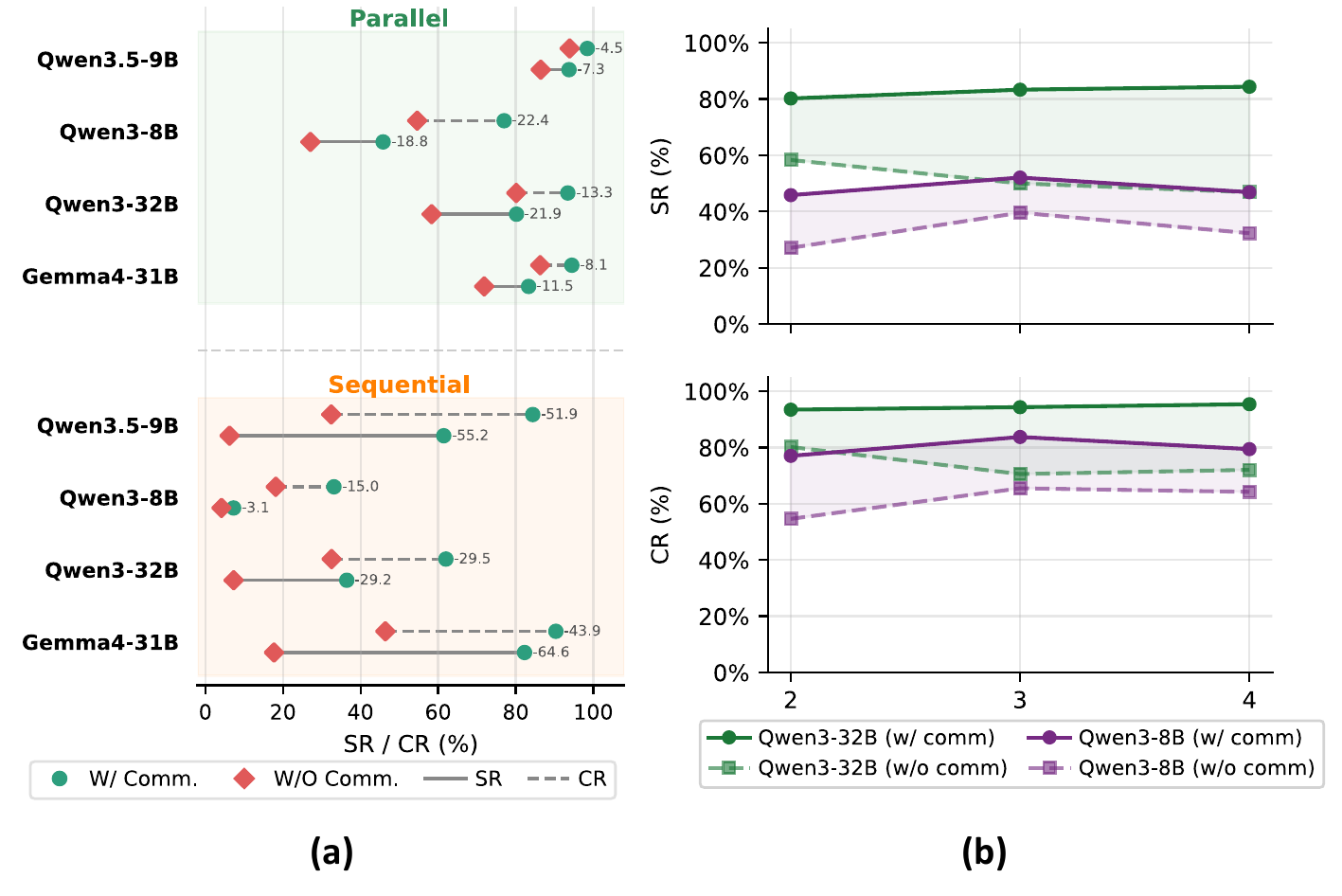}
    \caption{\textbf{Ablation study on communication.} (a) Performance change when remove communication. (b) Performance across different team sizes.}
    \label{fig:role_of_communication}
\vspace{-4pt}
\end{figure}
We ablate communication from the broadcast protocol under both parallel and sequential settings. As shown in Figure~\ref{fig:role_of_communication}, disabling communication consistently degrades performance across models, with a much larger drop on sequential tasks. When the number of agents grows, the gap between communication and no communication widens. These results indicate that \textbf{communication is essential for effective collaboration}, especially when tasks require stronger coordination or involve larger teams. More details in Appendix~\ref{app:comm-role}.
\subsection{Is Textual Communication Enough?}

While explicit textual communication is essential, it may be insufficient for embodied collaboration, where agents must share evolving task states and visually grounded observations. We therefore explore two alternative communication mechanisms: \textit{shared memory} and \textit{vision-augmented leadership}.

\begin{figure}[t]
\setlength{\abovecaptionskip}{0pt}
    \centering
        \includegraphics[width=\linewidth]{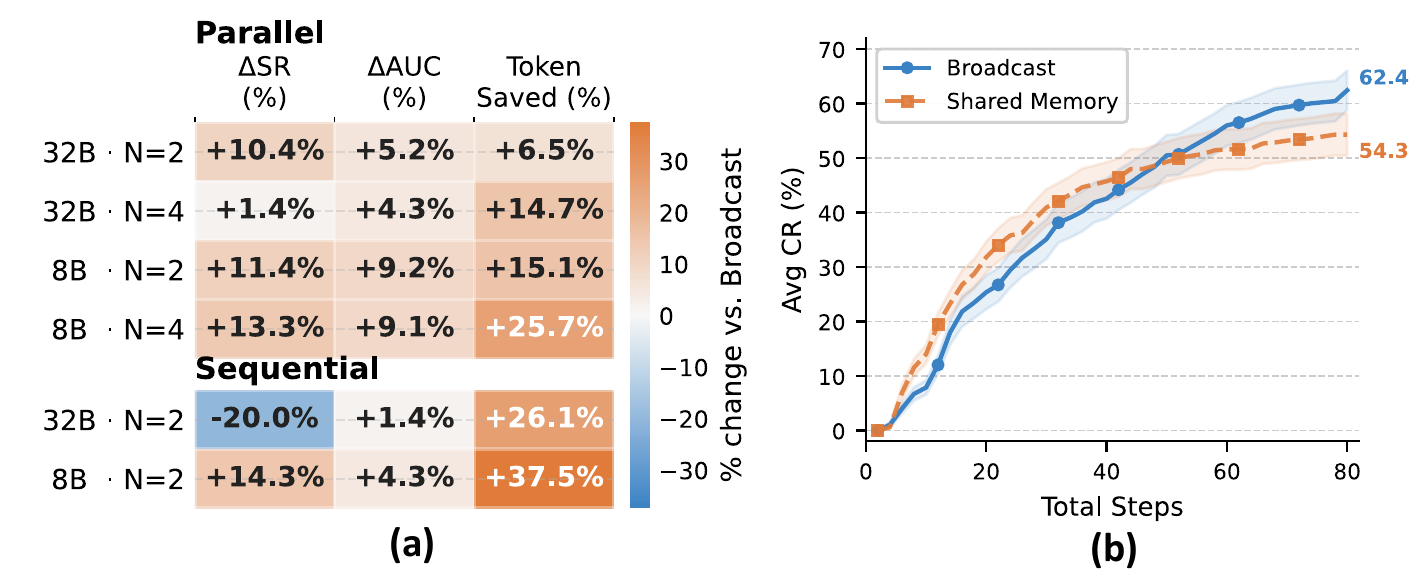}
    \caption{\textbf{Comparison between broadcast and shared-memory}. 
(a) Relative changes in effectiveness and efficiency over broadcast of Qwen3-VL. 
(b) Average completion progress over steps under broadcast and shared-memory of Qwen3-32B-VL.}
    \label{fig:3-1-shared-memory}
\vspace{-8pt}
\end{figure}

\paragraph{Shared memory leads to efficient and effective collaboration.}
In shared memory, agents skip explicit message exchange and coordinate through a shared memory block. As shown in Figure~\ref{fig:3-1-shared-memory}, this mechanism improves the performance and greatly reduces token consumption, indicating more efficient collaboration (absolute results in Figure~\ref{fig:d4}). It consistently improves success rate on parallel tasks, but degrades on sequential tasks, where precise timing and spatial alignment may still require explicit communication. Nevertheless, shared memory accelerates early progress, reducing more conflicts than broadcast under smaller step budgets.

\begin{figure}[h]
\setlength{\abovecaptionskip}{0pt}
    \centering
    \includegraphics[width=1\linewidth]{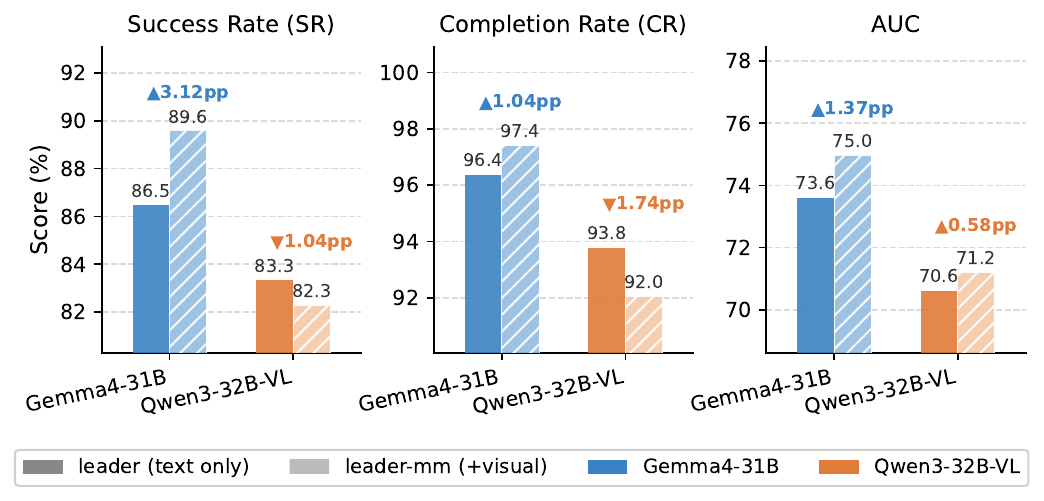}
    \caption{\textbf{Comparison in leader-based collaboration with and without visual augmentation.}}
    \label{fig:3-2-leader-mm}
\vspace{-8pt}
\end{figure}

\paragraph{Vision-augmented leadership helps when the leader is capable.}
To give the leader more grounded context, in vision-augmented leader-worker collaboration, we allow workers to send both textual reports and current visual observations to the leader. As shown in Figure~\ref{fig:3-2-leader-mm}, additional visual information improves AUC for both models, while its impact on SR and CR varies across models. This difference may stem from models' different abilities to effectively use visual cues. Models with stronger visual capabilities (Gemma4-31B), benefit from the additional visual input, whereas weaker models (Qwen3-32B-VL) drop.

\section{Is Multi-Agent Collaboration Robust?}
We examine robustness from two perspectives: the sensitivity of collaboration to
imperfect prior information, and the failure modes that arise under collaboration.
Additional prior information ablations and mixed-model team experiments are deferred to Appendix~\ref{app:mix-team} and \ref{para:info-ablation}.

\label{sec:exp-3}

\subsection{Does Collaboration Remain Effective Under Imperfect Information?}
At initialization, we provide all task-relevant object locations in the prior information $I$, which is hard to obtain in real scenarios. Therefore, we further investigate the collaboration under two realistic settings: (1) \textit{Missing location priors}: object locations are removed from the prior information, requiring agents to find targets through visual search. (2) \textit{Noisy location priors}: an additional 30\% false object locations are injected into the provided prior information. 
We extend the effective step budget to 80 steps to allow sufficient search, and apply parallel settings with basic broadcast protocol. 
\begin{figure}[t]
\setlength{\abovecaptionskip}{0pt}
    \centering
    \includegraphics[width=1\linewidth]{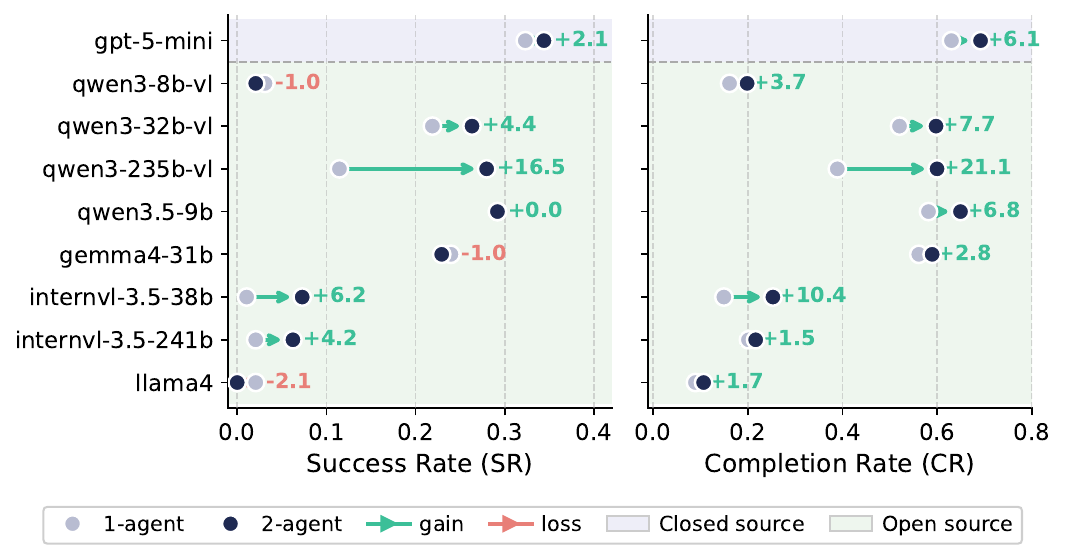}
    \caption{\textbf{Performance change under parallel settings without prior location information.}}
    \label{fig:4-1-no-loc}
\vspace{-8pt}
\end{figure}

\begin{figure}[t]
\setlength{\abovecaptionskip}{0pt}
    \centering
        \includegraphics[width=0.95\linewidth]{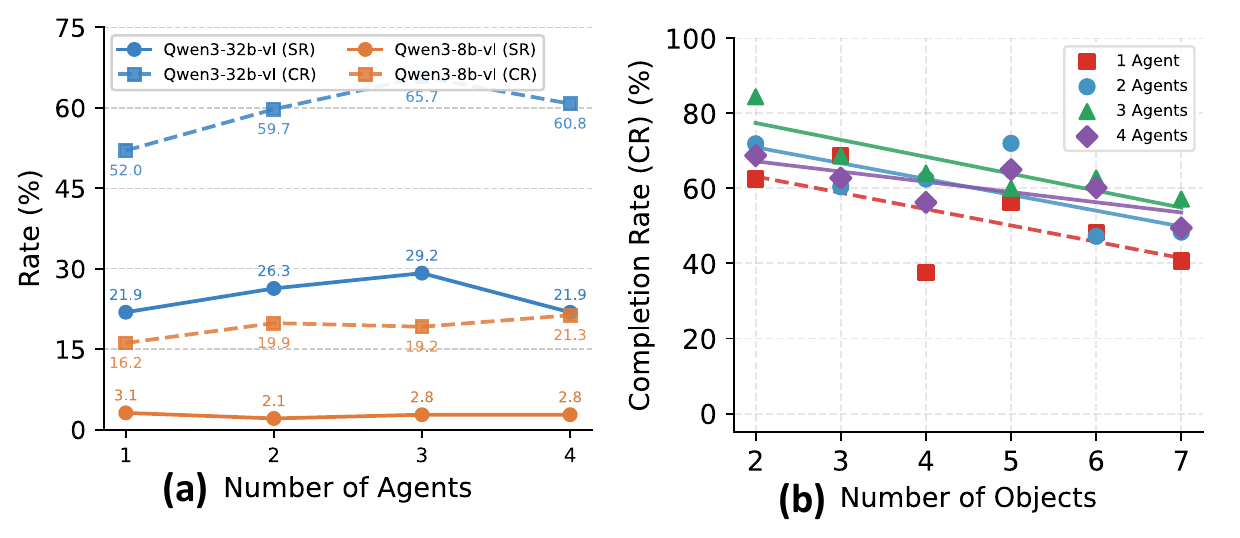}
    \caption{\textbf{Team size scaling effect without prior location information.} (a) Performance curve of different team size. (b) SR of Qwen3-32B-VL versus the \#objects, with fitted trend lines for different team sizes.}
    \label{fig:noloc-analysis}
\vspace{-8pt}
\end{figure}

As shown in Figure~\ref{fig:4-1-no-loc}, two-agent collaboration improves the success rate and reduces the conflicts, indicating that collaboration remains useful when task information is incomplete. 
When the number of agents is further increased, a similar inverted-U-shaped trend emerges, as shown in Figure~\ref{fig:noloc-analysis}(a), while performance remains more stable on complex tasks. Moreover, Table~\ref{tab:noise_no_location} demonstrates that collaboration achieves the largest relative performance gains when the information contains noise, suggesting that agents can compensate for misleading information through communication and distributed exploration. These results together validate that \textbf{multi-agent collaboration is robust to insufficient and noisy information}. 

\begin{table}[!t]
\centering
\setlength{\abovecaptionskip}{2pt}
\footnotesize
\setlength{\tabcolsep}{3pt}
\caption{
\textbf{Performance under noisy and incomplete task information.}
We compare three settings: noisy location priors (\textbf{W/ Noise}), clean location priors (\textbf{W/o Noise}), and no location priors (\textbf{W/o Loc.}). 
Each cell reports relative SR gain on top, with the corresponding 1-agent SR $\to$ 2-agent SR shown below.
}
\begin{tabular}{lccc}
\toprule
\textbf{Model} 
& \textbf{W/ Noise} 
& \textbf{W/o Noise} 
& \textbf{W/o Loc.} \\
\midrule
GPT-5-mini   
& \makecell{\textbf{+8.22\%}\\{\scriptsize 76.04$\to$82.29}}
& \makecell{\textbf{+1.16\%}\\{\scriptsize 90.62$\to$91.67}}
& \makecell{\textbf{+6.47\%}\\{\scriptsize 32.29$\to$34.38}} \\
Qwen3-32B
& \makecell{\textbf{+28.83\%}\\{\scriptsize 54.17$\to$69.79}}
& \makecell{\textbf{+28.57\%}\\{\scriptsize 58.33$\to$75.00}}
& \makecell{\textbf{+18.88\%}\\{\scriptsize 21.88$\to$26.04}} \\
Qwen3.5-9B   
& \makecell{\textbf{+4.01\%}\\{\scriptsize 78.12$\to$81.25}}
& \makecell{\textbf{+3.52\%}\\{\scriptsize 85.42$\to$88.54}}
& \makecell{\textbf{+0.00\%}\\{\scriptsize 29.17$\to$29.17}} \\
\bottomrule
\end{tabular}
\label{tab:noise_no_location}
\vspace{-8pt}
\end{table}

\subsection{Where Does Collaboration Break Down?}
\label{sec:fail}
We identify two major failure modes in multi-agent collaboration. These failures suggest that robust multi-agent embodied collaboration requires conflict resolution and belief verification mechanisms. Examples and more statistics in Appendix~\ref{app:fail}.

\textbf{Multiple agents can introduce conflicts}, including duplicate grabs, repeated labor, and redundant exploration. Although communication mitigates these conflicts for most models, InternVL3.5-241B still exhibits the highest duplicate-grab rates among 2-agent setup of 21.9\%, partly explaining its performance drop in Figure~\ref{fig:1-1-gain}. This also explains why performance drop with large team size: for Qwen3-32B-VL, scaling to five agents increases duplicate grabs to 57.4\%.

\textbf{Hallucinated beliefs can propagate across agents.} When one agent falsely believes the task is complete and broadcasts it, the team may stop prematurely. Although rare (1\%-1.5\%), such hallucinated completion is highly destructive, causing a 73.7\% SR drop in parallel tasks and a 40.9\% drop in sequential tasks. Object confusion can similarly cause agents to skip unfinished subtasks.

\section{Related Work}

\paragraph{Multi-Agent System}
Inspired by human society, multi-agent systems (MAS) have been studied as a way to improve task efficiency and effectiveness~\cite{Stone2000}. Recent work has explored LLM-based MAS across diverse domains, including automated research~\cite{schmidgall-etal-2025-agent,su-etal-2025-many}, programming~\cite{ICLR2024_6507b115,islam-etal-2025-codesim}, medical decision-making~\cite{fan-etal-2025-ai,Chen2025MAC}, and social simulation~\cite{piao2025agentsociety}. 
Other studies further examine their scalability and emergent collaborative behaviors~\cite{ICLR2025_66a026c0,ICLR2024_578e65cd}. Although recent work has begun to explore embodied multi-agent scenarios~\cite{ICLR2025_7d03c6bf,10504634}, multimodal embodied collaboration remains underexplored~\cite{yu2024minelandsimulatinglargescalemultiagent}.

\paragraph{Embodied Agent}
Recent studies have explored LLM- or MLLM-based embodied agents for a wide range of tasks, including object manipulation~\cite{Yang_2024_CVPR,monwilliams2025embodied}, navigation~\cite{zhang2024navid}, complex perception-interaction tasks~\cite{szot2024large}, and embodied human-AI collaboration~\cite{ICLR2025_a3cf318f}. 
Several benchmarks have also been proposed to evaluate embodied agents across diverse environments and capabilities~\cite{NEURIPS2022_27c546ab,Savva_2019_ICCV,yang2025embodiedbench,cheng2025embodiedevalevaluatemultimodalllms}. However, these benchmarks predominantly focus on single-agent settings, providing limited support for studying collaboration among multiple embodied agents.

\paragraph{Embodied Multi-Agent Collaboration Benchmarks} 
Several benchmarks have been proposed to evaluate embodied multi-agent collaboration~\cite{NEURIPS2025_ab16cc8f,puig2021watchandhelp,10.1007/978-3-030-58558-7_28,NEURIPS2019_f5b1b89d,agashe-etal-2025-llm}. Many are built on 2D games~\cite{sun-etal-2025-collab,yang2026coop2definingobservingrepairing}, offering limited support for physical interaction and spatial reasoning. Existing 3D benchmarks often convert visual observations into textual descriptions~\cite{ICLR2024_54b8b4e0,10610855,ICLR2025_a3cf318f,dong-etal-2024-villageragent,white2025collaboratingactionactionmultiagent}, decoupling visual grounding and exploration from collaborative decision-making and overlooking behaviors unique to multimodal embodied interaction. 
Some recent benchmarks support multimodal observations~\cite{NEURIPS2025_ab16cc8f,NEURIPS2025_6b8cb6b2,zha2026aircopbench}, but mainly evaluate visual question answering or offline reasoning rather than closed-loop interaction. TeamCraft~\cite{long2024teamcraftbenchmarkmultimodalmultiagent} takes an initial step toward evaluating embodied MLLM collaboration, but does not systematically study how communication mechanisms, coordination modes, and team size shape collaboration.

\section{Conclusion}
We empirically investigate how MLLM-based agents collaborate in embodied environments by constructing MECoBench, a multimodal embodied cooperation benchmark. 
Through controlled experiments and analyses, our results demonstrate that effective coordination can substantially improve both task performance and efficiency. By systematically examining collaboration modes and communication mechanisms, we reveal how these factors shape collaborative performance. We further evaluate collaboration under noisy and incomplete prior information, demonstrating its robustness in more challenging settings. Beyond aggregate performance, we identify emergent collaborative behaviors and failure patterns, explaining both the benefits of collaboration and the new challenges it introduces. Overall, our study provides a deeper understanding of MLLM-based embodied collaboration and introduces a benchmark for evaluating the collaborative capabilities of future MLLMs.

\section*{Limitations}
In this work, we propose MECoBench and conduct systematic experiments to study cooperative behaviors of MLLM-based agents in embodied environments. Our analysis reveals several non-trivial patterns in multi-agent collaboration, but there remain several limitations for future work.

\paragraph{Expanding tasks and scenarios.} MECoBench and corresponding evaluation platform is currently built on VirtualHome, an indoor simulator focused on household scenarios. As a result, our tasks are mainly limited to everyday indoor activities and object interactions. Future work can extend the benchmark to more diverse environments, such as open-world or outdoor scenes, and incorporate broader task types such as navigation and long-horizon exploration.

\paragraph{Data scale.} MECoBench contains 96 cases from 8 task types under two major task setups. This scale is sufficient for controlled evaluation and for revealing key collaboration patterns, but it remains relatively small compared with the diversity and complexity of real-world embodied tasks. Future work can expand the number of task instances, object categories, scene layouts, and collaboration requirements.

\paragraph{Agent scaling.}
Due to spatial constraints in the environment and the complexity of task execution, our experiments scale the team size up to 5 agents. Larger teams and more complex tasks are therefore left unexplored. 

\bibliography{custom}
\begin{table*}[!t]
\centering
\small
\setlength{\tabcolsep}{8pt}
\renewcommand{\arraystretch}{1.15}
\caption{Task templates used in our benchmark. Each task is defined by a set of target predicates, where \texttt{ON(obj, rec)} and \texttt{IN(obj, rec)} denote placing a surface on or inside a container, respectively.}
\begin{tabular}{ll}
\toprule
\textbf{Task Name} & \textbf{Predicate Set} \\
\midrule

Self-caring
& \makecell[l]{
ON(toothbrush, bathroomcounter), ON(toothpaste, bathroomcounter),\\
ON(towel, bathroomcounter), ON(barsoap, bathroomcounter)
} \\

\midrule
Work on desk
& \makecell[l]{
ON(keyboard, desk), ON(mouse, desk),\\
ON(cellphone, desk), ON(mug, desk)
} \\

\midrule
Gaming setup
& \makecell[l]{
ON(boardgame, coffeetable), ON(remotecontrol, coffeetable),\\
ON(magazine, coffeetable), ON(toy, coffeetable),\\
ON(book, coffeetable)
} \\

\midrule
Prepare afternoon tea
& \makecell[l]{
ON(cupcake, coffeetable), ON(juice, coffeetable),\\
ON(wine, coffeetable), ON(pudding, coffeetable),\\
ON(apple, coffeetable)
} \\

\midrule
Wash dishes
& \makecell[l]{
IN(plate, dishwasher), IN(waterglass, dishwasher),\\
IN(wineglass, dishwasher), IN(cutleryfork, dishwasher)
} \\

\midrule
Prepare food
& \makecell[l]{
ON(cupcake, kitchentable), ON(juice, kitchentable),\\
ON(pancake, kitchentable), ON(poundcake, kitchentable),\\
ON(wine, kitchentable), ON(pudding, kitchentable),\\
ON(apple, kitchentable), ON(coffeepot, kitchentable)
} \\

\midrule
Put food inside fridge
& \makecell[l]{
IN(cupcake, fridge), IN(juice, fridge),\\
IN(pancake, fridge), IN(poundcake, fridge),\\
IN(wine, fridge), IN(pudding, fridge),\\
IN(apple, fridge)
} \\

\midrule
Setup dinner table
& \makecell[l]{
ON(plate, kitchentable), ON(waterglass, kitchentable),\\
ON(wineglass, kitchentable), ON(cutleryfork, kitchentable)
} \\

\bottomrule
\end{tabular}
\label{tab:task_templates}
\end{table*}

\newpage
\
\newpage

\appendix

\section{MECoBench}
\label{app:benchmark}
This section presents detailed information about MECoBench, covering its task definitions, benchmark statistics, and construction details.

\subsection{Task definition}
\begin{figure}[!h]
    \centering
    \includegraphics[width=0.9\linewidth]{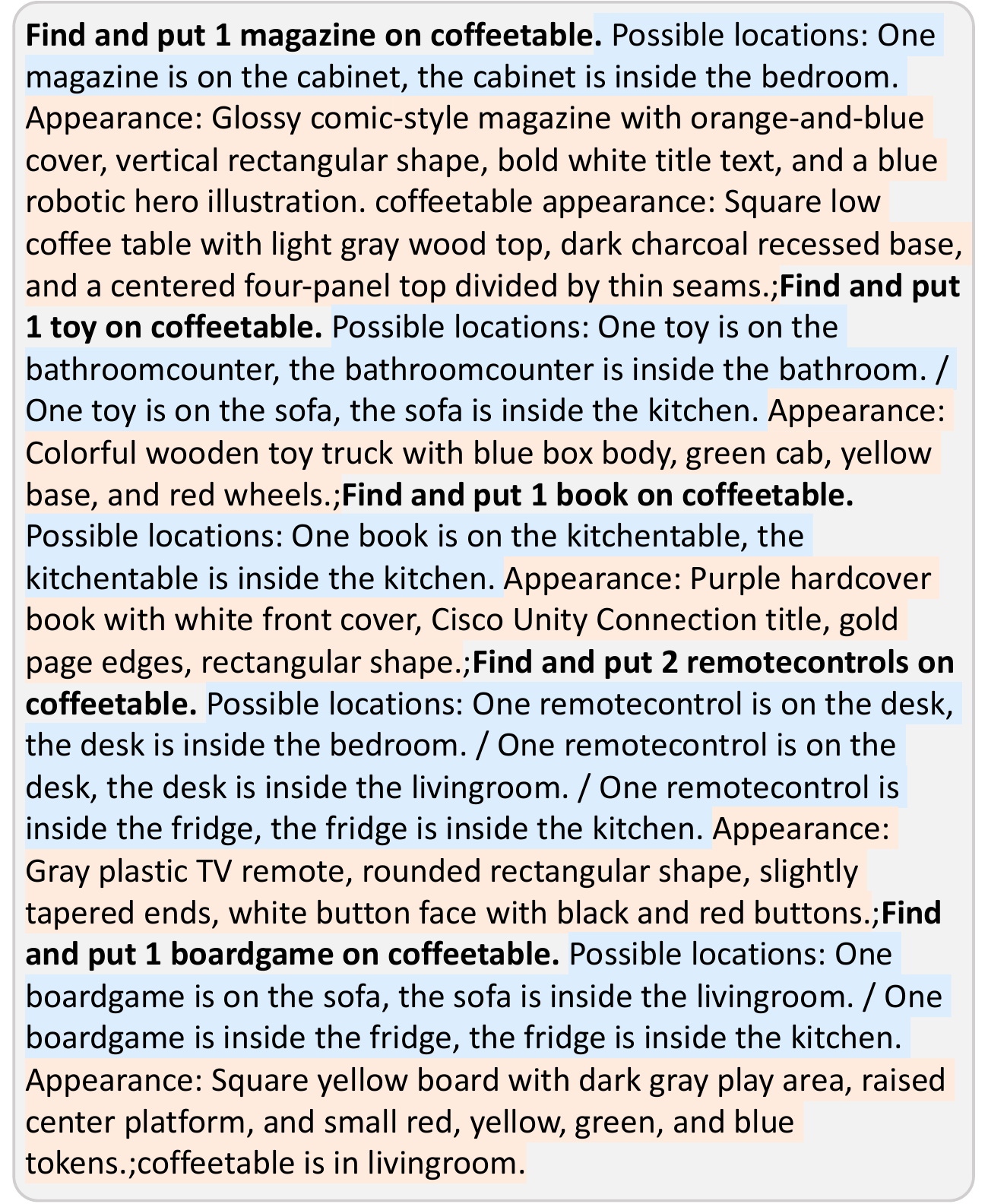}
    \caption{Example of task goal content. The goal specifies target objects, their possible locations highlighted in blue, and appearance descriptions highlighted in orange.}
    \label{fig:task-example}
\end{figure}

Table~\ref{tab:task_templates} shows the detailed templates of the eight task types designed for MECoBench. \textit{Self-caring}, \textit{work-on-desk}, and \textit{gaming setup} are three newly introduced types, while \textit{prepare afternoon tea}, \textit{wash dishes}, \textit{prepare food}, \textit{put food inside the fridge}, and \textit{setup dinner table} are modified from WAH. Figure~\ref{fig:task-example} gives an example of final task goal of gaming setup.

\subsection{Benchmark construction details}
The construction of MECoBench tasks follows a two-step pipeline. In step one, task templates are instantiated into concrete scenes with randomized object placement and validated for feasibility.

\paragraph{Goal Sampling.}  
For a given task type, we sample subgoals from predefined object-count ranges, specifying the target objects, required counts, and optional room or surface constraints.  To mitigate visibility issues such as occlusion, we add 0--2 redundant instances for each target object category during scene instantiation. For large objects that occupy more surface space, such as keyboards, boardgames, and magazines, the redundancy is limited to 0--1 instance.

\paragraph{Scene Instantiation and Object Placement.}  
Once a template is sampled, the target objects are assigned to legal initial locations based on the predefined legal placement (Table~\ref{tab:legal-placement}). Objects are placed randomly while respecting constraints:
\begin{itemize}
    \item No object is placed on the same surfaces or containers that are reserved for other goal locations.
    \item Surface space is checked to ensure sufficient area for the object.
    \item Containers are verified for available slots using their bounding boxes.
\end{itemize}
This procedure ensures that the initial scene is diverse yet consistent with the task template.

\label{appendix:data-pp}
\begin{table}[ht]
\centering
\small
\setlength{\tabcolsep}{4pt}
\caption{Legal placement locations used for randomized object initialization in MECoBench.}
\label{tab:legal-placement}
\begin{tabular}{ll}
\toprule
\textbf{Type} & \textbf{Legal Locations} \\
\midrule
Container 
& \makecell[l]{fridge, stove, dishwasher, microwave,\\
cabinet, bookshelf} \\
\addlinespace[2pt]
Surface 
& \makecell[l]{sofa, cabinet, bathroomcounter, nightstand,\\
bench, bed, desk, tvstand, kitchencounter,\\ coffeetable, stove, kitchentable} \\
\bottomrule
\end{tabular}
\end{table}

\paragraph{Validation and Goal Specification.}  
After placement, the VirtualHome environment is used to verify scene feasibility:
\begin{itemize}
    \item The environment graph is updated to reflect the actual object positions.
    \item Target object counts are checked to ensure that all task objects exist in the scene.
    \item Predicates from the template are instantiated with the real node IDs in the scene.
    \item Tasks failing these checks (e.g., insufficient instances or invalid placements) are discarded, and a new attempt is generated.
\end{itemize}

\subsection{Benchmark Statistic}
\label{appendix:data}
\begin{figure}[!h]
    \centering
    \includegraphics[width=0.8\linewidth]{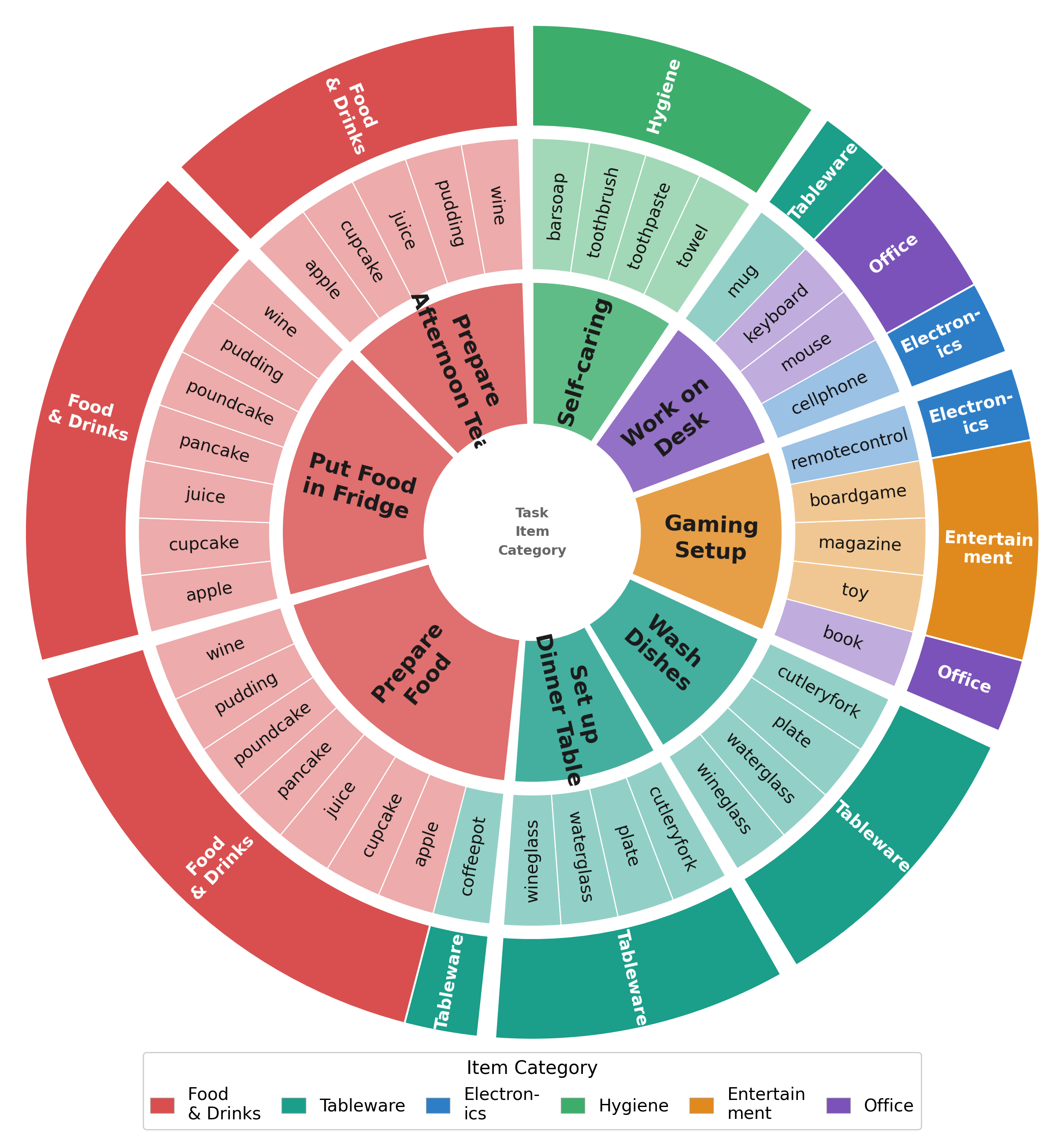}
    \caption{Three-layer sunburst chart illustrating the composition of eight household tasks used in our evaluation. The inner ring represents the eight task types; the middle ring shows the objects involved in each task; the outer ring indicates the object category.}
    \label{fig:task-type-statistic}
\end{figure}

\begin{figure}[!h]
\setlength{\abovecaptionskip}{0pt}
    \centering
    \includegraphics[width=1\linewidth]{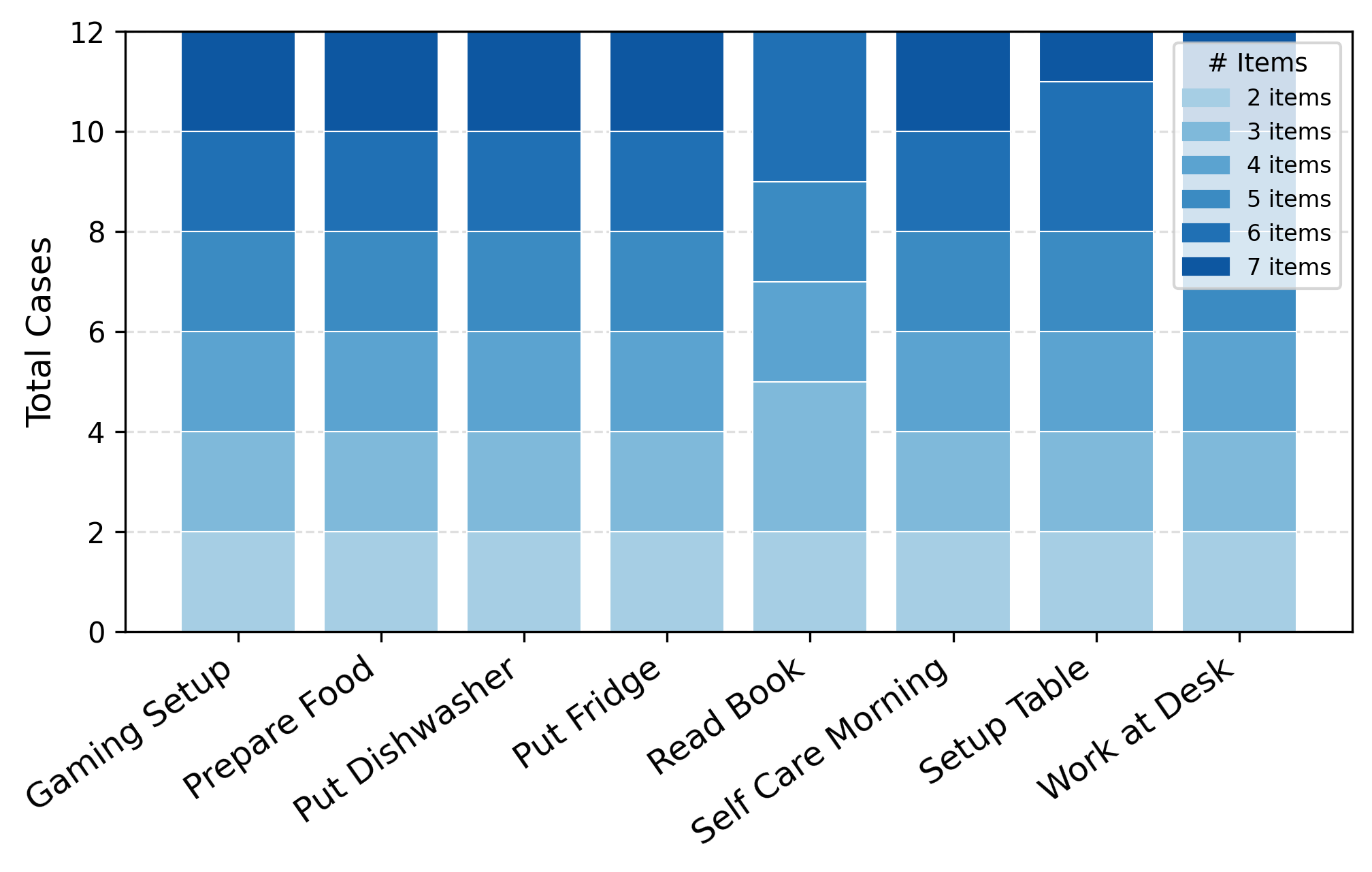}
    \caption{Data distribution grouped by task type and number of subgoals.}
    \label{fig:data-statistic}
\end{figure}

MECoBench involves 25 common everyday objects, which are grouped into six categories: food and drinks, tableware, electronics, hygiene items, entertainment items, and office items. Figure~\ref{fig:task-type-statistic} illustrates the correspondence between the eight task types and their associated object categories.

Figure~\ref{fig:data-statistic} shows the data distribution of MECoBench across different task types and task complexity levels, where task complexity is measured by the number of involved objects.

\section{Evaluation}

\subsection{Evaluation Setup}
\label{app:eval-setup}
\begin{figure}[ht]
    \centering
    \includegraphics[width=1\linewidth]{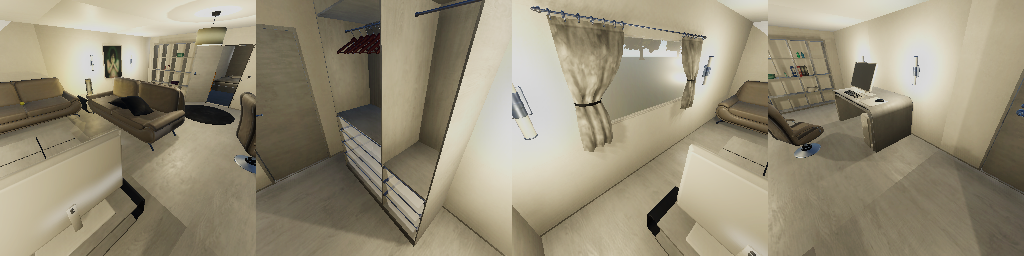}
    \caption{Example observation image.}
    \label{fig:view-exp}
\end{figure}

Each observation view is rendered at $256 \times 256$ pixels, resulting in a four-view concatenated observation of $256 \times 1024$. The four views are arranged in the order: front, back, left, and right. Both horizontal and vertical fields of view are set to $90^\circ$. The agent model height is 1.8\,m, and the camera is mounted at the same height with a forward offset of 0.15\,m relative to the agent's position. The camera is pitched downward by $30^\circ$ to capture more of the nearby surfaces and objects. Figure~\ref{fig:view-exp} shows an example of the resulting concatenated observation.

\begin{table*}[!t]
\centering
\small
\setlength{\tabcolsep}{10pt}
\renewcommand{\arraystretch}{1.05}
\caption{Full names of MLLMs used in our evaluation.}
\label{tab:model_versions}
\begin{tabular}{lll}
\toprule
\textbf{Model Name} & \textbf{Creator} & \textbf{Full Name / Identifier} \\
\midrule
GPT-5-mini & OpenAI & \texttt{gpt-5.4-mini} \\
GPT-5.4 & OpenAI & \texttt{gpt-5.4} \\
Gemini-3.1-Pro & Google & \texttt{gemini-3.1-pro} \\
\midrule
Qwen3-8B-VL & Qwen & \texttt{Qwen/Qwen3-VL-8B-Instruct} \\
Qwen3-32B-VL & Qwen & \texttt{Qwen/Qwen3-VL-32B-Instruct} \\
Qwen3-235B-VL & Qwen & \texttt{Qwen/Qwen3-VL-235B-A22B-Instruct} \\
Qwen3.5-9B & Qwen & \texttt{Qwen/Qwen3.5-9B} \\
Qwen3.5-27B & Qwen & \texttt{Qwen/Qwen3.5-27B} \\
\midrule
Gemma4-26B & Google & \texttt{google/gemma-4-26b-it} \\
Gemma4-31B & Google & \texttt{google/gemma-4-31b-it} \\
\midrule
InternVL-3.5-38B & OpenGVLab & \texttt{OpenGVLab/InternVL3\_5-38B} \\
InternVL-3.5-241B & OpenGVLab & \texttt{OpenGVLab/InternVL3\_5-241B-A28B} \\
\midrule
Llama-4 Scout & Meta & \texttt{meta-llama/Llama-4-Scout-17B-16E-Instruct} \\
\midrule
GLM-4.6V & Z.ai & \texttt{zai-org/GLM-4.6V} \\
GLM-4.6V-Flash & Z.ai & \texttt{zai-org/GLM-4.6V-Flash} \\
\bottomrule
\end{tabular}
\end{table*}

For all evaluated models, we adopt the recommended generation configurations whenever available. For closed-source models, we use the official APIs with their default generation settings. For open-source models, we follow the generation configurations specified in the corresponding model cards or vLLM~\cite{kwon2023efficient} documentation to ensure strong and reliable performance.

\subsection{Model Versions}
\label{app:model}
Table~\ref{tab:model_versions} summarizes the model versions and full identifiers used in our experiments. Proprietary models are accessed through their official APIs, while open-source models are locally deployed using vLLM~\cite{kwon2023efficient}.

\subsection{Action Space}
\label{app:action}
\begin{table*}[ht]
\centering
\small
\caption{Action space for agents in MECoBench. Parallel tasks include basic navigation and object manipulation actions. Sequential tasks introduce cross-room handover actions.}
\label{tab:action-space}
\begin{tabular}{lll}
\toprule
\textbf{Action} & \textbf{Description} & \textbf{Parameters} \\
\midrule
\multicolumn{3}{c}{\textit{Parallel Task Actions}} \\
\midrule
\texttt{walk} & Move towards a visible object, furniture, or door & object \\
\texttt{walk\_to\_room} & Move directly to another room & bedroom / livingroom / kitchen / bathroom \\
\texttt{grab} & Pick up a visible and grabbable object & object \\
\texttt{open} & Open a container & container \\
\texttt{put\_on} & Place the held object on a visible surface & object (holding), surface \\
\texttt{put\_in} & Place the held object inside a visible container & object (holding), container \\
\midrule
\multicolumn{3}{c}{\textit{Sequential Task Additional Actions}} \\
\midrule
\texttt{walk\_to\_door} & Move to a specific door in the environment topology & door\_id \\
\texttt{handover} & Give a held object to another agent when close (<2m) & object(holding), recipient agent name \\
\texttt{receive} & Wait to receive an object & none \\
\bottomrule
\end{tabular}
\end{table*}

Table~\ref{tab:action-space} lists the action space for agents when evaluation for both parallel and sequential tasks. Parallel tasks include basic navigation and object manipulation actions, while sequential tasks introduce cross-room handover actions.

\paragraph{Single-Agent Action Legality}
Each agent action must satisfy the following legality constraints:
\begin{itemize}
    \item \textbf{Hand occupancy:} Each agent can hold at most two objects. When both hands are occupied, the agent cannot execute \texttt{grab} or \texttt{open}. For \texttt{put\_*} and \texttt{handover}, the object to be placed or handed over must already be held by the acting agent.
    
    \item \textbf{Visibility:} The target objects of \texttt{walk}, \texttt{grab}, and \texttt{put\_*} actions must be visible in the agent's current observation.
    
    \item \textbf{Container state:} An agent cannot execute \texttt{open} on a container that is already open.
    
    \item \textbf{Room accessibility:} For \texttt{walk\_*}, \texttt{grab}, and \texttt{put\_*} actions, the target must be located in one of the rooms accessible to the agent.
    
    \item \textbf{Invalid actions:} Any action name not listed in Table~\ref{tab:action-space} is considered invalid.
\end{itemize}

\paragraph{Multi-Agent Interaction Constraints}
When multiple agents act simultaneously, additional interaction constraints are applied:
\begin{itemize}
    \item \textbf{Conflict resolution:} The conflict set includes \texttt{grab} and \texttt{open}. If multiple agents attempt the same conflict action on the same object at the same time, one agent is randomly selected as the \emph{winner}, while the others remain idle.
    
    \item \textbf{Handover distance:} A \texttt{handover} action is valid only when the giver and receiver are within 2 meters of each other.
\end{itemize}

\subsection{Action Decoding and Grounding}
\label{sec:action_decoding}
\begin{figure}[ht]
    \centering
    \includegraphics[width=0.9\linewidth]{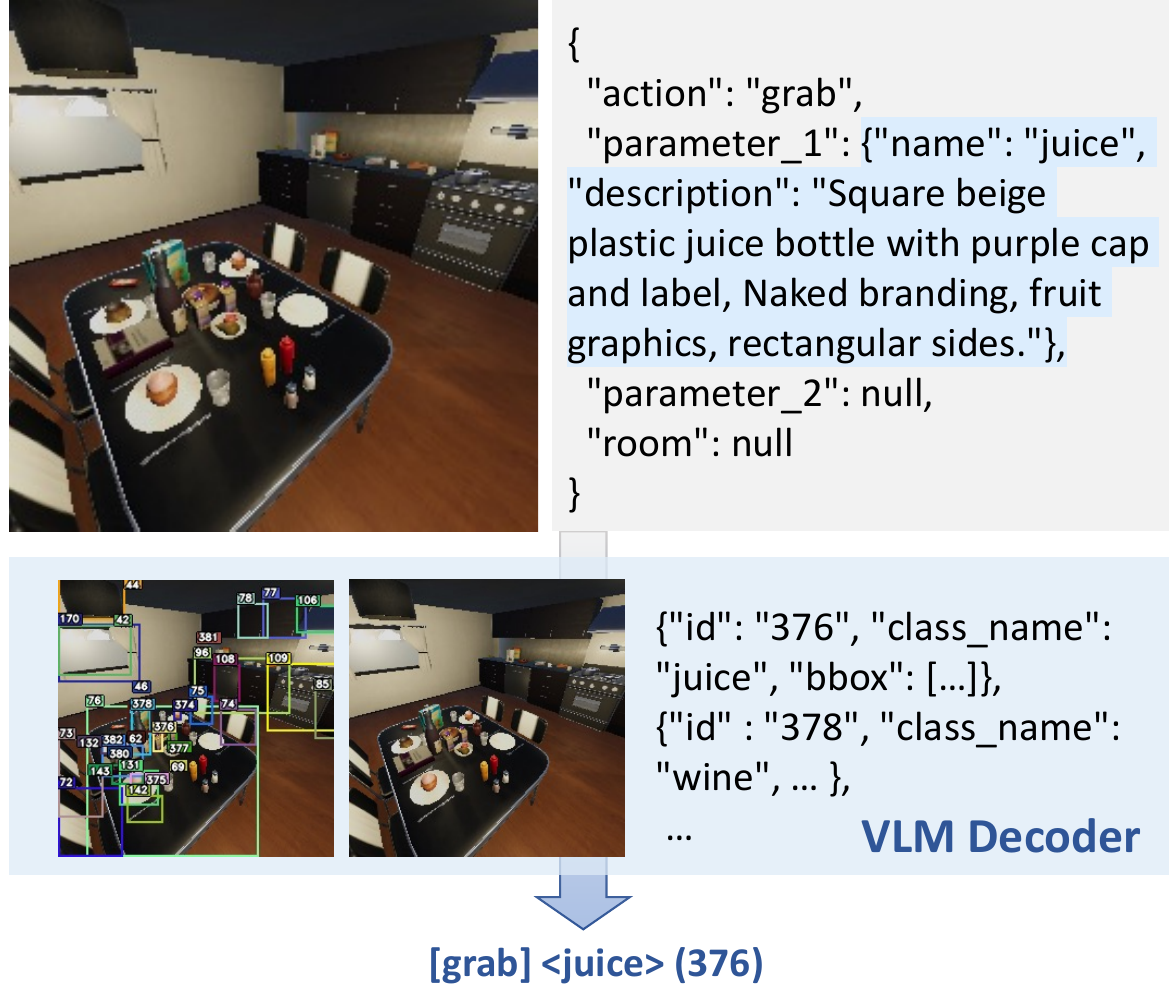}
    \caption{Example of action decoding and grounding.}
    \label{fig:action_decode_example}
\end{figure}
Agents output semantic actions based on their observation rather than executable simulator commands. 
Thus, each generated action is decoded and grounded before execution. 
Actions that do not require object grounding, such as \texttt{receive}, \texttt{walk\_to\_room/door}, and \texttt{handover}, are resolved by rule-based mappings. 
For object-centric actions, including \texttt{walk}, \texttt{grab}, \texttt{open}, \texttt{put\_on}, and \texttt{put\_in}, semantic object descriptions are grounded to concrete simulator object identifiers.

As shown in Fig.~\ref{fig:action_decode_example}, the top panels present the agent's first-person observation and generated semantic action. 
The decoder then uses visible object candidates, their bounding boxes, and metadata, together with both the original and bbox-annotated views, to select the target simulator ID and produce the final executable action.

\section{Core Prompt Templates}
\label{app:core-prompts}
This section presents the core prompt templates designed for agents in the MECoBench evaluation platform.

\subsection{Communicate Phase}
\begin{figure}[!ht]
    \centering
    \includegraphics[width=\linewidth]{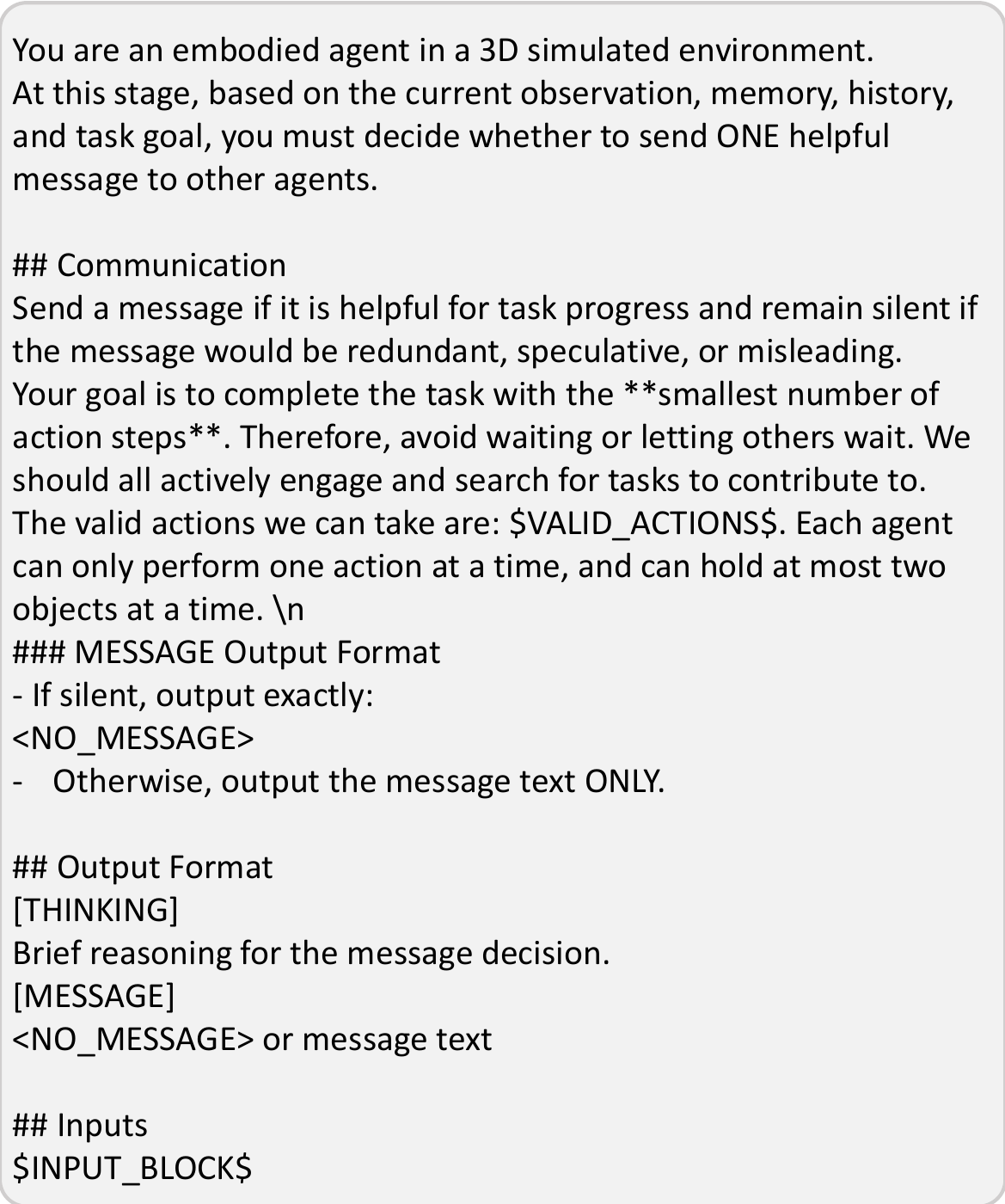}
    \caption{Broadcast protocol communication prompt. This module asks each agent to decide
    whether to send a concise, useful message before physical action planning.}
    \label{fig:prompt-base}
\end{figure}

\begin{figure}[!t]
    \centering
    \includegraphics[width=\linewidth]{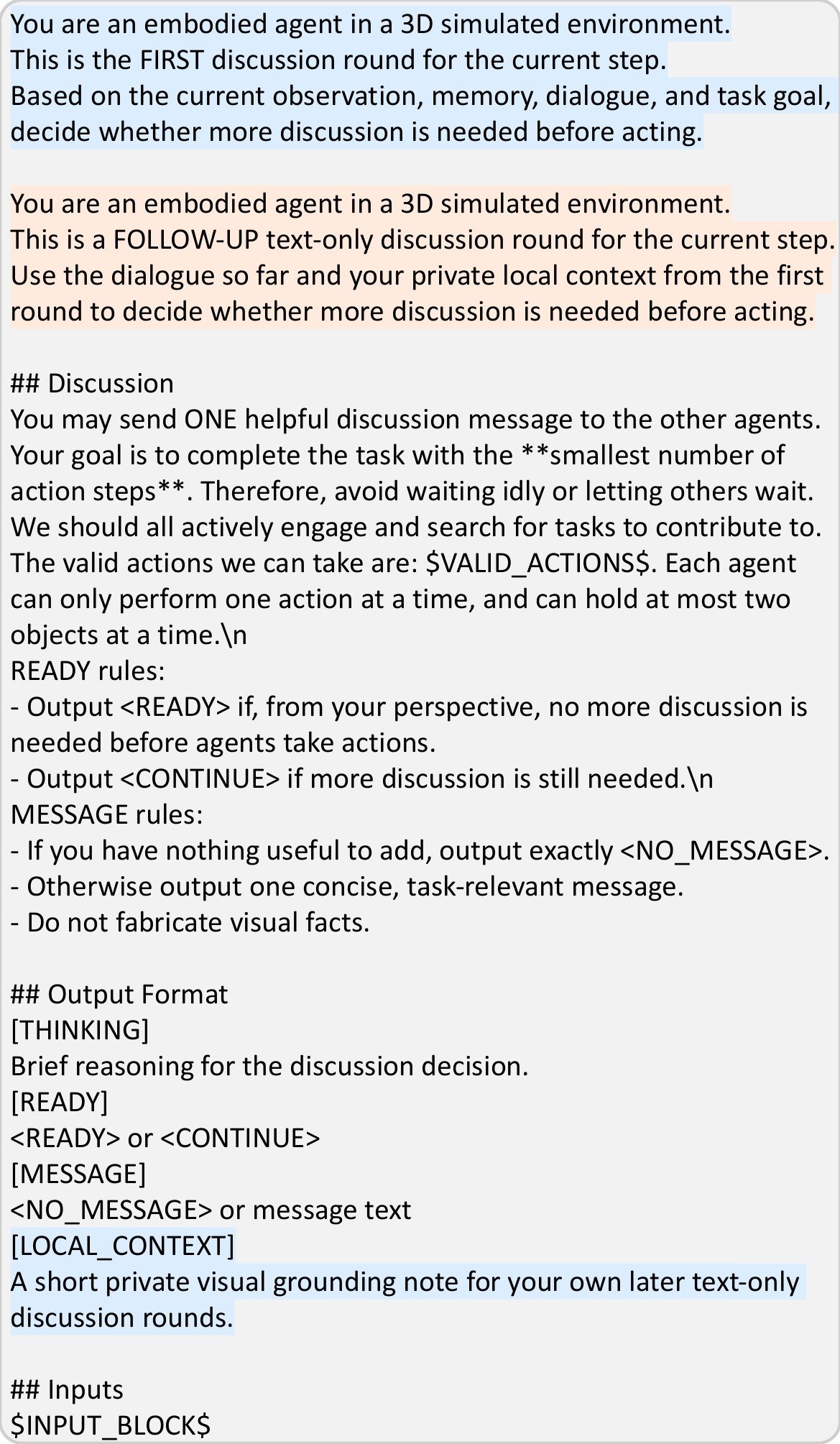}
    \caption{Discussion protocol prompt. The first round prompt is highlighted in blue, and the follow-up rounds prompt is highlighted in orange.}
    \label{fig:prompt-discussion}
\end{figure}

\begin{figure}[!t]
    \centering
    \includegraphics[width=\linewidth]{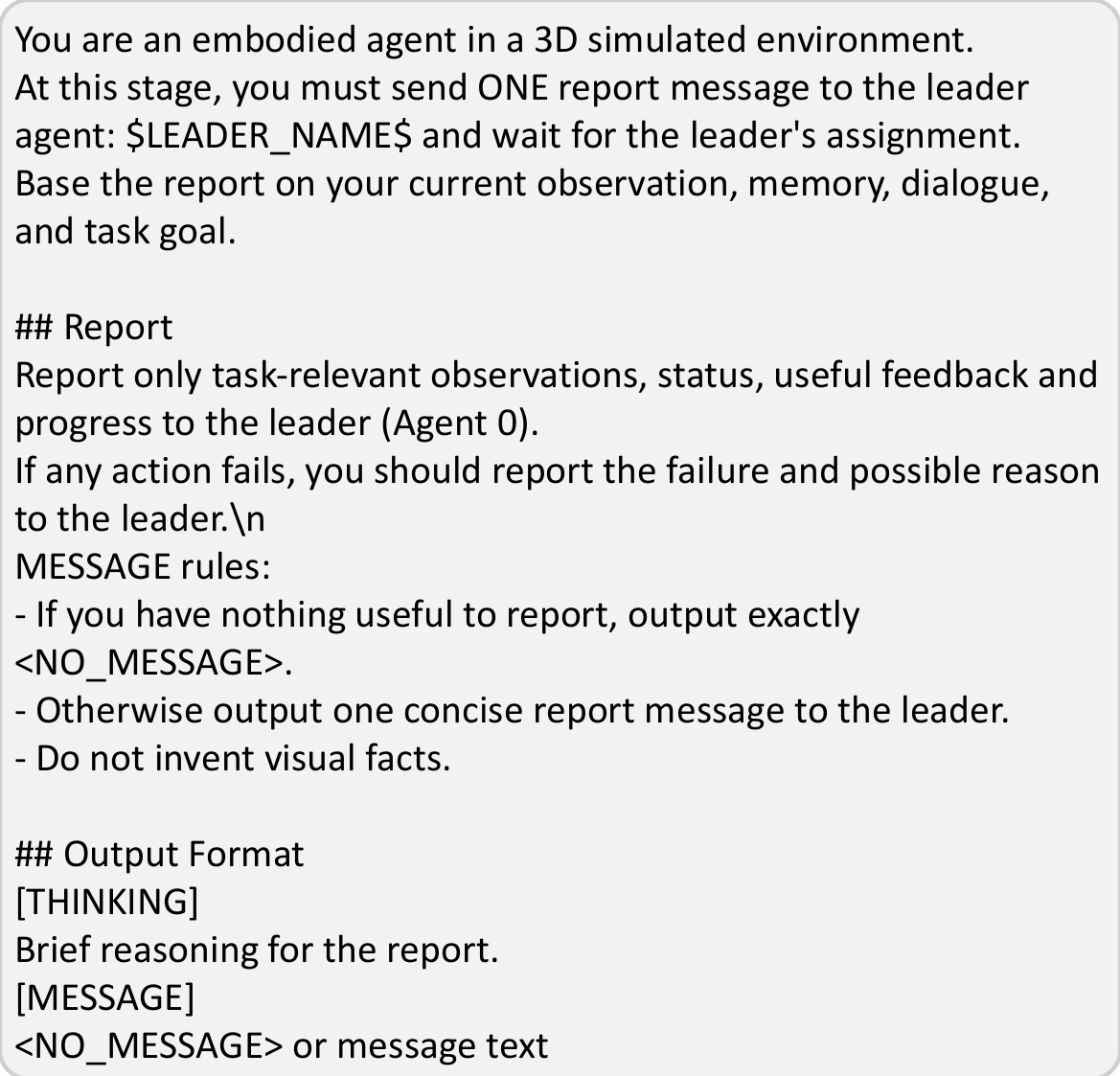}
    \caption{Centralized mode: worker report prompt.}
    \label{fig:prompt-worker}
\end{figure}

\begin{figure}[!t]
    \centering
    \includegraphics[width=\linewidth]{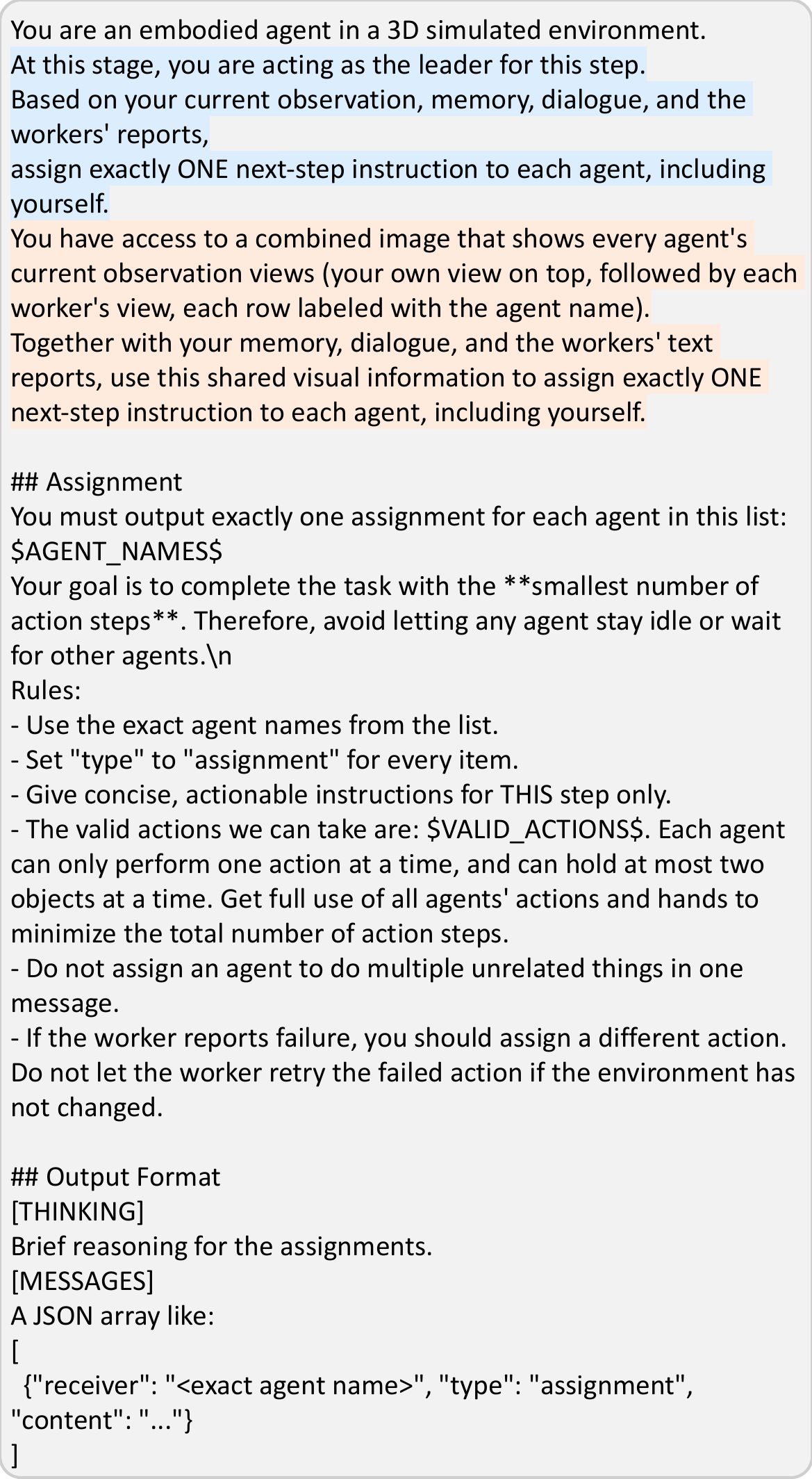}
    \caption{Centralized mode: leader assignment prompt. Text-only communication is highlighted in blue, and the vision-augmented is highlighted in orange.}
    \label{fig:prompt-leader}
\end{figure}

\begin{figure}[!t]
    \centering
    \includegraphics[width=\linewidth]{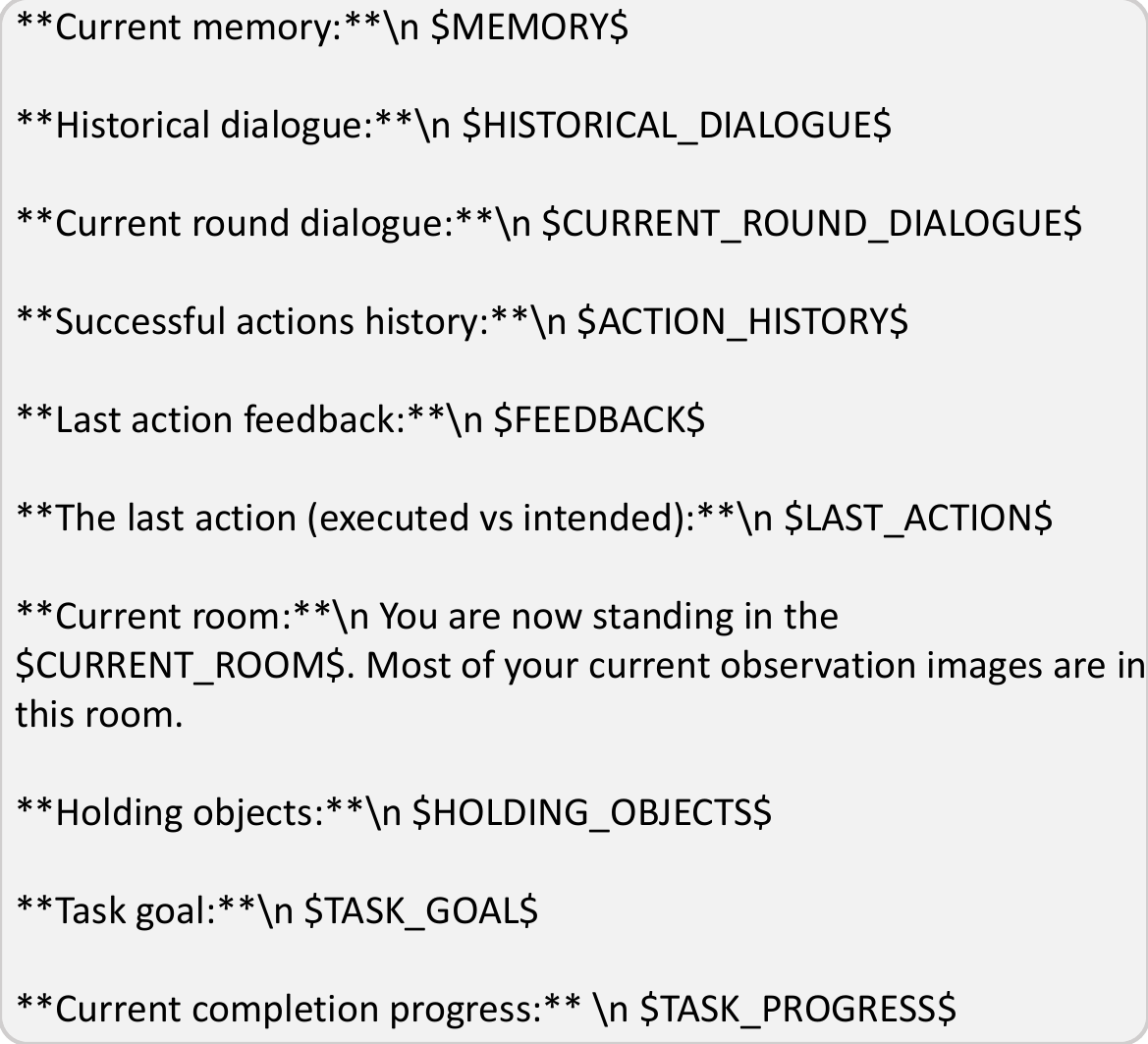}
    \caption{Input block for communicate phase.}
    \label{fig:prompt-input-comm}
\end{figure}

Below we list prompts for broadcast, discuss and leader-based communication protocol (Figure~\ref{fig:prompt-base},~\ref{fig:prompt-discussion},~\ref{fig:prompt-worker},~\ref{fig:prompt-leader},~\ref{fig:prompt-input-comm}).

\subsection{Act Phase}
\begin{figure}[!h]
    \centering
    \includegraphics[width=\linewidth]{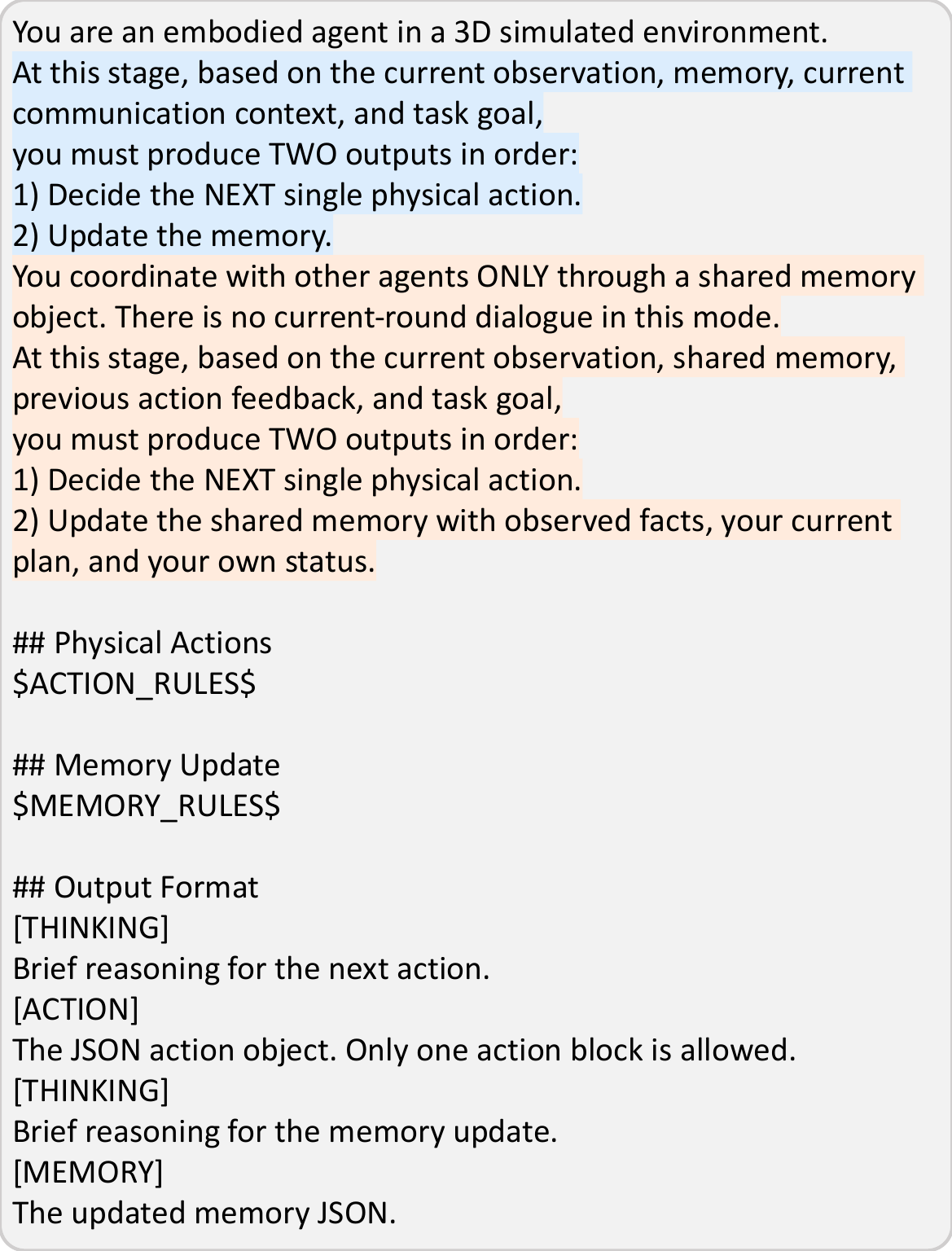}
    \caption{Action and memory update prompt. Normal memory is highlighted in blue and shared memory is highlighted in orange.}
    \label{fig:prompt-action-memory}
\end{figure}

\begin{figure}[!h]
    \centering
    \includegraphics[width=\linewidth]{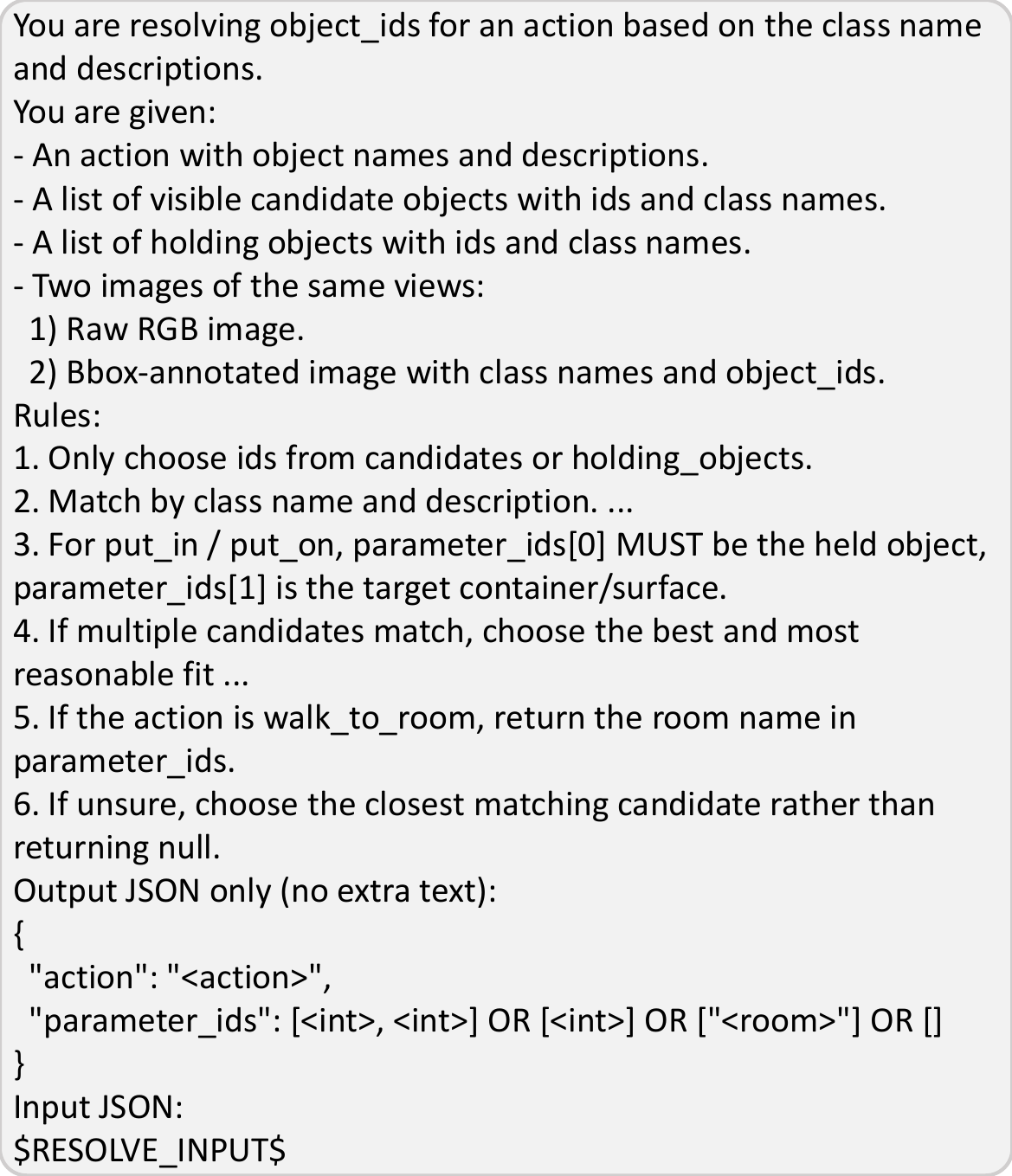}
    \caption{Prompt for action decoder.}
    \label{fig:prompt-action-decode}
\end{figure}

\begin{figure}[!h]
    \centering
    \includegraphics[width=\linewidth]{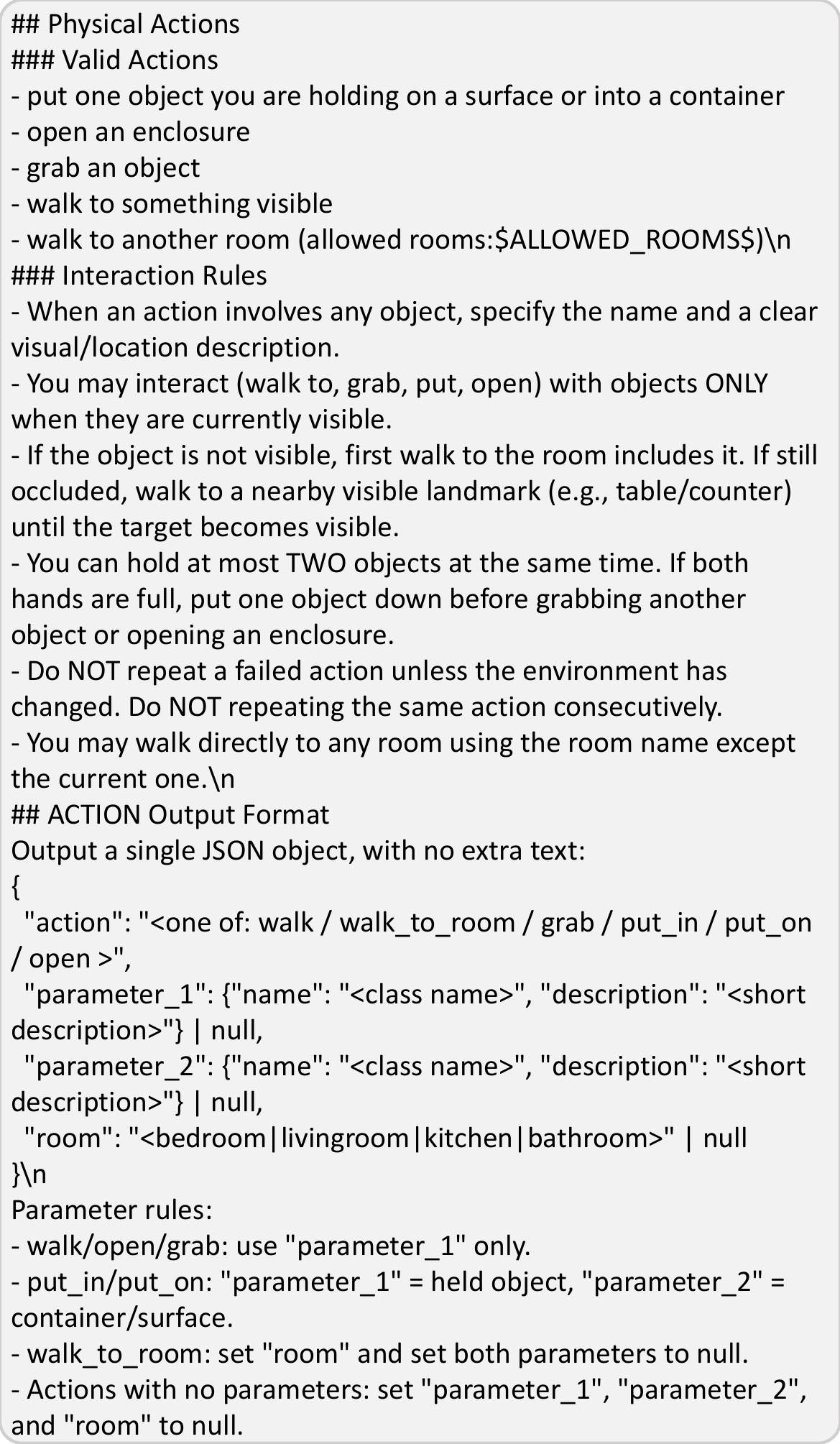}
    \caption{Action rules block.}
    \label{fig:prompt-action-rule}
\end{figure}

\begin{figure}[!h]
    \centering
    \includegraphics[width=\linewidth]{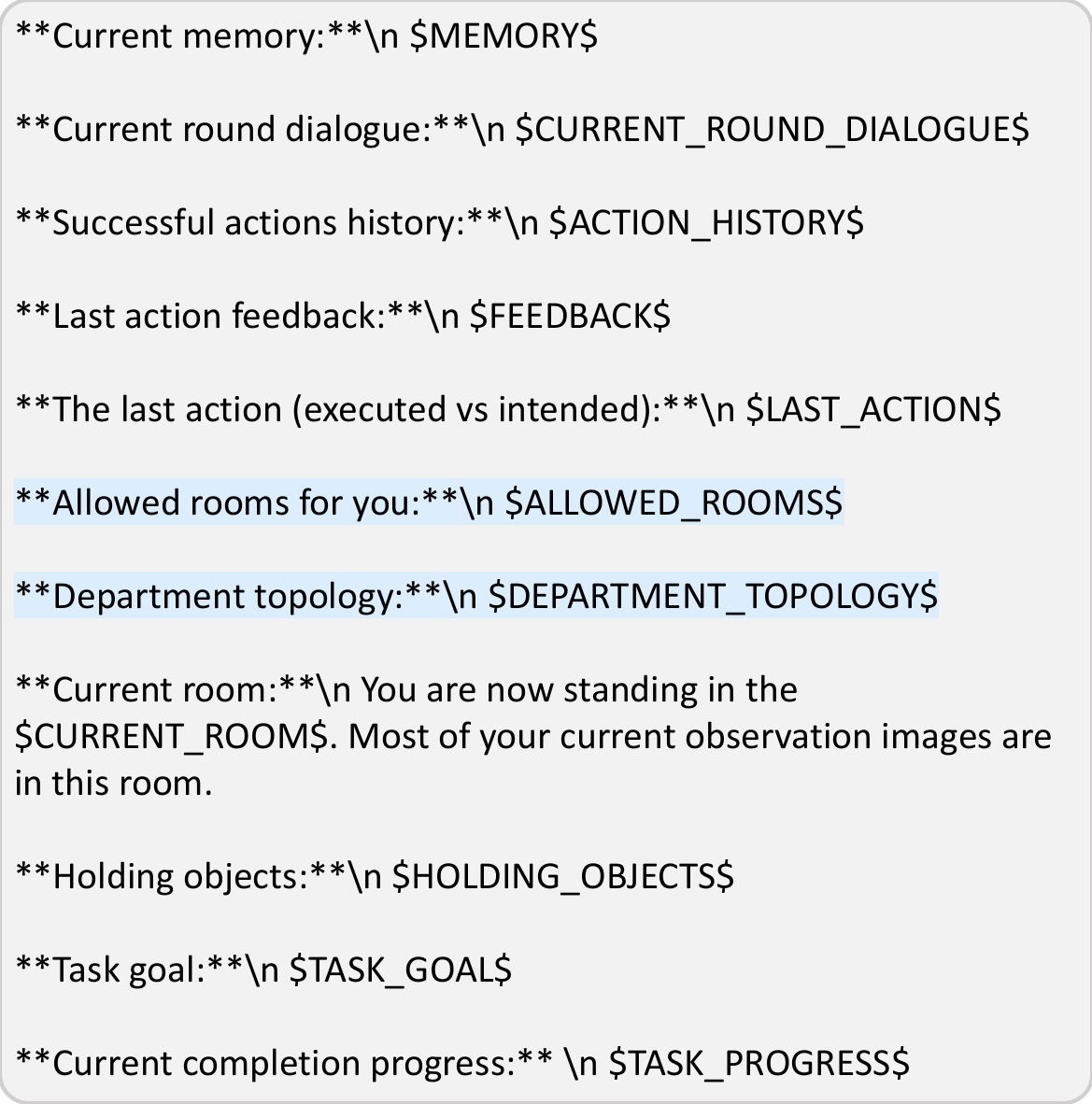}
    \caption{Input block for the act phase. For sequential tasks, the two additional parts are highlighted in blue.}
    \label{fig:prompt-input-act}
\end{figure}

\begin{figure}[!h]
    \centering
    \includegraphics[width=\linewidth]{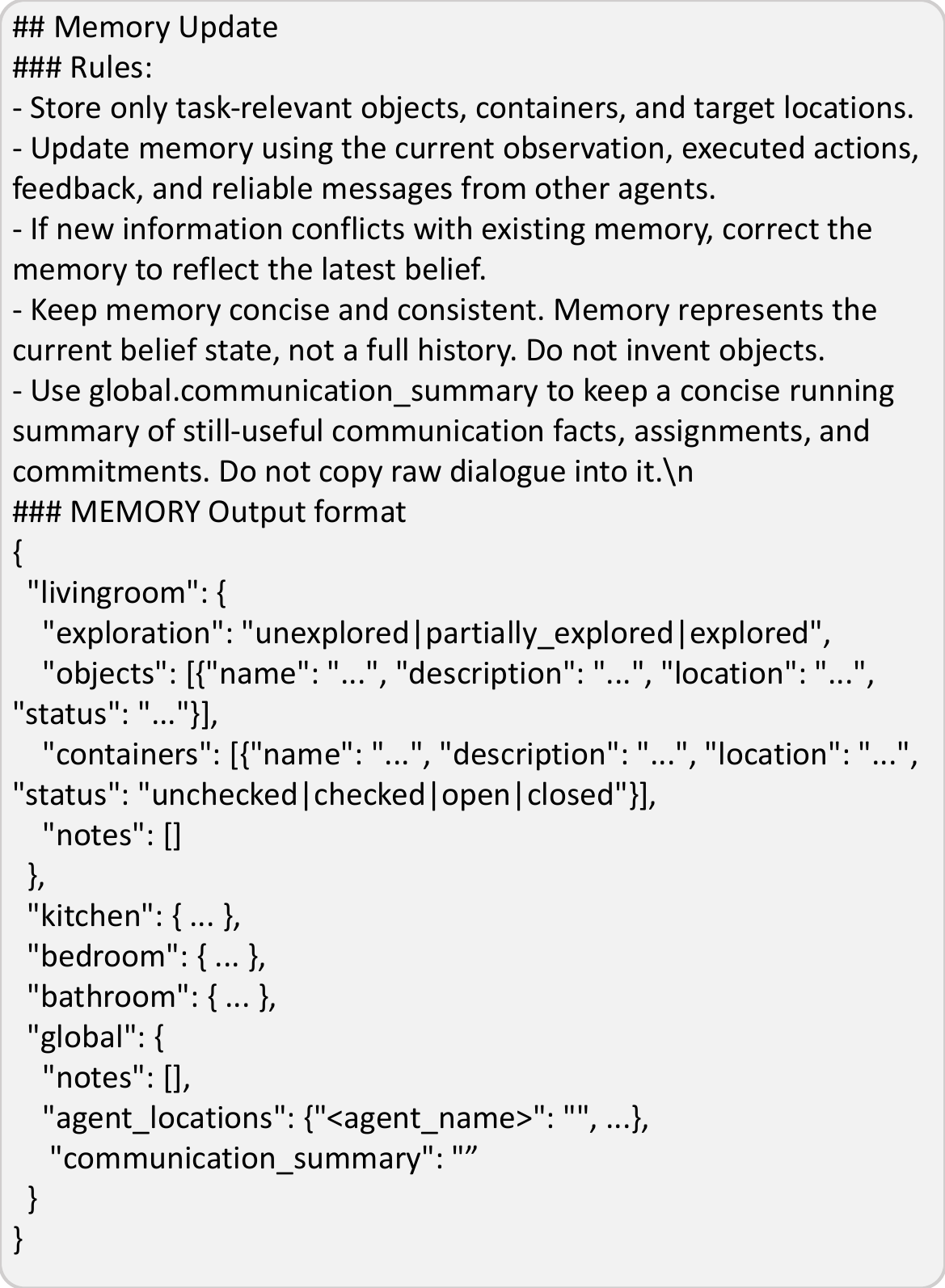}
    \caption{Memory update rules block.}
    \label{fig:prompt-memory-rule}
\end{figure}

\begin{figure}[!h]
    \centering
    \includegraphics[width=\linewidth]{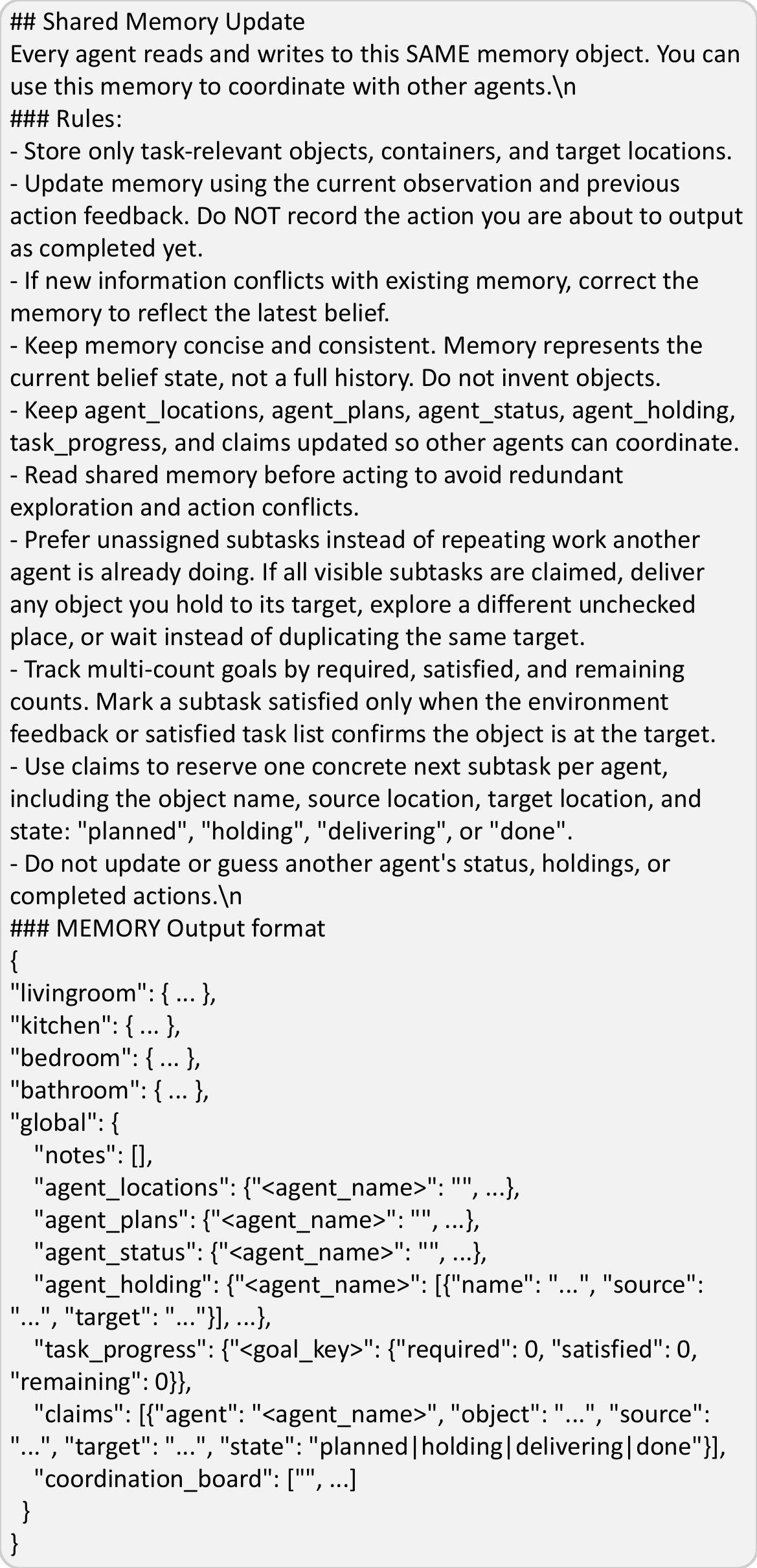}
    \caption{Shared memory update rules block.}
    \label{fig:prompt-shared-memory-rule}
\end{figure}

This section presents the prompts used for action decision-making, memory update and action decoding (Figure~\ref{fig:prompt-action-memory},~\ref{fig:prompt-action-decode},~\ref{fig:prompt-action-rule},~\ref{fig:prompt-input-act},~\ref{fig:prompt-memory-rule},~\ref{fig:prompt-shared-memory-rule}).

\section{Experiment Results}
In this section, we provide detailed results for the experiments in Sections~\ref{sec:exp-1},~\ref{sec:exp-2}, and~\ref{sec:exp-3}, as well as some additional experimental results.
\subsection{Performance of Different Models in Parallel and Sequential Tasks}
\label{app:main_results}

\begin{table*}[t]
\centering
\small
\newcommand{\graycell}[1]{\cellcolor{gray!10}#1}
\caption{Performance comparison under parallel and sequential normal settings. Avg. SR and Avg. CR denote the average SR and CR across parallel and sequential tasks under the 2-agent setup. Best results are shown in bold, and second-best results are underlined, respectively for 2-agent and 1-agent setup.}
\label{tab:main-1}
\begin{tabular}{lc ccc ccc cc}
\toprule
\textbf{Model} & \textbf{\#Agents}
& \multicolumn{3}{c}{\cellcolor{blue!10}\textbf{Parallel Task}}
& \multicolumn{3}{c}{\cellcolor{orange!15}\textbf{Sequential Task}}
& \multicolumn{2}{c}{\textbf{Avg.}} \\
\cmidrule(lr){3-5} \cmidrule(lr){6-8} \cmidrule(lr){9-10}
& & \textbf{SR} & \textbf{CR} & \textbf{AUC}
& \textbf{SR} & \textbf{CR} & \textbf{AUC}
& \textbf{SR} & \textbf{CR} \\
\midrule
\multicolumn{10}{c}{\textit{Closed-source MLLMs}} \\
\midrule
\multirow{2}{*}{GPT-5-mini}
& 1 & \textbf{0.906} & \textbf{0.974} & \textbf{0.793} & -- & -- & -- & -- & -- \\
& \graycell{2} & \graycell{\underline{0.917}} & \graycell{\underline{0.982}} & \graycell{\textbf{0.776}} & \graycell{0.719} & \graycell{0.901} & \graycell{\underline{0.624}} & \graycell{0.818} & \graycell{\textbf{0.942}} \\
\multirow{2}{*}{GPT-5.4}
& 1 & \underline{0.854} & \underline{0.960} & \underline{0.788} & -- & -- & -- & -- & -- \\
& \graycell{2} & \graycell{0.896} & \graycell{0.969} & \graycell{0.731} & \graycell{\underline{0.771}} & \graycell{\textbf{0.914}} & \graycell{\textbf{0.655}} & \graycell{\textbf{0.834}} & \graycell{\textbf{0.942}} \\
\multirow{2}{*}{Gemini-3.1-Pro}
& 1 & 0.802 & 0.927 & 0.645 & -- & -- & -- & -- & -- \\
& \graycell{2} & \graycell{0.771} & \graycell{0.937} & \graycell{0.606} & \graycell{0.573} & \graycell{0.824} & \graycell{0.472} & \graycell{0.672} & \graycell{0.881} \\
\midrule
\multicolumn{10}{c}{\textit{Open-source MLLMs}} \\
\midrule
\multirow{2}{*}{Qwen3-8B-VL}
& 1 & 0.354 & 0.620 & 0.419 & -- & -- & -- & -- & -- \\
& \graycell{2} & \graycell{0.458} & \graycell{0.770} & \graycell{0.535} & \graycell{0.073} & \graycell{0.332} & \graycell{0.208} & \graycell{0.266} & \graycell{0.551} \\
\multirow{2}{*}{Qwen3-32B-VL}
& 1 & 0.583 & 0.753 & 0.542 & -- & -- & -- & -- & -- \\
& \graycell{2} & \graycell{0.802} & \graycell{0.934} & \graycell{0.694} & \graycell{0.365} & \graycell{0.620} & \graycell{0.393} & \graycell{0.584} & \graycell{0.777} \\
\multirow{2}{*}{Qwen3-235B-VL}
& 1 & 0.656 & 0.855 & 0.648 & -- & -- & -- & -- & -- \\
& \graycell{2} & \graycell{0.667} & \graycell{0.896} & \graycell{0.649} & \graycell{0.354} & \graycell{0.675} & \graycell{0.451} & \graycell{0.511} & \graycell{0.786} \\
\multirow{2}{*}{Qwen3.5-9B}
& 1 & \underline{0.854} & 0.924 & 0.721 & -- & -- & -- & -- & -- \\
& \graycell{2} & \graycell{\textbf{0.938}} & \graycell{\textbf{0.985}} & \graycell{\underline{0.764}} & \graycell{0.615} & \graycell{0.844} & \graycell{0.558} & \graycell{0.777} & \graycell{0.915} \\
\multirow{2}{*}{Qwen3.5-27B}
& 1 & 0.844 & 0.950 & 0.760 & -- & -- & -- & -- & -- \\
& \graycell{2} & \graycell{\textbf{0.938}} & \graycell{0.969} & \graycell{0.726} & \graycell{0.698} & \graycell{0.872} & \graycell{0.587} & \graycell{0.818} & \graycell{0.921} \\
\multirow{2}{*}{Gemma4-26B}
& 1 & 0.760 & 0.867 & 0.636 & -- & -- & -- & -- & -- \\
& \graycell{2} & \graycell{0.812} & \graycell{0.944} & \graycell{0.644} & \graycell{0.688} & \graycell{0.839} & \graycell{0.556} & \graycell{0.750} & \graycell{0.892} \\
\multirow{2}{*}{Gemma4-31B}
& 1 & 0.833 & 0.921 & 0.703 & -- & -- & -- & -- & -- \\
& \graycell{2} & \graycell{0.833} & \graycell{0.945} & \graycell{0.688} & \graycell{\textbf{0.823}} & \graycell{\underline{0.904}} & \graycell{0.620} & \graycell{\underline{0.828}} & \graycell{\underline{0.925}} \\
\multirow{2}{*}{InternVL-3.5-38B}
& 1 & 0.677 & 0.875 & 0.659 & -- & -- & -- & -- & -- \\
& \graycell{2} & \graycell{0.594} & \graycell{0.830} & \graycell{0.553} & \graycell{0.135} & \graycell{0.429} & \graycell{0.280} & \graycell{0.365} & \graycell{0.630} \\
\multirow{2}{*}{InternVL-3.5-241B}
& 1 & 0.833 & 0.946 & 0.728 & -- & -- & -- & -- & -- \\
& \graycell{2} & \graycell{0.719} & \graycell{0.914} & \graycell{0.594} & \graycell{0.031} & \graycell{0.267} & \graycell{0.180} & \graycell{0.375} & \graycell{0.591} \\
\multirow{2}{*}{Llama-4}
& 1 & 0.250 & 0.467 & 0.345 & -- & -- & -- & -- & -- \\
& \graycell{2} & \graycell{0.250} & \graycell{0.543} & \graycell{0.380} & \graycell{0.042} & \graycell{0.225} & \graycell{0.163} & \graycell{0.146} & \graycell{0.384} \\
\multirow{2}{*}{GLM-4.6V}
& 1 & 0.594 & 0.825 & 0.501 & -- & -- & -- & -- & -- \\
& \graycell{2} & \graycell{0.719} & \graycell{0.895} & \graycell{0.600} & \graycell{0.010} & \graycell{0.283} & \graycell{0.188} & \graycell{0.365} & \graycell{0.589} \\
\multirow{2}{*}{GLM-4.6V-Flash}
& 1 & 0.094 & 0.351 & 0.271 & -- & -- & -- & -- & -- \\
& \graycell{2} & \graycell{0.125} & \graycell{0.456} & \graycell{0.316} & \graycell{0.000} & \graycell{0.090} & \graycell{0.060} & \graycell{0.063} & \graycell{0.273} \\
\bottomrule
\end{tabular}
\end{table*}

Table~\ref{tab:main-1} shows the detailed result of figure~\ref{fig:1-1-overall} and \ref{fig:1-1-gain}. Overall, GPT-5.4 achieves the best performance. GPT-5-mini and Gemma4-31B are the second-tier models, showing consistently strong results across the two collaboration structures. In addition, Qwen3.5, Gemma4-26B, and Gemini-3.1-Pro also exhibit relatively strong performance. Qwen3-VL-32B and Qwen3-VL-235B fall into the middle-performance group, with relatively balanced results across parallel and sequential tasks. Qwen3-VL-8B is a relatively weak model, but its performance remains comparatively balanced between the two coordination structures. In contrast, InternVL-3.5 and GLM-4.6V struggle particularly on sequential tasks, where precise collaboration is required. Finally, Llama-4-Scout and GLM-4.6V-Flash perform poorly under both collaboration structures.

\paragraph{Model scaling effects.} Overall, except for InternVL-3.5, larger models generally perform better within the same family. However, scaling tends to amplify existing task-specific patterns rather than change them. Some families remain balanced across parallel and sequential tasks, while others are consistently stronger in single-agent or parallel settings, or particularly weak on sequential tasks requiring precise coordination. For example, Gemma4 benefits from scaling especially on sequential tasks, whereas InternVL-3.5 achieves stronger single-agent parallel performance at larger scale but degrades in sequential collaboration. This suggests that stronger individual capability does not necessarily translate into better communication and coordination.

Based on model performance and cost-effectiveness, we select Qwen3-VL-32B and Qwen3-VL-8B as two representative models for studying team-size scaling and related factors in subsequent experiments. For strong models, we conduct most experiments with GPT-5-mini and Gemma4-31B, and include GPT-5.4 when necessary.

\begin{table}[htbp]
\centering
\small
\caption{Results on Parallel tasks for Qwen3-VL models with one to five agents. Comparison of performance with (broadcast) and without communication. The best results of both models are shown in bold.}
\label{tab:qwen3_parallel_normal_agents}
\setlength{\tabcolsep}{4pt}
\begin{tabular}{c ccc ccc}
\toprule
\textbf{\#Ag.}
& \multicolumn{3}{c}{\cellcolor{blue!10}\textbf{With comm.}}
& \multicolumn{3}{c}{\cellcolor{orange!15}\textbf{Without Comm.}} \\
\cmidrule(lr){2-4} \cmidrule(lr){5-7}
& \textbf{SR} & \textbf{CR} & \textbf{AUC}
& \textbf{SR} & \textbf{CR} & \textbf{AUC} \\
\midrule
\multicolumn{7}{c}{\textit{Qwen3-VL-8B}} \\
\midrule
1 & 0.354 & 0.620 & 0.419 & 0.354 & 0.620 & \textbf{0.419} \\
2 & 0.458 & 0.770 & 0.535 & 0.271 & 0.546 & 0.339 \\
3 & \textbf{0.521} & \textbf{0.837} & \textbf{0.558} & \textbf{0.396} & \textbf{0.655} & 0.411 \\
4 & 0.469 & 0.794 & 0.501 & 0.323 & 0.642 & 0.409 \\
5 & 0.458 & 0.794 & 0.468 & --    & --    & --    \\
\midrule
\multicolumn{7}{c}{\textit{Qwen3-VL-32B}} \\
\midrule
1 & 0.583 & 0.753 & 0.542 & \textbf{0.583} & 0.753 & 0.542 \\
2 & 0.802 & 0.934 & \textbf{0.694} & \textbf{0.583} & \textbf{0.802} & \textbf{0.567} \\
3 & 0.833 & 0.943 & 0.656 & 0.500 & 0.705 & 0.481 \\
4 & \textbf{0.844} & \textbf{0.954} & 0.658 & 0.469 & 0.720 & 0.468 \\
5 & 0.740 & 0.892 & 0.567 & --    & --    & --    \\
\bottomrule
\end{tabular}
\end{table}

\begin{table*}[!t]
\centering
\small
\caption{Success rate and completion rate by number of objects for Qwen3-VL models on Parallel tasks. Best results for each model are shown in bold, and second-best results are underlined.}
\label{tab:qwen_obj_vs_agents}
\setlength{\tabcolsep}{3pt}
\begin{tabular}{l c cc cc cc cc cc}
\toprule
\textbf{Model} & \textbf{\#Objects}
& \multicolumn{2}{c}{\textbf{1 Agent}}
& \multicolumn{2}{c}{\textbf{2 Agents}}
& \multicolumn{2}{c}{\textbf{3 Agents}}
& \multicolumn{2}{c}{\textbf{4 Agents}}
& \multicolumn{2}{c}{\textbf{5 Agents}} \\
\cmidrule(lr){3-4} \cmidrule(lr){5-6} \cmidrule(lr){7-8} \cmidrule(lr){9-10} \cmidrule(lr){11-12}
& & \textbf{SR} & \textbf{CR}
& \textbf{SR} & \textbf{CR}
& \textbf{SR} & \textbf{CR}
& \textbf{SR} & \textbf{CR}
& \textbf{SR} & \textbf{CR} \\
\midrule
\multirow{6}{*}{Qwen3-32B-VL}
& 2 & 0.875 & 0.938 & 0.875 & 0.938 & \underline{0.938} & \underline{0.969} & \underline{0.938} & \underline{0.969} & \textbf{1.000} & \textbf{1.000} \\
& 3 & 0.765 & 0.863 & \textbf{1.000} & \textbf{1.000} & \textbf{1.000} & \textbf{1.000} & \underline{0.941} & \underline{0.980} & 0.875 & 0.938 \\
& 4 & 0.625 & 0.812 & \underline{0.938} & \underline{0.984} & \textbf{1.000} & \textbf{1.000} & \textbf{1.000} & \textbf{1.000} & 0.750 & 0.922 \\
& 5 & 0.438 & 0.650 & 0.688 & 0.838 & 0.688 & 0.888 & \textbf{0.812} & \textbf{0.938} & \underline{0.800} & \underline{0.867} \\
& 6 & 0.389 & 0.630 & \underline{0.667} & \underline{0.935} & \textbf{0.778} & \textbf{0.944} & \underline{0.667} & \underline{0.935} & 0.500 & 0.880 \\
& 7 & 0.385 & 0.604 & 0.615 & 0.901 & 0.583 & 0.905 & \textbf{0.750} & \textbf{0.964} & \underline{0.667} & \underline{0.917} \\
\midrule
\multirow{6}{*}{Qwen3-8B-VL}
& 2 & 0.750 & 0.875 & \underline{0.875} & \underline{0.938} & \textbf{0.938} & \textbf{0.969} & 0.812 & 0.906 & \underline{0.875} & \underline{0.938} \\
& 3 & 0.588 & 0.784 & \underline{0.824} & \underline{0.922} & \textbf{0.882} & \textbf{0.941} & \underline{0.824} & \underline{0.922} & 0.706 & 0.863 \\
& 4 & 0.312 & 0.562 & \textbf{0.500} & \textbf{0.828} & \textbf{0.500} & \textbf{0.828} & \textbf{0.500} & \textbf{0.828} & \underline{0.438} & \underline{0.766} \\
& 5 & 0.250 & 0.562 & 0.375 & 0.713 & \underline{0.438} & \underline{0.800} & 0.333 & 0.747 & \textbf{0.500} & \textbf{0.787} \\
& 6 & 0.111 & 0.444 & 0.056 & 0.685 & \underline{0.167} & \underline{0.713} & \textbf{0.278} & \textbf{0.741} & 0.111 & 0.704 \\
& 7 & \underline{0.077} & \underline{0.472} & \underline{0.077} & \underline{0.483} & \textbf{0.154} & \textbf{0.769} & 0.000 & 0.637 & \underline{0.077} & \underline{0.692} \\
\bottomrule
\end{tabular}
\end{table*}

\begin{table*}[!t]
\centering
\small
\newcommand{\graycell}[1]{\cellcolor{gray!10}#1}
\caption{Results for 2-agent broadcast and no-communication protocols on Parallel and Sequential tasks. NA indicates that no handover attempt was made.}
\label{tab:coord_base_nocomm_parallel_seq}
\setlength{\tabcolsep}{3pt}
\begin{tabular}{ll cccc cccc}
\toprule
\textbf{Model} & \textbf{Mode}
& \multicolumn{4}{c}{\cellcolor{blue!10}\textbf{Parallel Task}}
& \multicolumn{4}{c}{\cellcolor{orange!15}\textbf{Sequential Task}} \\
\cmidrule(lr){3-6} \cmidrule(lr){7-10}
& & \textbf{SR} & \textbf{CR} & \textbf{AUC} & \textbf{CAR}
& \textbf{SR} & \textbf{CR} & \textbf{AUC} & \textbf{HFR} \\
\midrule
\multirow{2}{*}{Gemma4-31B}
& Broadcast    & 0.833 & 0.945 & 0.689 & 0.019 & 0.823 & 0.904 & 0.620 & 0.322 \\
& \graycell{No Comm} & \graycell{0.719} & \graycell{0.863} & \graycell{0.558} & \graycell{0.023} & \graycell{0.177} & \graycell{0.464} & \graycell{0.308} & \graycell{0.486} \\
\multirow{2}{*}{Qwen3-32B-VL}
& Broadcast    & 0.802 & 0.934 & 0.694 & 0.020 & 0.365 & 0.620 & 0.393 & 0.294 \\
& \graycell{No Comm} & \graycell{0.583} & \graycell{0.802} & \graycell{0.567} & \graycell{0.028} & \graycell{0.073} & \graycell{0.325} & \graycell{0.232} & \graycell{0.733} \\
\multirow{2}{*}{Qwen3.5-9B}
& Broadcast    & 0.938 & 0.985 & 0.764 & 0.008 & 0.615 & 0.844 & 0.558 & 0.400 \\
& \graycell{No Comm} & \graycell{0.865} & \graycell{0.939} & \graycell{0.680} & \graycell{0.036} & \graycell{0.062} & \graycell{0.325} & \graycell{0.231} & \graycell{0.903} \\
\multirow{2}{*}{Qwen3-8B-VL}
& Broadcast    & 0.458 & 0.770 & 0.535 & 0.010 & 0.073 & 0.332 & 0.208 & 0.662 \\
& \graycell{No Comm} & \graycell{0.274} & \graycell{0.552} & \graycell{0.342} & \graycell{0.017} & \graycell{0.042} & \graycell{0.182} & \graycell{0.132} & \graycell{NA} \\
\bottomrule
\end{tabular}
\vspace{-12pt}
\end{table*}

\begin{figure}[!t]
    \centering
    \includegraphics[width=1\linewidth]{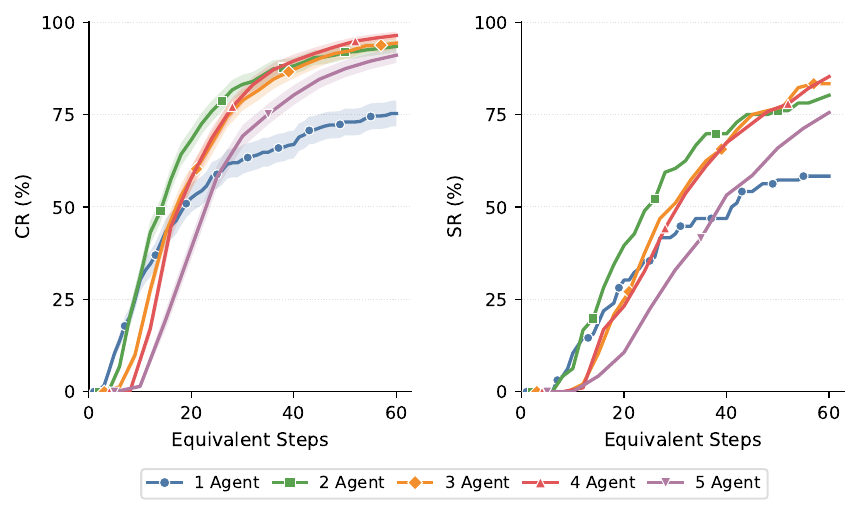}
    \caption{Task progress over steps for 1 to 5 agents under the broadcast protocol on the parallel task. Results are obtained with Qwen3-VL-32B.}
    \label{fig:agent-num-curve}
    \vspace{-12pt}
\end{figure}

\paragraph{Team size scaling effects.} Table~\ref{tab:qwen3_parallel_normal_agents} shows the detailed result for Figure~\ref{fig:agent-number-complexity} (a) and \ref{fig:role_of_communication}(b). The results show that: (1) Simply increasing the number of agents does not necessarily improve performance. Without communication, adding more agents can even introduce disorder and redundant actions, leading to lower efficiency and worse task completion. However, when agents are allowed to communicate, multi-agent collaboration becomes more effective in coordinating actions and completing tasks.
(2) More agents are not always better. Qwen3-VL-32B benefits from additional agents up to four agents, while Qwen3-VL-8B reaches its best performance with three agents and drops afterward. This suggests that weaker models are more affected by coordination overhead. 
(3) Collaboration can partially compensate for limited individual model capability, but cannot fully remove the capability gap. With communication, three Qwen3-VL-8B agents approach the single-agent performance of Qwen3-VL-32B, but the final performance remains constrained by the underlying model capability. Table~\ref{tab:qwen_obj_vs_agents} shows the detailed results in figure~\ref{fig:agent-number-complexity}(b). This indicate that multi-agent collaboration is especially helpful for more complex tasks with more target objects. For 7-object tasks, the SR nearly doubles for both models: from 0.385 to 0.750 for Qwen3-VL-32B, and from 0.077 to 0.154 for Qwen3-VL-8B. A task progress comprison of different team size shown in Figure~\ref{fig:agent-num-curve}.

\subsection{Role of communication}
\label{app:comm-role}
Table~\ref{tab:coord_base_nocomm_parallel_seq} shows the detailed results in figure~\ref{fig:role_of_communication}(a) when ablating communication from collaboration. Removing communication degrades both task performance and collaboration quality. In parallel tasks with a shared workspace, it generally increases CAR, indicating more conflicts actions between agents.  In sequential tasks, the impact is more severe, as agents fail to coordinate cross-region handovers, causing HFR to rise substantially.

\subsection{Comparison of Different Collaboration Modes}
\label{app:collab-mode}

\begin{table*}[ht]
\centering
\small
\caption{Results of different communication protocol under parallel and sequential tasks across models. \textit{Spatial Err} denotes the frequency of spatial errors, where an agent attempts to move to a location outside its assigned region. Best results within each model are shown in bold, and cross-model best results are highlighted in yellow.}
\label{tab:mode_radar_data}
\setlength{\tabcolsep}{3pt}
\begin{tabular}{ll cccccc cccccc}
\toprule
\textbf{Model} & \textbf{Mode}
& \multicolumn{6}{c}{\cellcolor{blue!10}\textbf{Parallel}}
& \multicolumn{6}{c}{\cellcolor{orange!15}\textbf{Sequential}} \\
\cmidrule(lr){3-8} \cmidrule(lr){9-14}
& & \textbf{SR} & \textbf{CR} & \textbf{AUC} & \textbf{DOL} & \textbf{Tokens} & \textbf{CAR}
& \textbf{SR} & \textbf{CR} & \textbf{AUC} & \textbf{Spatial Err} & \textbf{Tokens} & \textbf{HFR} \\
\midrule
\multicolumn{14}{c}{\textit{\textbf{Closed-source MLLMs}}} \\
\midrule
\multirow{3}{*}{GPT-5.4}
& Broadcast & \cellcolor{yellow!25}\textbf{0.900} & \textbf{0.961} & 0.700 & 0.773 & \textbf{4331.9} & 0.015 & \textbf{0.767} & 0.904 & 0.640 & \textbf{0.017} & \textbf{7328.1} & 0.228 \\
& Discuss   & 0.767 & 0.932 & 0.730 & 0.717 & 6266.6 & \cellcolor{yellow!25}\textbf{0.001} & \textbf{0.767} & \textbf{0.911} & \cellcolor{yellow!25}\textbf{0.652} & 0.021 & 7619.3 &\textbf{0.177} \\
& Leader    & 0.784 & 0.934 & \cellcolor{yellow!25}\textbf{0.752} & \cellcolor{yellow!25}\textbf{0.883} & 6558.6 & 0.009 & \textbf{0.767} & 0.870 & 0.639 & 0.023 & 8085.5 & 0.223 \\
\midrule
\multicolumn{14}{c}{\textit{\textbf{Open-source MLLMs}}} \\
\midrule
\multirow{4}{*}{Gemma4-31B}
& Broadcast & 0.833 & 0.945 & 0.689 & 0.794 & 5381.8 & 0.019 & 0.823 & 0.904 & 0.620 & 0.038 & 8042.4 & 0.322 \\
& Discuss   & 0.802 & 0.855 & 0.638 & 0.731 & 6852.5 & \cellcolor{yellow!25}\textbf{0.001} & 0.812 & 0.922 & 0.625 & 0.025 & 8978.3 & 0.224 \\
& Leader    & \textbf{0.865} & \cellcolor{yellow!25}\textbf{0.964} & \textbf{0.736} & \textbf{0.840} & 6863.0 & 0.003 & \cellcolor{yellow!25}\textbf{0.854} & \cellcolor{yellow!25}\textbf{0.963} & \cellcolor{yellow!25}\textbf{0.652} & \cellcolor{yellow!25}\textbf{0.016} & 8723.5 & \cellcolor{yellow!25}\textbf{0.138} \\
& No Comm   & 0.719 & 0.863 & 0.558 & 0.706 & \textbf{3577.4} & 0.023 & 0.177 & 0.464 & 0.308 & 0.297 & \textbf{4539.7} & 0.486 \\
\midrule
\multirow{4}{*}{Qwen3-32B-VL}
& Broadcast & 0.802 & 0.934 & 0.694 & 0.802 & 4682.0 & 0.020 & 0.365 & 0.620 & 0.393 & 0.166 & 6749.4 & \textbf{0.294} \\
& Discuss   & 0.808 & 0.929 & 0.694 & 0.818 & 6351.7 & \textbf{0.003} & 0.333 & 0.583 & 0.374 & 0.100 & 8324.3 & 0.304 \\
& Leader    & \textbf{0.833} & \textbf{0.938} & \textbf{0.706} & \textbf{0.855} & 5964.7 & 0.012 & \textbf{0.396} & \textbf{0.622} & \textbf{0.408} & \textbf{0.046} & 7516.5 & 0.378 \\
& No Comm   & 0.583 & 0.802 & 0.567 & 0.629 & \cellcolor{yellow!25}\textbf{3017.7} & 0.028 & 0.073 & 0.325 & 0.232 & 0.451 & \textbf{3559.0} & 0.733 \\
\midrule
\multirow{4}{*}{Qwen3-8B-VL}
& Broadcast & 0.458 & 0.770 & 0.535 & 0.676 & 4585.4 & 0.010 & \textbf{0.073} & \textbf{0.332} & 0.207 & 0.312 & 6788.2 & 0.662 \\
& Discuss   & 0.458 & 0.783 & 0.494 & \textbf{0.702} & 7206.1 & 0.011 & 0.021 & 0.283 & 0.195 & 0.319 & 8934.5 & \textbf{0.601} \\
& Leader    & \textbf{0.490} & \textbf{0.785} & \textbf{0.548} & 0.642 & 5701.5 & \textbf{0.005} & 0.052 & \textbf{0.332} & \textbf{0.227} & \textbf{0.223} & 7161.4 & 0.658 \\
& No Comm   & 0.274 & 0.552 & 0.342 & 0.357 & \textbf{3310.1} & 0.017 & 0.042 & 0.182 & 0.132 & 0.613 & \cellcolor{yellow!25}\textbf{3475.1} & NA \\
\bottomrule
\end{tabular}
\end{table*}

\begin{figure*}[!t]
    \centering
    \begin{minipage}[t]{0.64\linewidth}
        \centering
        \includegraphics[width=\linewidth]{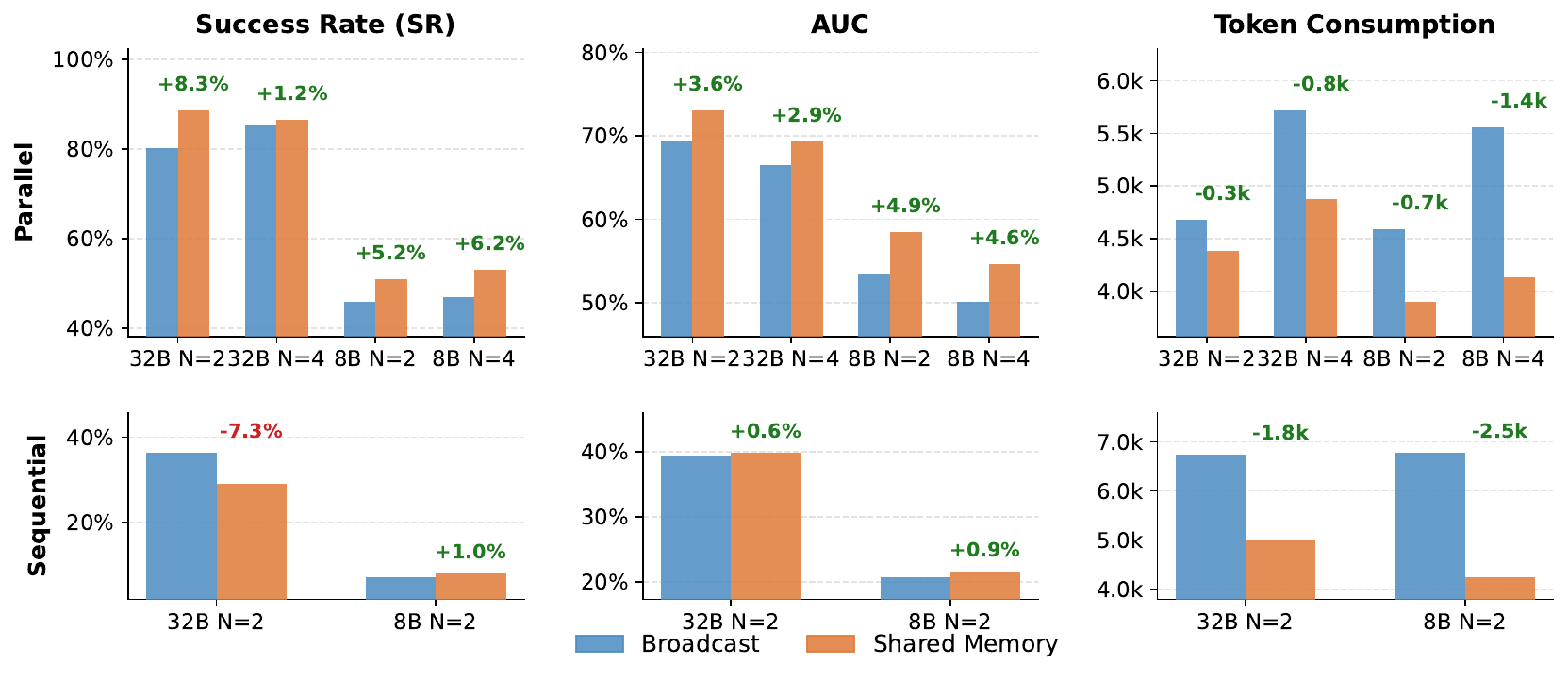}
        \caption{Comparison of explicit broadcast and implicit shared memory protocol. The models are from the Qwen3-VL family.}
        \label{fig:d4}
    \end{minipage}
    \hfill
    \begin{minipage}[t]{0.34\linewidth}
        \centering
        \includegraphics[width=\linewidth]{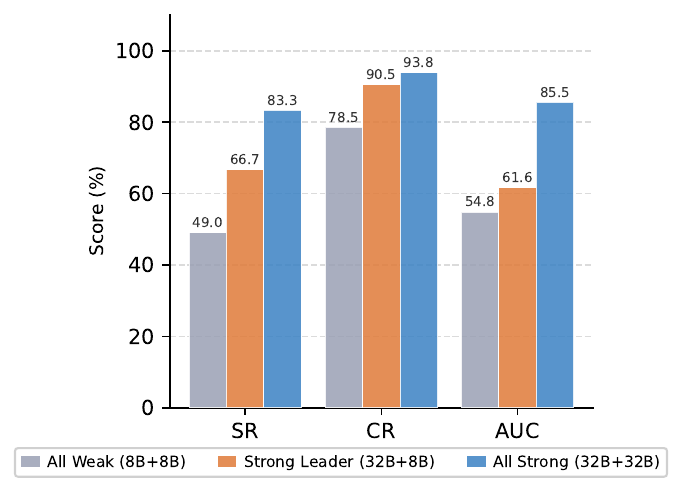}
        \caption{Effect of leader capability in centralized collaboration. The models are from the Qwen3-VL family.}
        \label{fig:strong-leader}
    \end{minipage}
    \vspace{-12pt}
\end{figure*}

\begin{figure}[!t]
    \centering
    \includegraphics[width=0.95\linewidth]{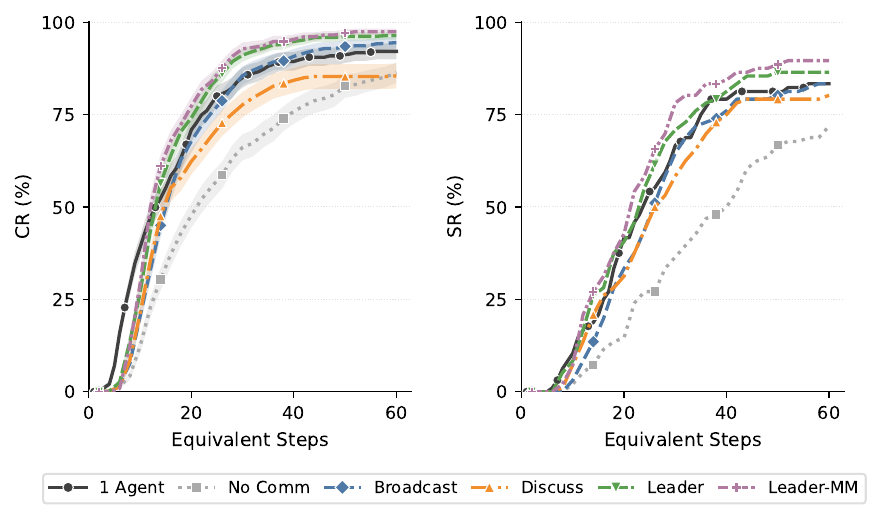}
    \caption{Task progress over equivalent steps under different communication protocols on parallel tasks.
Results are obtained with Gemma4-31B.
Leader-MM denotes the vision-augmented leadership protocol.}
    \label{fig:comm-curve}
    \vspace{-12pt}
\end{figure}

Table~\ref{tab:mode_radar_data} gives the detailed results of Figure~\ref{fig:comm-mode-analysis}(a). The results show that centralized collaboration generally improves division of labor and task efficiency, reflected by higher DOL and AUC. In contrast, the discuss protocol is more effective at reducing conflicts, leading to lower CAR. For weaker models, centralized collaboration also helps reduce spatial cognition errors in sequential tasks, suggesting that leader-based coordination can provide useful guidance when agents have limited planning or spatial reasoning ability. Figure~\ref{fig:comm-curve} shows the comparison of task progress under single agent and different communication protocol in two-agent team.

\paragraph{Shared Memory.} Figure~\ref{fig:d4} reports the absolute values of success rate, AUC, and token consumption for broadcast and shared memory, providing a more detailed version of Figure~\ref{fig:3-1-shared-memory}.

\begin{table*}[!t]
\centering
\small
\caption{Results under different prior-information settings on parallel tasks. Full prior includes both object appearance and location information. Without appearance removes object appearance description, without location removes prior location information, and no prior removes both.}
\label{tab:parallel_prior_ablation}
\setlength{\tabcolsep}{3pt}
\begin{tabular}{l ccc ccc ccc ccc}
\toprule
\textbf{Model}
& \multicolumn{3}{c}{\cellcolor{blue!10}\textbf{Full Prior}}
& \multicolumn{3}{c}{\cellcolor{orange!15}\textbf{Without Appearance}}
& \multicolumn{3}{c}{\cellcolor{green!10}\textbf{Without Location}}
& \multicolumn{3}{c}{\cellcolor{gray!15}\textbf{No Prior}} \\
\cmidrule(lr){2-4} \cmidrule(lr){5-7} \cmidrule(lr){8-10} \cmidrule(lr){11-13}
& \textbf{SR} & \textbf{CR} & \textbf{AUC}
& \textbf{SR} & \textbf{CR} & \textbf{AUC}
& \textbf{SR} & \textbf{CR} & \textbf{AUC}
& \textbf{SR} & \textbf{CR} & \textbf{AUC} \\
\midrule
GPT-5-mini
& 0.917 & 0.982 & 0.776
& 0.865 & 0.955 & 0.730
& 0.344 & 0.691 & 0.436
& 0.271 & 0.609 & 0.431 \\
Gemma4-31B
& 0.833 & 0.945 & 0.688
& 0.833 & 0.937 & 0.662
& 0.229 & 0.589 & 0.347
& 0.302 & 0.570 & 0.356 \\
Qwen3-32B-VL
& 0.802 & 0.934 & 0.694
& 0.802 & 0.923 & 0.655
& 0.260 & 0.591 & 0.365
& 0.344 & 0.688 & 0.456 \\
\bottomrule
\end{tabular}
\end{table*}

\begin{table}[t]
\centering
\small
\caption{Results on parallel tasks without location information. Each cell reports SR/CR/AUC. Qwen-* is short for Qwen3-VL, Inter-* is short for InterVL3.5.}
\label{tab:parallel_no_location}
\setlength{\tabcolsep}{3pt}
\begin{tabular}{l cc}
\toprule
\textbf{Model}
& \cellcolor{blue!10}\textbf{1-Agent}
& \cellcolor{orange!15}\textbf{2-Agent} \\
\midrule
\multicolumn{3}{c}{\textit{Closed-source MLLMs}} \\
\midrule
GPT-5-mini        
& \textbf{0.323 / 0.630 / 0.436}
& \textbf{0.344 / 0.691 / 0.436} \\
\midrule
\multicolumn{3}{c}{\textit{Open-source MLLMs}} \\
\midrule
Qwen-8B       
& 0.031 / 0.162 / 0.103
& 0.021 / 0.199 / 0.101 \\
Qwen-32B      
& 0.219 / 0.520 / 0.344
& 0.260 / 0.591 / 0.365 \\
Qwen-235B     
& 0.115 / 0.389 / 0.264
& 0.271 / 0.581 / 0.332 \\
Qwen3.5-9B        
& \underline{0.292 / 0.582 / 0.389}
& \underline{0.292 / 0.649 / 0.435} \\
Gemma-31B        
& 0.240 / 0.561 / 0.358
& 0.229 / 0.589 / 0.347 \\
Intern-38B  
& 0.010 / 0.146 / 0.087
& 0.073 / 0.253 / 0.144 \\
Intern-241B 
& 0.021 / 0.201 / 0.113
& 0.062 / 0.217 / 0.114 \\
Llama-4           
& 0.021 / 0.089 / 0.058
& 0.000 / 0.107 / 0.056 \\
\bottomrule
\end{tabular}
\end{table}

\subsection{Mix Team: Strong Model as Leader.}
\label{app:mix-team}
We further investigate the role of leader in centralized mode. As shown in Figure~\ref{fig:strong-leader}, using a stronger leader improves the mixed team, but does not make it approach the all-strong setting. The mixed team nearly matches the all-strong team in CR, but its SR is only slightly above the average of the all-weak and all-strong teams, and its AUC is even below this average. This indicates that a strong leader mainly improves task coverage, but efficient and overall success still depends heavily on the worker's capability.

\subsection{Performance Under Imperfect Information.}

\begin{figure}[!t]
\setlength{\abovecaptionskip}{0pt}
    \centering
        \includegraphics[width=\linewidth]{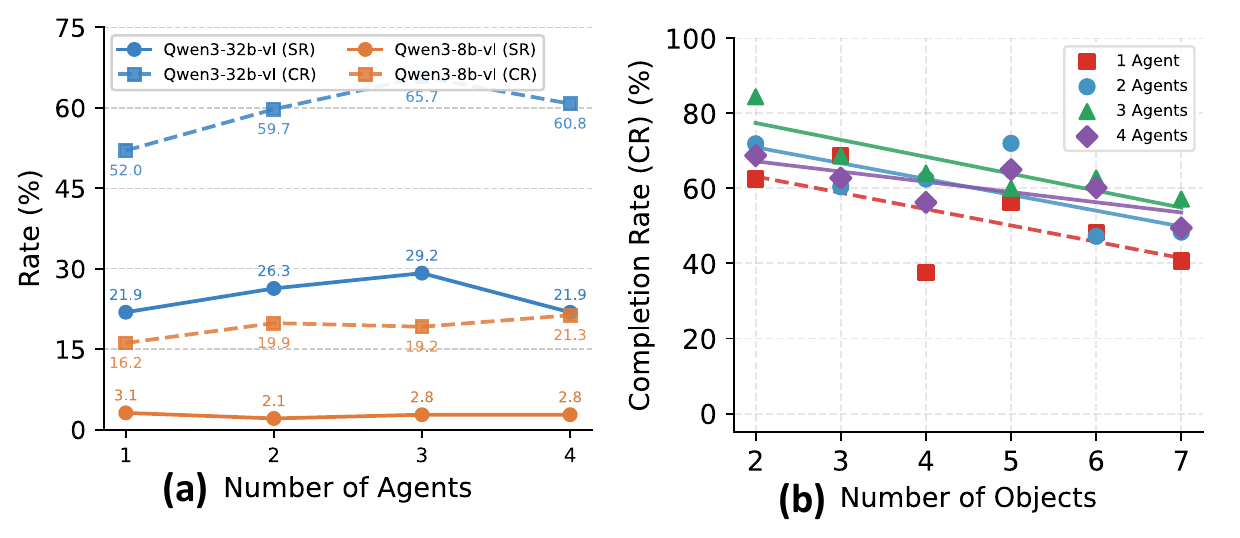}
    \caption{Team size scaling effect without location prior. (a) SR and CR change as the number of agents increases. (b) SR versus the number of target objects, with fitted trend lines for different team sizes.}
    \label{fig:app-noloc-analysis}
    \vspace{-8pt}
\end{figure}

\paragraph{Missing location priors} Table~\ref{tab:parallel_no_location} reports detailed results under the no-prior-location setting on parallel tasks across different models. Figure~\ref{fig:app-noloc-analysis} shows the team size scaling effect under stronger exploration demands, where prior location information is removed. Compared with the full-information setting, Qwen3-32B-VL exhibits a similar but weaker inverted-U-shaped trend, suggesting that moderate team scaling remains helpful. For Qwen3-8B-VL, SR remains low across team sizes, indicating a capability bottleneck. Moreover, when grouping cases by object count, larger teams still show a milder performance drop.
More detailed results in Table~\ref{tab:qwen3_parallel_no_location_agents}.

\begin{table}[!h]
\centering
\small
\caption{Results on Parallel tasks without location prior for Qwen3-VL models with one to four agents, under the broadcast protocol. The best results of each model is highlighted in bold.}
\label{tab:qwen3_parallel_no_location_agents}
\setlength{\tabcolsep}{4pt}
\begin{tabular}{lcccc}
\toprule
\textbf{Model} & \textbf{\#Agents} & \textbf{SR} & \textbf{CR} & \textbf{AUC} \\
\midrule
Qwen3-8B-VL  & 1 & 0.031 & 0.162 & 0.103 \\
             & 2 & 0.021 & 0.199 & 0.101 \\
             & 3 & \textbf{0.094} & 0.298 & \textbf{0.168} \\
             & 4 & 0.062 & \textbf{0.307} & 0.163 \\
\midrule
Qwen3-32B-VL & 1 & 0.219 & 0.520 & 0.344 \\
             & 2 & 0.260 & 0.591 & 0.365 \\
             & 3 & \textbf{0.292} & \textbf{0.657} & \textbf{0.404} \\
             & 4 & 0.219 & 0.608 & 0.374 \\
\bottomrule
\end{tabular}
\end{table}

\begin{figure}[!t]
    \centering
    \includegraphics[width=0.8\linewidth]{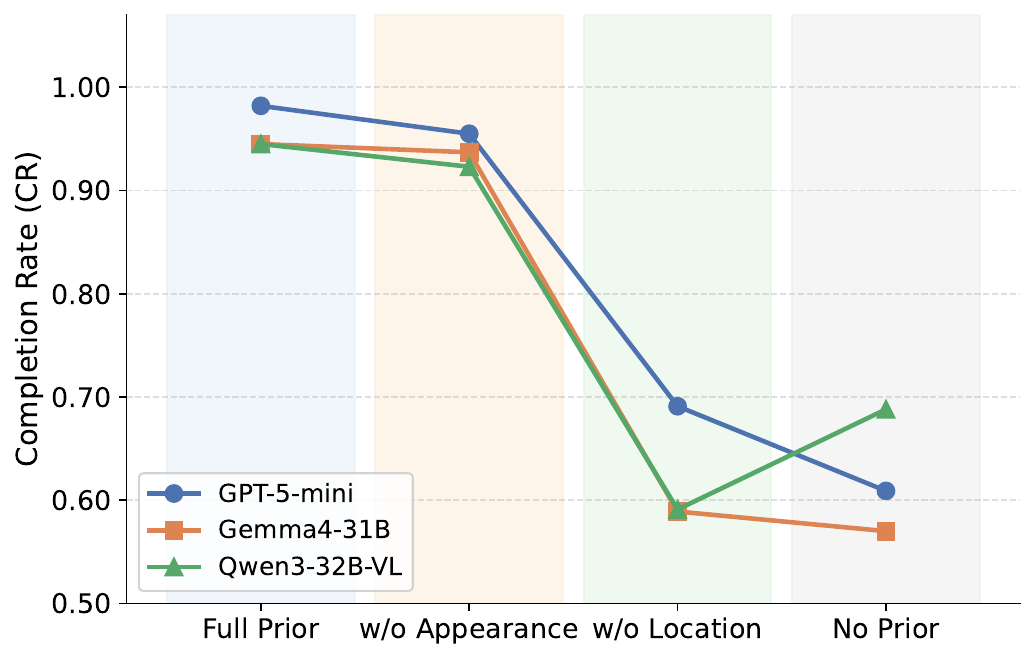}
    \caption{Completion rate under different prior-information settings on parallel tasks.}
    \label{fig:info_cr}
    \vspace{-8pt}
\end{figure}

\begin{figure*}[!t]
    \centering
    \includegraphics[width=0.9\linewidth]{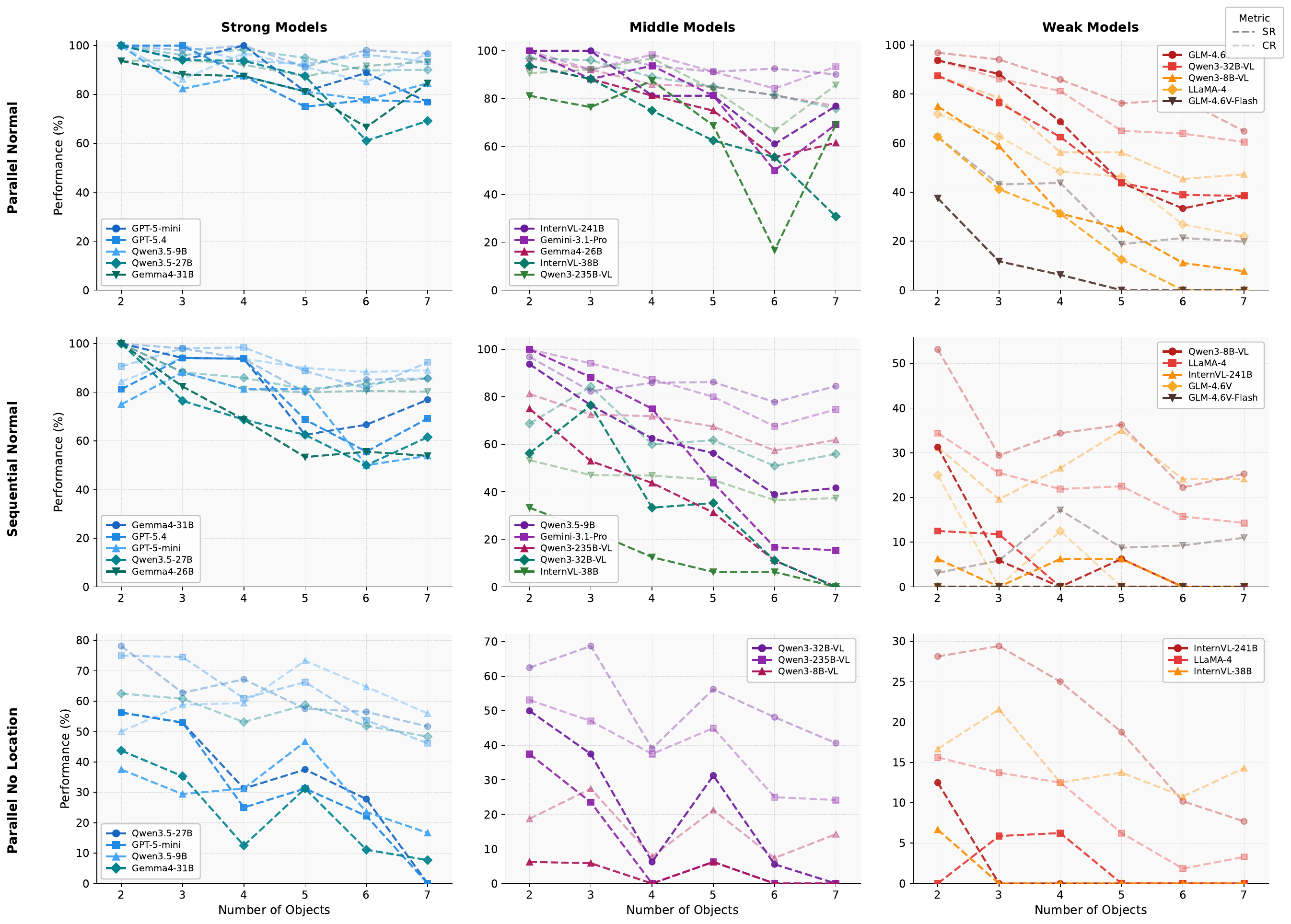}
    \caption{Performance by number of objects across task conditions and model groups.}
    \label{fig:all-task-obj}
    \vspace{-8pt}
\end{figure*}

\paragraph{Ablation on prior information}
\label{para:info-ablation}
To investigate how different types of prior information affect agent performance, we conduct an ablation study by removing object appearance descriptions and prior location information. As shown in figure~\ref{fig:info_cr}, prior location information is critical for task completion, as it helps agents quickly locate target objects and reduces inefficient exploration. Removing location information leads to a substantial performance drop across all models. In contrast, the effect of appearance descriptions is model-dependent: GPT-5-mini benefits from appearance cues, Gemma4-31B shows only minor changes, and Qwen3-32B-VL performs worse with appearance-only priors than with no prior. A possible reason is that models with weaker fine-grained visual grounding are more susceptible to misleading appearance descriptions and may confuse visually similar objects. Detailed results are reported in Table~\ref{tab:parallel_prior_ablation}.

\begin{table}[!h]
\centering
\small
\caption{Results on Parallel tasks with noise location prior cross models.}
\newcommand{\graycell}[1]{\cellcolor{gray!10}#1}
\label{tab:parallel_noise_agents}
\setlength{\tabcolsep}{4pt}
\begin{tabular}{lcccc}
\toprule
\textbf{Model} & \textbf{\#Agents} & \textbf{SR} & \textbf{CR} & \textbf{AUC} \\
\midrule
\multirow{2}{*}{GPT-5-mini}
 & 1 & 0.760 & 0.902 & 0.719 \\
 & \graycell{2} & \graycell{0.823} & \graycell{0.950} & \graycell{0.709} \\

\multirow{2}{*}{Qwen3.5-9B}
 & 1 & 0.781 & 0.911 & 0.665 \\
 & \graycell{2} & \graycell{0.812} & \graycell{0.903} & \graycell{0.663} \\

\multirow{2}{*}{Qwen3-32B-VL}
 & 1 & 0.542 & 0.750 & 0.526 \\
 & \graycell{2} & \graycell{0.698} & \graycell{0.878} & \graycell{0.620} \\
\bottomrule
\end{tabular}
\end{table}

\paragraph{Noisy location priors.}
Table~\ref{tab:parallel_noise_agents} reports detailed results on parallel tasks where 30\% of the prior location information is noisy, comparing single-agent and two-agent settings.

\subsection{Additional Performance Analysis}
\label{app:obj-num}
Figure~\ref{fig:all-task-obj} reports performance by the number of target objects across model groups. As each target object corresponds to a subgoal, a larger number of target objects indicates higher task complexity. Overall, both completion rate and success rate decline as task complexity increases.

\section{Emergent collaborative behavior}
By analyzing the communication and action trajectories under the basic broadcast protocol, we identify several emergent collaborative behavior patterns. In addition to the analysis in Section~\ref{sec:how-agent-collab}, this section provides representative examples and further details.
\begin{table*}[t]
\centering
\small
\setlength{\tabcolsep}{5pt}
\renewcommand{\arraystretch}{1.15}
\caption{Examples of collaborative behaviors emerge in the agent interactions in MECoBench under broadcast protocol.}
\label{tab:collab_behavior_examples}
\begin{tabularx}{\textwidth}{@{}p{2.0cm} p{1.9cm} X@{}}
\toprule
\textbf{Category} & \textbf{Behavior} & \textbf{Example} \\
\midrule
\multirow{3}{2.0cm}{\parbox{3.0cm}{\textit{Information \\Sharing}}}
& \textit{Info-Plan}
& ``\hl{I will grab the two puddings and check the fridge for the wine.}
  Could you go to the living room to find the cupcake?'' \\[2pt]
& \textit{Info-Loc}
& ``\hl{I am in the bathroom} and will head to the kitchen to look for the
  pudding and cupcake.'' \\[2pt]
& \textit{Info-Obj}
& ``\hl{I see the cupcake on the desk.} I will grab it and then head to the
  living room.'' \\
\cmidrule(l){1-3}
\multirow{2}{2.0cm}{\textit{Task \\Coordination}}
& \textit{Act-Del}
& ``\hl{Please go to the kitchen to grab the second apple,} and then we can
  meet at the coffee table.'' \\[2pt]
& \textit{Task-Assign}
& ``\hl{I'll handle the kitchen and the bathroom.}'' \\
\cmidrule(l){1-3}
\multirow{4}{2.0cm}{\textit{Alignment \&\\Correction}}
& \textit{Coord-Q.}
& \textbf{State confirmation:}
  ``\hl{Do you have the cupcake and the second apple?} If not, please check
  the fridge.''\\[3pt]
& & \textbf{Location confirmation:}
  ``I'm at door~287 holding the cupcake. \hl{Are you at door~287 and ready
  to receive?} Please confirm.'' \\[4pt]
& \textit{Correct.}
& \textbf{Self-correction:}
  ``\hl{I accidentally grabbed a waterglass instead of the wineglass.}
  I will put it back and then get the wineglass.''\\[3pt]
& & \textbf{Peer-correction:}
  ``Agent~0, \hl{I see you're holding milk, but we need the juice.}
  Please drop the milk and grab the correct juice from the kitchen table.'' \\
\bottomrule
\end{tabularx}
\end{table*}

\subsection{Examples}
\label{app:4.3}
Table~\ref{tab:collab_behavior_examples} presents representative examples of collaborative behaviors, with the relevant text spans highlighted in yellow.

\subsection{Detailed Analysis}
\label{app:4.3-1}
\begin{table}[h]
\centering
\footnotesize
\setlength{\tabcolsep}{3.2pt}
\renewcommand{\arraystretch}{1.12}
\caption{Coordination queries are associated with harder coordination
states. \textit{Q.} denotes coordination query.}
\label{tab:coord_q_analysis}
\begin{tabular}{llccccc}
\toprule
\textbf{Task} & \textbf{Group} & \textbf{\#Case}
& \textbf{Avg.} & \textbf{SR} & \textbf{Stuck}
& \textbf{Neg.} \\
& & & \textbf{\#Obj.} & \textbf{(\%)}
& \textbf{(\%)} & \textbf{(\%)} \\
\midrule
Parallel
& w/ Q.  & 370  & 4.70 & 64.6 & 8.4 & 11.1 \\
& w/o Q. & 1727 & 4.35 & 74.7 & 2.0 & 4.5  \\
\midrule
Sequential
& w/ Q.  & 591 & 4.51 & 32.3 & 3.9 & 6.6 \\
& w/o Q. & 627 & 4.30 & 47.8 & 1.6 & 3.2 \\
\bottomrule
\end{tabular}
\end{table}

\begin{table}[!t]
\centering
\small
\setlength{\tabcolsep}{3.2pt}
\renewcommand{\arraystretch}{1.08}
\caption{Overall and wrong-grab-conditional effects of correction.
\textit{Corr.} and \textit{WG} denote correction and wrong grab.
\textit{WG-Cond}. denotes cases where at least one wrong grab happened.}
\label{tab:correction_overall}
\begin{tabular}{lllrrr}
\toprule
\textbf{Task} & \textbf{Cond.} & \textbf{Group}
& \textbf{SR} & \textbf{CR} & \textbf{\#WG} \\
& & & \textbf{(\%)} & \textbf{(\%)} & \\
\midrule

\multirow{7}{*}{Parallel}
& \multirow{3}{*}{Overall}
& w/ Corr.  & 67.3 & 90.1 & 3.05 \\
&& w/o Corr. & 74.1 & 88.5 & 1.37 \\
&& $\Delta$  & $-6.7$ & $+1.6$ & -- \\
\cmidrule(lr){2-6}
& \multirow{4}{*}{WG}
& w/ Corr.  & 61.1 & 88.3 & 3.79 \\
&& w/o Corr. & 66.2 & 87.0 & 2.84 \\
&& Raw $\Delta$ & $-5.1$ & $+1.3$ & -- \\
&& Adj. $\Delta$ & $\mathbf{+2.4}$ & $\mathbf{+3.5}$ & -- \\

\midrule

\multirow{7}{*}{Sequential}
& \multirow{3}{*}{Overall}
& w/ Corr.  & 29.8 & 67.4 & 2.68 \\
&& w/o Corr. & 41.1 & 59.8 & 0.76 \\
&& $\Delta$  & $-11.3$ & $+7.7$ & -- \\
\cmidrule(lr){2-6}
& \multirow{4}{*}{WG}
& w/ Corr.  & 26.9 & 67.2 & 4.33 \\
&& w/o Corr. & 21.1 & 53.4 & 3.02 \\
&& Raw $\Delta$ & $+5.8$ & $+13.8$ & -- \\
&& Adj. $\Delta$ & $\mathbf{+11.8}$ & $\mathbf{+15.4}$ & -- \\

\bottomrule
\end{tabular}
\vspace{-8pt}
\end{table}

\paragraph{Coordination Queries Indicate Collaboration Challenges.}
In the main text (Table~\ref{tab:micro_behavior}), coordination query correlates negatively with SR and CR. To further investigate the underlying cause, we introduce two behavioral indicators of collaboration difficulty: negative report (\textit{Neg.}), where an agent reports failing to locate a target object at a specific location, and \textit{stuck}, where an agent reports being stuck in middle.
As shown in Table~\ref{tab:coord_q_analysis}, cases with coordination query involve more objects on average and exhibit substantially higher stuck and negative report rate. This pattern holds across both task types, suggesting that coordination query serves as a reactive signal to genuine collaboration obstacles.

\begin{table}[t]
\centering
\small
\setlength{\tabcolsep}{2.6pt}
\renewcommand{\arraystretch}{1.08}
\caption{Correction effects in parallel tasks stratified by the
number of wrong grabs.}
\label{tab:parallel_correction_strata}
\begin{tabular}{clrrrr}
\toprule
\textbf{\#WG} & \textbf{Group} & \textbf{\#Case}
& \textbf{SR} & \textbf{CR}
& \textbf{$\Delta$SR/CR} \\
& & & \textbf{(\%)} & \textbf{(\%)} & \\
\midrule
\multirow{2}{*}{1}
& Corr.    & 70  & 82.9 & 95.2 & \multirow{2}{*}{$+2.2/+3.7$} \\
& No Corr. & 351 & 80.6 & 91.5 & \\
\midrule
\multirow{2}{*}{2}
& Corr.    & 53  & 69.8 & 89.3 & \multirow{2}{*}{$-6.2/-1.5$} \\
& No Corr. & 200 & 76.0 & 90.8 & \\
\midrule
\multirow{2}{*}{3}
& Corr.    & 44 & 68.2 & 91.9 & \multirow{2}{*}{$+6.9/+4.1$} \\
& No Corr. & 80 & 61.3 & 87.8 & \\
\midrule
\multirow{2}{*}{$4+$}
& Corr.    & 118 & 41.5 & 82.4 & \multirow{2}{*}{$+7.4/+6.8$} \\
& No Corr. & 208 & 34.1 & 75.6 & \\
\midrule
\multicolumn{3}{l}{Adjusted effect}
& \multicolumn{2}{r}{$\Delta$SR/CR}
& $\mathbf{+2.4/+3.5}$ \\
\bottomrule
\end{tabular}
\end{table}

\begin{table*}[t]
\centering
\small
\setlength{\tabcolsep}{6pt}
\renewcommand{\arraystretch}{1.3}
\caption{Representative examples of multi-agent collaboration failure modes.}
\label{tab:failure_mode_examples}
\begin{tabularx}{\textwidth}{p{3.0cm} p{2.5cm} X}
\toprule
\textbf{Failure Mode} & \textbf{Subtype} & \textbf{Representative Trace Excerpt} \\
\midrule

\textbf{Agent-induced Conflict}
  & Duplicate Grab
  & \textbf{Agent~1:} \textit{``I found the wine on the coffee table.
      I will grab it and bring it to the living room.''}
    \newline\smallskip
    \textbf{Agent~4:} \textit{``I am grabbing the wine from the coffee table.
      I will take it to the living room.''}
    \newline\smallskip
    $\Rightarrow$ Both agents execute \texttt{[grab]~<wine>}.
  \\[2pt]

\midrule

\multirow{2}{3.0cm}{\raggedright\textbf{Hallucinated Belief Propagation}}
  & False Completion
  & \textit{``All items have been successfully placed on the bathroom counter.
      The task is complete.''}
    \newline\smallskip
    $\Rightarrow$ All agents execute \texttt{[wait]} until the end.
  \\[2pt]

\cmidrule(lr){2-3}

  & Object Misidentification
  & \textbf{Agent~3:} \textit{``I have found and will now grab
      the \textbf{wineglass} from the bedroom cabinet.''}
    \newline\smallskip
    Agent~3 executes \texttt{[grab]~<dishbowl>} (misidentifying dishbowl as wineglass),
    then \texttt{[put]~<dishbowl>} on the kitchentable.
    \newline\smallskip
    \textbf{Agent~0 \& 2:} \textit{``Agent~3 has already grabbed the
      wineglass---I will focus on the plate / waterglass.''}
    \newline\smallskip
    $\Rightarrow$ All agents stop searching for the actual wineglass.
    Task ends with \texttt{wineglass} unsatisfied.
  \\

\bottomrule
\end{tabularx}
\vspace{-8pt}
\end{table*}

\paragraph{Correction Helps Recover from Errors}
Section~\ref{sec:how-agent-collab} shows that correction correlates negatively with SR but positively with CR, suggesting a potential role in error recovery. To examine this effect, we analyze its co-occurrence with wrong grabs (WGs), defined as actions that grasp task-irrelevant objects. Table~\ref{tab:correction_overall} shows that episodes with correction contain substantially more WGs than those without, indicating that correction is mainly triggered by execution errors. Since WG episodes with correction also exhibit greater error severity, direct comparison may underestimate its recovery benefit. We therefore stratify WG episodes by error count and compute a weighted adjusted effect using shared stratum weights for both groups (Table~\ref{tab:parallel_correction_strata}). After adjustment, correction consistently improves both SR and CR across task types, confirming its effectiveness in error recovery. Stratum-level results further show larger gains under higher WG counts, suggesting that correction is especially useful when errors are more severe.

\section{Failure Mode Analysis}
\label{app:fail}
In this section, we provide additional examples and a more detailed analysis of the multi-agent collaboration failure modes presented in Section~\ref{sec:fail}.

\subsection{Examples}
Table~\ref{tab:failure_mode_examples} presents representative examples of different failure modes.

\subsection{Detailed Analysis}

\begin{table}[!h]
\setlength{\abovecaptionskip}{2pt}
\centering
\small
\setlength{\tabcolsep}{3pt}
\renewcommand{\arraystretch}{1.08}
\caption{\textbf{Failure mode in collaboration.} \emph{Freq.} denotes the fraction of episodes containing the behavior; \emph{lift} is the ratio of its rate in failed episodes to that in successful episodes.}
\label{tab:failure-behaviors}
\begin{tabular}{llrrrr}
\toprule
\textbf{Behavior} & \textbf{Task} & \textbf{Freq.} & \textbf{Lift} & \textbf{SR w/} & \textbf{$\Delta$SR} \\
\midrule
Hallucinated & Parallel   & 1.0\%  & $\infty$     &  0.0\% & $-$73.7 \\
completion   & Sequential & 1.5\%  & $\infty$     &  0.0\% & $-$40.9  \\
\midrule
Grab conflict & Parallel  & 22.9\% &  1.5$\times$ & 71.4\% &  $-$9.0 \\
\bottomrule
\end{tabular}
\vspace{-8pt}
\end{table}

\begin{figure}[h]
    \centering
    \includegraphics[width=1\linewidth]{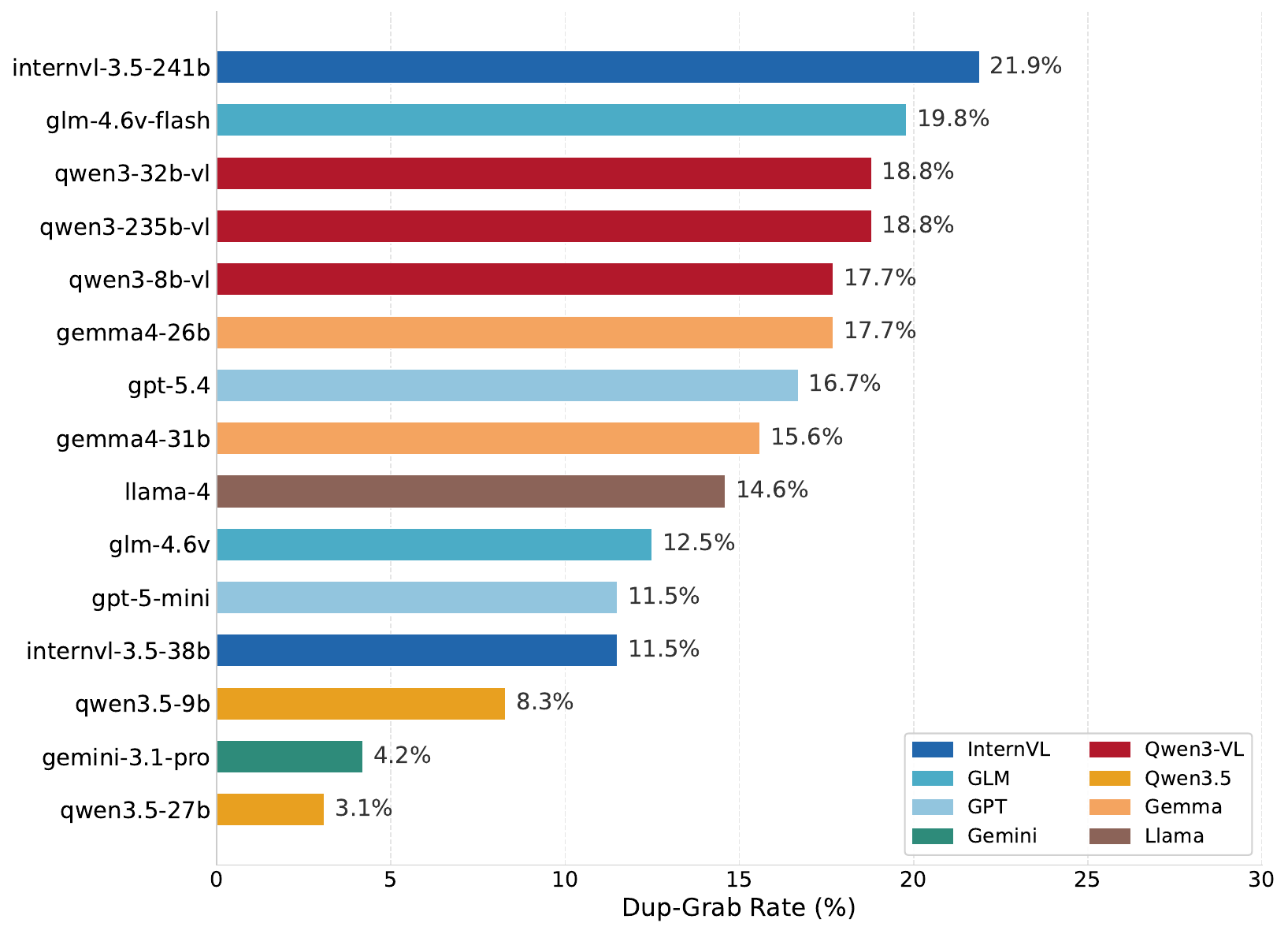}
    \caption{Duplicate grab rate of different models in parallel task under 2-agent team and broadcast protocol.}
    \label{fig:2-agent-grab}
\end{figure}
Table~\ref{tab:failure-behaviors} summarizes the occurrence frequency of different failure modes and their impact on success rate (SR), aggregated over all models and team sizes under the broadcast protocol. Figure~\ref{fig:2-agent-grab} compares duplicate-grab rates across models in two-agent teams.

\end{document}